\documentclass[12pt]{article}
\usepackage{amsfonts}
\usepackage[mathscr]{eucal}
\usepackage[dvips]{color}

\newcommand{\be}{\begin{equation}}
\newcommand{\ee}{\end{equation}}
\newcommand{\bea}{\begin{eqnarray}}
\newcommand{\eea}{\end{eqnarray}}
\newcommand{\ba}{\begin{array}}
\newcommand{\ea}{\end{array}}
\newcommand{\nn}{\nonumber \\}

\newcommand{\lb}{\label}
\newcommand{\p}[1]{(\ref{#1})}

\newcommand{\bpsi}{{\bar\psi}{}}
\newcommand{\e}{\epsilon}
\newcommand{\eps}{\varepsilon}

\newcommand{\R}{\mathbb R}

\def\pa{\partial}

\def\sfrac#1#2{{\textstyle\frac{#1}{#2}}}
\def\={\ =\ }
\def\und{\qquad\textrm{and}\qquad}
\def\im{{\rm i}}
\def\ep{{\rm e}}

\begin{document}

\begin{titlepage}
\begin{flushright}
ITP-UH-11/11\\ JINR E2-2011-125
\end{flushright}

\vspace*{2cm}

\begin{center}
\renewcommand{\thefootnote}{$\star$}

{\LARGE\bf Superconformal mechanics~\footnote{
\,\,Invited review by Journal of Physics A: Mathematical and Theoretical}}

\vspace{2cm}
\renewcommand{\thefootnote}{$\times$}

{\large\bf Sergey~Fedoruk}${\,}^1$\footnote{
\,\,On leave of absence from V.N.\,Karazin Kharkov National University, Ukraine},\,\,\,
{\large\bf Evgeny~Ivanov}${\,}^1$,\,\,\,
{\large\bf Olaf~Lechtenfeld}${\,}^2$ \vspace{1cm}

${}^1${\it Bogoliubov  Laboratory of Theoretical Physics, JINR,}\\
{\it 141980 Dubna, Moscow region, Russia} \\
\vspace{0.1cm}

{\tt fedoruk,eivanov@theor.jinr.ru}\\
\vspace{0.5cm}

${}^2${\it Institut f\"ur Theoretische Physik,
Leibniz Universit\"at Hannover,}\\
{\it Appelstra{\ss}e 2, D-30167 Hannover, Germany} \\
\vspace{0.1cm}

{\tt lechtenf@itp.uni-hannover.de}\\
\vspace{0.3cm} \setcounter{footnote}{0}

\end{center}
\vspace{0.2cm} \vskip 0.6truecm  \nopagebreak

\begin{abstract}
\noindent
We survey the salient features and problems of conformal and superconformal
mechanics and portray some of its developments over the past decade.
Both classical and quantum issues of single- and multiparticle systems are covered.
\end{abstract}

\vspace{5cm}

{\small

\phantom{Keywords: }multi-particle models
}

\newpage
\end{titlepage}

\vspace*{1.5cm}
\tableofcontents

\newpage

\setcounter{footnote}{0}
\setcounter{equation}0

\section{Introduction}

The stable interest in conformal mechanics
and its various superconformal extensions
is caused by several interconnected reasons.

Superconformal mechanics models are characterized by a simple
non-compact non-Abelian group of dynamical symmetry, namely ${\rm SL}(2,\mathbb{R})$
or one of its supersymmetric extensions. Although the unitary representations of
unimodular real groups were constructed by Bargmann~\cite{Barg} in
the forties, physical applications of these group-theoretical results in
(super)conformal mechanics models became apparent only much later.

One of the first descriptions of a simple conformal mechanical
system (with an additional oscillator potential) appears already in
the textbook of Landau and Lifshitz~\cite{LanLif}. Later on, at the
end of the sixties and in the seventies, interest in models with
$d{=}1$ conformal symmetry was supported from two independent
sources. First, conformal symmetry characterizes an important class
of integrable many-particle systems discovered by Calogero in his
pioneering papers~\cite{C1,C2}. Second, models of conformal
mechanics were studied as examples of one-dimensional ($d{=}1$)
field theories. One of the forerunners in this direction
was~\cite{BBFFT}.\footnote{See also~\cite{Kastrup}.} However, the beginning of concrete studies and
applications of conformal mechanics, at the classical and the
quantum level, is marked by the seminal paper of de~Alvaro, Fubini
and Furlan~\cite{AFF}. Their model can be recovered from the
two-particle Calogero model~\cite{C1,C2} after eliminating the
center-of-mass coordinate, establishing a link between these two
early forays into applications of $d{=}1$ conformal symmetry.

The eighties brought supersymmetry to the models of conformal
mechanics. The models of superconformal mechanics are important particular
cases of supersymmetric quantum mechanics \cite{WIT}.
The papers~\cite{AP,FR} found the ${\cal N}{=}\,2$
supersymmetric generalization of the de~Alvaro--Fubini--Furlan
model. Two different (but closely related) ${\cal N}{=}\,4$
superconformal mechanics models based on ${\rm SU}(1,1|2)$ were
constructed in~\cite{FR,IKL2}. Approximately at the same time,
\cite{FM} constructed the ${\cal N}{=}\,2$ supersymmetric extension
of the multi-particle Calogero system.

The interest in conformal and superconformal mechanics rose higher in connection
with the AdS/CFT correspondence~\cite{Mal,GuKlPol,Wit}, when it was discovered that such models
describe (super)particles moving in near-horizon (AdS) geometries
of black-hole solutions to supergravities in diverse dimensions.
Namely, it was suggested in~\cite{CDKKTP} that the radial motion of a
massive charged particle near the horizon of an extremal
Reissner--Nordstr\"om black hole is described by some ``relativistic'' type of conformal
mechanics, which reduces to that of~\cite{AFF} in a ``non-relativistic'' limit.
The target variable of this conformal mechanics is the radial AdS coordinate
of an ${\rm AdS}_2\times S^2$ background. The latter is the bosonic body of
the maximally supersymmetric near-horizon extremal Reissner--Nordstr\"om
solution of ${\cal N}{=}\,2$ $d{=}4$ supergravity~\cite{CDKKTP}, with the
full isometry supergroup being ${\rm SU}(1,1|2)$. This observation led
\cite{CDKKTP} to conjecture that the full dynamics of a superparticle
in the near-horizon geometry of an extremal Reissner--Nordstr\"om black hole
is governed by ${\cal N}{=}\,4$ superconformal mechanics.

This tight relation to the AdS/CFT correspondence spurred further works
on ${\cal N}{=}\,4$ superconformal mechanics with ${\rm SU}(1,1|2)$ symmetry
at the end of the nineties.
Such models were constructed in the framework of nonlinear realizations
and transferred to the black-hole context in~\cite{AIPT}, partly
recreating earlier results of~\cite{IKL2}.
In~\cite{GT}, it was then argued that the large-$n$ limit of an $n$-particle
generalization of ${\rm SU}(1,1|2)$ superconformal mechanics may provide
a microscopic description of the near-horizon dynamics in a {\it multi-center\/}
extremal Reissner--Nordstr\"om black-hole geometry.
Further evidence in favor of the proposal of~\cite{CDKKTP} was adduced
in~\cite{IKNie,BGIK} by a canonical transformation linking
the radial motion of a (super)particle on ${\rm AdS}_2\times S^2$
to ${\cal N}{=}0\,$, ${\cal N}{=}2\,$~\cite{IKNie} and
${\cal N}{=}\,4$~\cite{BGIK} superconformal mechanics.

First attempts~\cite{W,BGK,BGL} to construct {\it multi-\/}particle systems with
${\cal N}{=}\,4$ superconformal invariance revealed a surprising rigidity, which
renders it a hard problem. In contrast to the ${\cal N}{=}\,2$ superconformal case
with a single prepotential~$U$, a second prepotential, $F$, appears. Both are
not only subject to homogeneity conditions, but must also solve a coupled system
of quadratic partial differential equations which are prominent in mathematical
physics: the WDVV equation (for~$F$)~\cite{WDVV1,WDVV2} and the associated
twisted-period equation (for~$U$)~\cite{Dub,FeiS}.
As documented in a number of works~\cite{GLP1,GLP2,GLP3,BKS1,KLP,LST,BKS2},
even taking the $F$~solution from the WDVV~literature,
it proved to be very tedious to explicitly solve the $U$~equation for more
than three particles in the case of ${\rm SU(1,1|2)}$ symmetry.

These technical difficulties are not the only reason to look beyond ${\rm SU}(1,1|2)$
to the most general ${\cal N}{=}\,4$ superconformal group in one dimension,
which is the exceptional one-parameter supergroup $D(2,1;\alpha)$ depending on a
real parameter~$\alpha$~\cite{Sorba}.
It reduces to ${\rm SU(1,1|2)}{\times\!\!\!\!\!\!\supset}{\rm SU(2)}$
at $\alpha{=}0$ and $\alpha{=}{-}1$.
In fact, the isometry supergroup of a near-horizon $M$-brane solution
of $d{=}11$ supergravity was determined as $D(2,1;\alpha){\times}D(2,1;\alpha)$
\cite{GMT-2}, and $D(2,1;\alpha)$ is physically realized for any value of
the parameter~$\alpha$ in near-horizon $M$-theory solutions~\cite{Town}.
The general ${\cal N}{=}\,4$ superconformal group may be relevant to the multi-black-hole system,
since the corresponding moduli spaces of $n$ black holes in four- and five-dimensional
supergravities are described by sigma-model-type multi-black-hole
mechanics respecting $D(2,1;\alpha)$ invariance~\cite{MS2,MS1,MSS,Britto}.

It turns out that, in order to go beyond $\alpha{=}0$ or $\alpha{=}{-}1$,
it is necessary to enlarge the degrees of freedom by a set of ``semi-dynamical''
harmonic isospin variables~\cite{FIL1,BK1,FIL2,FIL3}.
This slight generalization allows one to find a $U$~solution for many WDVV solutions~$F$, but only for
a particular value of~$\alpha$ depending on~$F$~\cite{kl}. The relation to previously-found
solutions in the special cases of ${\rm SU}(1,1|2)$~\cite{KLP} or ${\rm OSp}(4|2)$~\cite{FIL1}
remains unclear.

Another method of constructing many-particle systems with superconformal symmetry was proposed
in~\cite{FIL1,FIL2,FIL3}.
There, such systems arise from a superconformally-invariant gauging of certain supersymmetric
matrix mechanics models, which contain the semi-dynamical isospin degrees of freedom just mentioned, in addition
to the dynamical and purely auxiliary ones.
These semi-dynamical variables are described by a Chern--Simons (or Wess--Zumino) mechanical
action~\cite{ChS1,ChS2,Poly0} and, after quantization, represent isospin (or spin) degrees
of freedom. The resulting isospin-extended superconformal many-particle models exist for
any value of~$\alpha$, but the particle coordinates parametrize a non-flat target space,
except for $\alpha{=}{-}\frac12$, i.e.~the case of ${\rm OSp}(4|2)$.

Finally, any compact supersymmetric mechanics system can be enhanced to a superconformal one
by coupling it to a super dilaton~\cite{hkln,hlns}. In this way, the system to start with
provides the ``angular part'' and the dilaton yields the ``radial part'' of the combined
superconformal mechanics. In particular, any ${\cal N}{=}\,4$ extension of an angular system
can be lifted to a $D(2,1;\alpha)$ invariant mechanics~\cite{hkln}.

The present review provides an overview of recent results obtained
by us and other authors in their studies of $d{=}1$ superconformal
models, mainly classical but also quantum-mechanical, mainly
single-particle but also multi-particle. Given the large amount of
works devoted to the subject, we naturally focus mostly on those
which, in our opinion, have been most influential or prospective for
future investigations. Our review can be considered as being
complementary to already existing reviews (see, e.g.,~\cite{Britto},
\cite{Gosh}).

The plan of the review is as follows.

We start, in Section 2, with the detailed description of the $d{=}1$ conformal group
${\rm SO}(2,1)\simeq{\rm SU}(1,1)\simeq{\rm SL}(2,\mathbb{R})$, its representations
and the renowned de~Alvaro--Fubini--Furlan model of conformal mechanics together
with its multi-particle generalization, i.e.~the rational Calogero model.
Besides presenting some well known geometrical, classical and quantum results in this area,
we also include developments based on our recent papers. This concerns
a derivation of conformal mechanics and Calogero models from gauging specific matrix models
by non-propagating $d{=}1$ gauge fields as well as an extension of the standard conformal
mechanics by ``semi-dynamical'' isospin variables, which provides a new mechanism
for generating the conformal potential. These considerations form a prerequisite
for the methods used in subsequent sections covering superconformal mechanics.

Section 3 is devoted to ${\cal N}{=}\,1$ and ${\cal N}{=}\,2$ superextensions
of the conformal group and to the characterization of the corresponding
superconformal mechanics models, both in the superfield and in the component language.
Again, besides addressing previously known models
(like the Freedman--Mende  ${\cal N}{=}\,2$ super Calogero models),
we discuss more recently proposed kinds of such models.
We elaborate on new ${\cal N}{=}\,1$ and ${\cal N}{=}\,2$
superconformal extensions of the bosonic Calogero model obtained through supersymmetric
versions of the gauging procedure introduced in the section before.

${\cal N}{=}\,4$ superconformal models are the subject of Section 4.
This Section is the most extensive and largely based upon the results obtained
with participation of the authors. We start with presenting the necessary facts
about ${\cal N}{=}\,4$ superconformal algebras, which can all be treated as particular
cases of the general $D(2,1;\alpha)$ superalgebra.
Then we discuss various models of ${\cal N}{=}\,4$ superconformal mechanics, both
in the one- and in the multi-particle case, starting either from the ordinary
${\cal N}{=}\,4$ superspace formulation or from the component one.
Subsequently, we develop a formulation in harmonic ${\cal N}{=}\,4, d{=}1$ superspace,
which often makes more transparent the geometric properties of such models and
yields new types of them. In particular, we describe a new $D(2,1;\alpha)$ invariant
super Calogero model extending the so called ``spinning Calogero model''.
This model involves the semi-dynamical isospin variables and realizes
the ${\cal N}{=}\,4$ extension of the previously defined $d{=}1$ gauging procedure.
Finally we list the known applications of ${\cal N}{=}\,4$ superconformal
mechanics models within the AdS/CFT correspondence and in black-hole physics.

In Section 5 we summarize what is known to date about superconformal mechanics
with ${\cal N}$ larger than~4.

The Conclusion contains some final remarks and lists possible directions of further
investigations of the subjects discussed in the review.

\setcounter{equation}{0}

\section{Conformal mechanics}

\subsection{The one-dimensional conformal algebra and its representations}

The conformal algebra in one dimension is $o (2,1)$ spanned by the
Hermitian generators $H$, $K$ and $D$ satisfying the commutation relations
\begin{equation}\label{SL2R}
\left[\,D,H\,\right] = - iH \, , \qquad \left[\,K,H\,\right]=-2iD \, , \qquad \left[\,D,K\,\right]=iK\,.
\end{equation}

Let us define the dimensionless ${\rm O}(2,1)$ vector $T_{r}$, $r,s,t=0,1,2$,
\begin{equation}\label{T-vec-def}
T_0={\textstyle\frac12}\left(m {K}+ m^{-1} {H}\right),\qquad T_1={\textstyle\frac12}\left(m {K}- m^{-1} {H}\right),\qquad T_2=D\,,
\end{equation}
where $m$ is a constant of mass dimension.\footnote{
When considering physical applications of conformal symmetry,
the generators $H$ and $K$ have physical origin and possess opposite dimensions:
$[H]=L^{-1}$, $[K]=L$.}
Using these generators, one gets another representation of the same $o (2,1)$ algebra (\ref{SL2R})
\begin{equation}\label{o21-vec}
\left[\,T_{r}, T_{s}\,\right]=i\,\epsilon_{rst} \,T^{t} \,,
\end{equation}
with $\epsilon_{012}=+1$, $T^{r}=g^{rs}T_{s}$,
$g_{rs}{=}\,{\rm diag}(-++)$.

For applications to supersymmetric theories, it is important to be aware of a spinorial representation
of the $d{=}1$ conformal algebra. Introducing the ${\rm SU}(1,1)\simeq {\rm O}(2,1)$ bispinor $T_{\alpha\beta}=T_{\beta\alpha}$, $\alpha,\beta,\gamma=1,2$,
\begin{equation}\label{T-vec-spin}
{T}_{11}={H}\,,\qquad {T}_{22}={K}\,,\qquad {T}_{12}={D}\,,
\end{equation}
one gets the spinorial representation of the $o (2,1)$ commutation relations (\ref{SL2R})
\begin{equation}\label{o21-spin}
[\,{T}_{\alpha\beta}, {T}_{\gamma\delta}\,] =
i\left(\epsilon_{\alpha\gamma}{T}_{\beta\delta} +\epsilon_{\beta\delta}{T}_{\alpha\gamma}\right)\,,
\end{equation}
where $\epsilon_{12}=-\epsilon_{21}=+1$.

The second-order Casimir operator of the $o (2,1)$ algebra is given by the following expression
\begin{equation}\label{so21-Cas}
{T}^2 ={\textstyle\frac{1}{2}}\,\{{H},{K}\}-{D}^2=-{T}^{r}{T}_{r}= {\textstyle\frac{1}{2}}\,{T}^{\alpha\beta}{T}_{\alpha\beta} \,.
\end{equation}
The noncompact group ${\rm SL}(2,\mathbb{R})\simeq {\rm SU}(1,1)\simeq {\rm O}(2,1)$ has only infinite-dimensional
unitary representations. They are characterized by the eigenvalues of the Casimir operator and of the compact generator $T_0$.

The infinite-dimensional unitary representations of the discrete series of the universal
covering of the group SU(1,1) are labeled by positive numbers $r_0$ which can be integer or half-integer~\cite{Barg,Per}.
The basis functions of these representations are eigenvectors of the compact SU(1,1) generator  $T_0$. Its eigenvalues are
$r=r_0 +n$,
$n\in \mathbb{N}$~\cite{Barg,Per,AFF}.
On the same states, the Casimir operator  (\ref{so21-Cas}) takes the value
\begin{equation}
{T}^2=r_0(r_0-1)\,.
\end{equation}

\subsection{The de\,Alfaro--Fubini--Furlan model and its interpretation}

The de\,Alvaro--Fubini--Furlan (AFF) conformal mechanics is described by the action \cite{AFF}
\begin{equation}\label{ac-AFF}
S_0= \int dt \,\left(\,\dot x^2 -\gamma^2\, x^{-2}\,\right) \equiv \int dt \,{\cal L}_0\,,
\end{equation}
which implies the equation of motion
\begin{equation}\label{eq-rho}
\ddot x=\gamma^2\, x^{-3}\,.
\end{equation}
The canonical dimension of $x$ is $[x]=L^{1/2}$ whereas the constant
$\gamma$ is dimensionless, $[\gamma]=L^0$.

The action (\ref{ac-AFF}) is invariant under  the $d{=}1$ conformal transformations
\begin{equation}
\delta t = f(t)\,,\qquad \delta x={\textstyle\frac{1}{2}\,}\dot f\,x \qquad\textrm{with}\qquad \partial_t^3 f(t) = 0\,, \lb{Passive}
\end{equation}
whence
\begin{equation}
\delta S_0 =\int\!dt\,\dot\Lambda \qquad\textrm{with}\qquad
\Lambda={\textstyle\frac{1}{2}\,} x^2\,\ddot f\,.\lb{Passiv1}
\end{equation}
This ``passive'' form of $d{=}1$ conformal transformations treats the time variable $t$ and the field $x(t)$
on equal footing, which matches with the geometric interpretation of both as different parameters of the group
${\rm SL}(2,\mathbb{R})$ (see below). It is very convenient for checking invariance of various actions and will be
applied for this purpose in other cases, including the models of superconformal mechanics. On the other hand, for deriving
the conserved currents more convenient is the ``active'' form of the same transformations,
\be
\delta t = 0\,,\qquad \delta x={\textstyle\frac{1}{2}\,}\dot f\,x - f \dot x\,,
\ee
under which
\be
\delta S_0 =\int\!dt\,\dot{\tilde{\Lambda}} \qquad\textrm{with}\qquad
\tilde{\Lambda}={\textstyle\frac{1}{2}\,} x^2\,\ddot f - f {\cal L}_0\,.\lb{Active2}
\ee
Identifying the constant parameters $a, b$ and $c$ of the $d{=}1$ translations, dilatations and conformal boosts with the coefficients
in the $t$-expansion of $f(t)$ as
\be
f(t) = a + b\,t + c\,t^2\,,
\ee
it is easy to read off from \p{Active2} the corresponding conserved Noether charges
\begin{equation}\label{charges-cl}
H = {\textstyle\frac{1}{4}}\, p^2+ \frac{\gamma^2}{ x^{2}} \,,\qquad
D =- {\textstyle\frac{1}{2}}\,xp+t H \,,\qquad
K = x^2-txp+t^2H \,,
\end{equation}
where $p = 2\,\dot x\,$. Note that in the Hamiltonian formalism the conservation is understood to be with respect
to the full time derivative, i.e.
\begin{equation}
\frac{d}{d t}\,K=\frac{\partial}{\partial t}\,K + \{K,H \}_{{}_P}=0\,,\qquad
\frac{d}{d t}\,D=\frac{\partial}{\partial t}\,D + \{D,H \}_{{}_P}=0\,.
\end{equation}
With respect to canonical Poisson brackets the charges (\ref{charges-cl}) form  the algebra $sl(2,\mathbb{R})$
\begin{equation}
\{H,D \}_{{}_P}=H,\qquad \{K,D \}_{{}_P}=-K,\qquad \{H,K \}_{{}_P}=2D\,, \lb{PoSL2R}
\end{equation}
which thus defines the symmetry of the model. Note that the explicit
dependence on $t$ in the generators $D$ and $K$ can be absorbed into
the similarity transformation $(D, K) = e^{tH}(D_0, K_0)e^{-tH}$,
where the commutators are understood as Poisson brackets, and $D_0
=- {\textstyle\frac{1}{2}}\,xp\,, \; K_0 = x^2$ are the
$t$-independent parts of $D$ and $K\,$. Together with $H$ they
satisfy the same Poisson-bracket algebra \p{PoSL2R}. In what
follows, we will basically consider only these main terms in the
generators of the conformal algebra and its superextensions.

The  expressions for the Noether charges  (\ref{charges-cl}) imply the following value
for the classical Casimir operator (\ref{so21-Cas}),
\begin{equation}\label{so21-Cas-cl0}
{T}^2 =\gamma^2\,,
\end{equation}
and the Hamiltonian can be rewritten as
\begin{equation}\label{charges-cl-H}
H = {\textstyle\frac{1}{4}}\, p^2+ \frac{T^2}{ x^{2}} \,.
\end{equation}
As was noticed in~\cite{hkln,hlns}, this expression provides a general classical Hamiltonian
of conformal mechanics models,
including multi-dimensional and supersymmetric extensions of the one-dimensional AFF mechanics. The canonical
variables $p$ and $x$ represent the radial degree of freedom, while the additional angular and fermionic
coordinates and their canonical momenta are hidden in the conformal Casimir~$T^2$ which, on its own,
can be treated as a Hamiltonian of some `angular' mechanical system.

The AFF conformal mechanics \cite{AFF} admits a nice geometric interpretation. As was demonstrated in \cite{IKL1},
it can be obtained by applying the Maurer-Cartan (or nonlinear realizations) method to the algebra $o (2,1)$.
We choose the exponential parametrization for the  element of the group
${\rm SO}(2,1)$\,:
\begin{equation}\label{group-el-cm}
G_0 = e^{itH}\, e^{izK}\, e^{iuD}\,,
\end{equation}
and construct the following left-covariant Maurer-Cartan (MC)  one-forms
\begin{equation}\label{om-def0}
G_0^{-1}dG_0 = i\Big(\omega_H H+ \omega_K K+ \omega_D D \Big)\,,
\end{equation}
with
\begin{equation}\label{om-cm}
\omega_H = e^{-u}dt\,,\qquad
\omega_K = e^{u}\left(dz+z^2 dt\right)\,,\qquad
\omega_D = du-2zdt\,.
\end{equation}

In the conformal mechanics \cite{AFF}, like in the construction of unitary representations of the group SO(2,1),
one is led to choose the basis  (\ref{T-vec-def}) in the $o(2,1)$ algebra.
The MC one-forms associated with the generators $T_{r}$ are, respectively,
\begin{equation}\label{new-o21-CM}
\omega_0= m^{-1}\omega_K+ m\, \omega_H\,, \qquad
\omega_1= m^{-1}\omega_K- m\, \omega_H\,,\qquad
\omega_2= \omega_D\,.
\end{equation}

The dynamics of the AFF conformal mechanics is obtained by imposing the following
constraints on the one-dimensional coset fields $z(t)$ and $u(t)$~\cite{IKL1}
\begin{equation}\label{IKL-const}
\mbox{(a)}\,\,\,\,\,\omega_1=0\,,\qquad \qquad \mbox{(b)}\,\,\,\,\,\omega_{2}=0\,.
\end{equation}
Equation (\ref{IKL-const}b) is the inverse Higgs \cite{IO} constraint trading the field $z(t)$ for the time derivative of the dilaton $u(t)$,
\begin{equation}\label{z-rez}
z={\textstyle\frac12}\,\dot u\,,
\end{equation}
while (\ref{IKL-const}a) is the dynamical constraint  which leads to the equation of motion (\ref{eq-rho})
for the newly introduced single independent variable
\be
x=\mu^{-1/2}e^{u/2}\,, \lb{xu}
\ee
where $\mu$ is the dimensionful part of $m$, $[\mu]=cm^{-1}$
and $m=\mu \gamma$, $[\gamma ]=cm^{0}$. Being constraints on the left-invariant Cartan 1-forms, eqs. (\ref{IKL-const})
enjoy manifest $d{=}1$ conformal SO(2,1)symmetry.

In the formalism of the MC one-forms the conformal mechanics action can be rewritten as~\cite{IKL1}
\begin{equation}\label{ac-cm-f}
S_0=-\gamma\int \omega_0 = -\int\,dt\Big[ \,\mu^{-1}e^u\left(\dot z + z^2 \right) +\mu\gamma^2 e^{-u} \,\Big]\,.
\end{equation}
We see that the action (\ref{ac-cm-f}) is specified by the remaining non-vanishing MC one-form in $o(2,1)$.
Both the kinematical constraint (\ref{z-rez})
($\omega_2=0$) and the dynamical
equation (\ref{eq-rho}) ($\omega_{1}=0$), follow from the action (\ref{ac-cm-f}) as the equations of motion.
Substituting the kinematical solution \p{z-rez} into \p{ac-cm-f} and passing to the variable $x$ by \p{xu}, we
recover the original AFF conformal mechanics action \p{ac-AFF}. Note that this equivalence is valid
modulo a total $t$-derivative under the integral, which explains why the action \p{ac-cm-f}
is exactly invariant with respect to the ``passive'' conformal transformations \p{Passive}, while \p{ac-AFF} is invariant up to a total
$t$-derivative, eq. \p{Passiv1}.

Note that (\ref{IKL-const}) define a class of geodesics on the ${\rm SO}(1,2)$ group manifold,
such that the geodesics are generated by the right action of the one-parameter compact subgroup with the generator $T_0$ \cite{IKL1}.
Only such a class leads
to the standard conformal mechanics with good quantum properties \cite{AFF}, as opposed to any other non-trivial choice
of the constraints (for example, the choice of $\omega_{0}=0$ instead of (\ref{IKL-const}a)).
This is the reason why in our considerations we stick just to the constraints (\ref{IKL-const}).
The coordinate $\tau$ associated with the generator $T_0$ in the exponential parametrization of this compact subgroup
is the natural parameter along the geodesic curve. In the black-hole interpretation of (super)conformal mechanics (see below),
$\tau$ plays the role of the test-particle proper time near the horizon of the extremal black hole.

Let us make a few comments on the Hamiltonian formulation of the model (\ref{ac-cm-f}).

The definition of the momenta yields the second-class constraints
\begin{equation} \label{constr-cm}
p_u  \approx 0\,, \qquad
p_z +\mu\,e^u \approx  0\,.
\end{equation}
These constraints allow one to eliminate the phase space variables $(p_z, p_u)\,$.
The Dirac brackets for the surviving pair of the phase space variables $(u,z)$ and the Hamiltonian take the form
\begin{equation}\label{tHam1-cm}
\{ u, z\}_{{}_D}=\mu\,e^{-u}\,,\qquad H= {\textstyle\frac{1}{4\mu}}\left(e^u z^2  +4\gamma^2\mu^2 e^{-u}\right).
\end{equation}
After introducing the variables $x=\mu^{-1/2}e^{u/2}$, $p= 2\mu^{-1/2}e^{u/2}z$ which possess standard Dirac brackets,
\begin{equation}\label{can-tr-cm}
\{ x, p\}_{{}_D}=1\,,
\end{equation}
we find that the system (\ref{ac-cm-f}) is described by the Hamiltonian
\begin{equation}\label{tHamAFF-cm}
H={\textstyle\frac{1}{4}}\, p^2+ \gamma^2\, x^{-2}\,,
\end{equation}
which also follows from the action (\ref{ac-AFF}).

The basic features of the quantization based on the standard Hamiltonian (\ref{tHamAFF-cm}) are as follows
\cite{AFF,FR,Britto}:
\begin{itemize}
\item The spectrum of ${H}$ is continuous: if \,\,${H}\,|E>=E\,|E>$, \,\, then\,\, ${H}\,e^{i\alpha D}\,|E>=e^{2\alpha}\,E\,|E>\,$.
\item The spectrum includes all $E{>}0$ eigenvalues of $H$;
for each of them exists a plane-wave normalizable state.
\item The state with $E{=}0$ is not even plane-wave normalizable and so it can not be chosen as the ground state.
\end{itemize}

The absence of the normalizable ground state implies that the description of quantum conformal mechanics  in terms of
the ${H}$ eigenstates is ill-defined.
The correct description of the model is achieved with choosing the compact operator $T_0$ as the Hamiltonian
with respect to which one should define the energy spectrum \cite{AFF}:
\begin{equation}\label{tHamAFF-T}
\tilde H:=2m T_0=H+m^2 K={\textstyle\frac{1}{4}}\, p^2+ \gamma^2\, x^{-2} +m^2 x^2 \,.
\end{equation}
It contains an oscillator-like term and so has a well defined ground state.
The corresponding Hilbert space is the space of functions which span infinite-dimensional unitary representations of SU(1,1), as
described in the previous subsection. The spectrum of the Hamiltonian (\ref{tHamAFF-T}) is discrete.
Its ground state is not invariant under  the whole SU(1,1),
but breaks it spontaneously~\cite{AFF,Rab}.

The passing to the phase with the new evolution operator $T_0$ and spontaneously broken SU(1,1)$\sim $SL(2,$\mathbb{R}$) symmetry
can be interpreted as a redefinition of the time coordinate \cite{AFF,CDKKTP,Britto,Rab}.
Namely, if we introduce a new evolution parameter $\tau$
and coordinate $q$ through the relations~\cite{AFF}
\begin{equation}\label{new-tau-q}
\tau:={\textstyle\frac{1}{m}}\arctan(mt)\,,\qquad
q(\tau):=x(t)/\sqrt{1+m^2t^2}\,,
\end{equation}
the action (\ref{ac-AFF}), up to boundary terms, takes the form
\begin{equation}\label{ac-AFF-m}
S_0= \int d\tau \left(\,\dot q^2 -\gamma^2 q^{-2}-m^2 q^{2}\,\right),\qquad
\dot q:=dq/d\tau\,,
\end{equation}
which is the action corresponding to the Hamiltonian (\ref{tHamAFF-T}).
Such a coordinate change has a nice geometrical interpretation in the application to
black-hole dynamics. The near-horizon geometry is that of
AdS$_2$ with ${\rm SL}(2,\mathbb{R})$ as the isometry group. The good global time coordinate
for the particle moving on this near-horizon AdS$_2$ is precisely the evolution parameter $\tau$
(for details see \cite{CDKKTP,Britto,Rab}). Note that the action \p{ac-AFF-m}, despite the presence of the
oscillator term, is still invariant under a particular non-linear realization of the underlying SL(2,$\mathbb{R}$)
group~\cite{pashnev}.

The quantum counterparts of the $o(2,1)$ generators \p{charges-cl} at $t=0$ read
\begin{equation}\label{charges-quant}
H = {\textstyle\frac{1}{4}}\, \hat p^2+ \frac{\gamma^2}{ \hat x^{2}} \,,\qquad
D =- {\textstyle\frac{1}{4}}\,\{\hat x ,\hat p\} \,,\qquad
K = \hat x^2 \,,
\end{equation}
where
\begin{equation}\label{can-qu-cm}
[ \hat x, \hat p]=i\,.
\end{equation}
The Casimir operator (\ref{so21-Cas}) takes the value
\begin{equation}\label{so21-Cas-0}
T^2=\gamma^2-{\textstyle\frac{3}{16}}\,.
\end{equation}
Therefore, the strength $\gamma$ of the conformal potential is related to the constant $r_0$
labeling the ${\rm SL}(2,\mathbb{R})$ representation as
\begin{equation}
r_0={\textstyle\frac{1}{2}} \left(1+\sqrt{4\gamma^2+{\textstyle\frac{1}{4}}}\right)\,.
\end{equation}
The choice of positive $r_0$ guarantees a good behavior of the ground state wave function at the
origin~\cite{AFF}.
Note that, using the quantum Casimir operator (\ref{so21-Cas-0}),
we can cast the quantum Hamiltonian (\ref{charges-quant}) in the form
\begin{equation}\label{charges-qu-H}
H = {\textstyle\frac{1}{4}}\, p^2+ \frac{T^2+\frac{3}{16}}{ x^{2}} \,.
\end{equation}
Comparing it with the classical expression (\ref{charges-cl-H}),
we observe the appearance of a constant shift in the numerator of the potential term.

\subsection{Conformal mechanics from $d{=}1$ gauging}

It is interesting that the AFF conformal mechanics action and its version \p{ac-AFF-m} with the oscillator-type term can be reproduced by
applying some $d{=}1$ gauging procedure to the very simple complex free-particle action \cite{sigma}. This example is a prototype of more
complicated (super)conformal models which can also be constructed by applying $d{=}1$ gauge procedure and its supersymmetric
generalizations (see Sections 2.5, 3.3.2, 4.3.2 and 4.3.3).

Consider a complex $d{=}1$ field $z(t), \bar z(t)$ with the following Lagrangian:
\begin{equation}
L_z = \dot{z}\,\dot{\bar z} +im \left(\dot z \bar z -  z\dot{\bar z}\right). \label{ex1}
\end{equation}
The first term is the kinetic energy, the second one is the simplest $d{=}1$ WZ term. One of the symmetries of this system is
the invariance under U(1) transformations:
\begin{equation}
z' = e^{-i\lambda} z\,, \quad \bar{z}' = e^{i\lambda} \bar{z}\,.
\end{equation}
Now we gauge this symmetry by promoting $\lambda \rightarrow \lambda(t)$. The gauge invariant action involves
the $d{=}1$ gauge field $A(t)$
\begin{equation}
L_{gauge} = ( \dot{z} + i A z)\,(\dot{\bar z} - i A \bar z) + im \left(\dot z \bar z -  z\dot{\bar z} + 2i A z\bar z\right) + 2\gamma\, A\,,
\quad
A' = A + \dot\lambda\,,
\end{equation}
where a ``Fayet-Iliopoulos term'' $\propto \gamma$ has been also added. This term is gauge invariant (up to a total derivative) by itself.

The next step is to choose the appropriate gauge in $L_{gauge}$:
\begin{equation}
z = \bar z \equiv q(t)\,.
\end{equation}
We substitute it into $L_{gauge}$ and obtain:
\begin{equation}
L_{gauge} = ( \dot{q} + i A q)\,(\dot{q} - i A q) + 2i m\, A q^2 + 2\gamma\, A =
(\dot{q})^2 + A^2 q^2 - 2m A q^2 + 2\gamma A\,. \label{ex2}
\end{equation}
The field $A(t)$ is the typical example of auxiliary field: it can be eliminated by its algebraic equation of motion:
\begin{equation}
\delta A: \qquad A = m  - \gamma\,q^{-2}\,.
\end{equation}
The final form of the gauge-fixed Lagrangian is as follows
\begin{equation}
L_{gauge} \Rightarrow (\dot{q})^2 - \left(m q - \gamma q^{-1}\right)^2. \label{ex3}
\end{equation}
Up to an additive constant $\sim m \gamma$, this Lagrangian coincides with \p{ac-AFF-m}. At $m =0$, one recovers the standard conformal mechanics:
\be
L_{gauge}^{(m =0)} = (\dot{q})^2 - \gamma^2\,q^{-2}\,.\lb{Lgauge10}
\ee
The initial action $S_z = \int dt L_z$ at $m=0$ is manifestly invariant under the conformal transformations $\delta t = f(t),
\delta z = \frac{1}{2}\dot{f}\, z\,, \; (\partial_t)^3 f=0\,$. The conformal invariance is preserved by the gauging procedure, provided
that the gauge field $A(t)$ transform as $\partial_t$, i.e. $\delta A(t) = - \dot{f}\, A(t)\,$. As a result, the gauge-fixed
action $\int dt L_{gauge}^{(m=0)}$ also respects the conformal invariance.

This $d{=}1$ gauging procedure can be interpreted as an off-shell Lagrangian analog of the well known Hamiltonian reduction.
In the present case, in the parametrization $z = q e^{i \varphi}$, the Hamiltonian reduction consists in imposing the constraints
$p_\varphi - 2\gamma \approx 0\,, \;\varphi \approx 0\,,$ upon which the Hamiltonian of the system (\ref{ex1}) is reduced to the AFF
Hamiltonian.

\subsection{Conformal mechanics with additional isospin degrees of freedom}

We can consider an extension of the AFF conformal mechanics model by additional isospin degrees of freedom.
These additional degrees of freedom are described by Chern--Simons (or, in other terminology,  Wess-Zumino)
mechanical action~\cite{ChS1,ChS2,Poly0,Plyu}.
Using them, together with gauging some isometries of some simple initial actions, one can recover the Lagrangian models constructed earlier by
other methods and to construct new dynamical systems \cite{DI1,DI2}.
Such an approach reveals an interesting deviation from the standard conformal quantum
mechanics: besides the standard dilatonic variable $x(t)$ with the conformal
potential, it also contains a fuzzy sphere \cite{Mad1,Mad2} described
by additional isospin variables. As a result, the  relevant wave
functions form non-trivial SU(2) multiplets, as
opposed to the SU(2) singlet wave function of the standard conformal mechanics. The
strength of the conformal potential proves to coincide with the
eigenvalue of the ${\rm SU}(2)$ Casimir operator (i.e.~``spin'') and so it is
quantized.

Let us consider the following action \cite{FIL2,FIL3}:
\begin{equation}\label{bose0}
\tilde S_0 =  \int dt \,\Big[\dot x\dot x  +  {\textstyle\frac{i}{2}}\left(\bar
z_k \dot z^k - \dot{\bar z}_k z^k\right)-\frac{\alpha^2(\bar z_k
z^{k})^2}{16x^2} -A \left(\bar z_k z^{k} -c \right) \Big] \,,
\end{equation}
where $\alpha$ is some dimensionless parameter (as we shall see below, it coincides with the parameter characterizing
the most general ${\cal N}{=}\,4$ superconformal group $D(2,1;\alpha)$).
This action is invariant under the local ${\rm U}(1)$ transformations
\be\label{0-ga}
A' = A - \dot{\lambda}_0\,, \quad z^i{}' = e^{i\lambda_0}z^i\,, \;
\bar{z}_i{}' =
e^{-i\lambda_0}\bar{z}_i\,.
\ee
The $d{=}1$ gauge connection
$A(t)$ in \p{bose0} is the Lagrange multiplier for the constraint
\begin{equation}\label{con0}
\bar z_k z^{k}=c\,.
\end{equation}
After varying with respect to $A$, the action \p{bose0} is gauge invariant
only with
taking into account
this algebraic constraint which is gauge invariant by itself. It is
convenient to
fully fix the residual
gauge freedom by choosing the phases of $z^1$ and $z^2$ opposite to
each other.
In this gauge,
the constraint \p{con0} is solved by~\footnote{
The $d{=}1$ field $\gamma(t)$ should not be confused with the constant $\gamma$
of the previous subsections.}
\begin{equation}\label{ga-u1}
z^{1} = \kappa \cos{\textstyle\frac{\gamma}{2}}\,e^{i\beta/2}\,,\qquad
z^{2}= \kappa \sin{\textstyle\frac{\gamma}{2}}\,e^{-i\beta/2}\,,\qquad
\kappa^2=c\,.
\end{equation}
In terms of the newly introduced  fields $x(t), \gamma(t)$ and $\beta(t)$ the action~(\ref{bose0}) takes
the form
\begin{equation}\label{bose01}
S_b =  \int dt \,\Big[\dot x\dot x  -\frac{\alpha^2c^2}{16x^2} - \frac{c}{2}\,
\cos\gamma\, \dot \beta  \Big] \,.
\end{equation}

The action~(\ref{bose01}) contains the ``true'' kinetic term only for one
bosonic component $x$
which also possesses the conformal potential, whereas two other fields
$\beta$ and $\gamma$ parametrizing the coset SU(2)/U(1)
are described by a WZ term and so become a sort of isospin degrees of
freedom (target SU(2) harmonics) upon
quantization. The conformal invariance of the WZ term is evident in the notation (\ref{bose01}), keeping in mind that the SU(2) fields
$\beta(t)$ and $\gamma(t)$ have zero conformal weights. Note that the fact of conformal invariance of $d{=}1$ WZ terms
was pointed out for the first time by Jackiw \cite{Jackiw}.

It should be pointed out that the considered model realizes a new mechanism of
generating conformal potential $\sim 1/x^2$ for
the field $x(t)$. Before eliminating auxiliary fields, the
component action contains no explicit term
of this kind. It arises as a result of varying with respect to
the Lagrange multiplier $A(t)$ and making use of the arising constraint \p{con0}. As we shall see
soon, in quantum theory this new mechanism entails a quantization
of the constant $c\,$.

The corresponding canonical Hamiltonian of the model~(\ref{bose0}) reads
\begin{equation}\label{bose-H0-s}
H_0 =  \frac{1}{4} \left[ p^2 + \frac{\alpha^2(\bar z_k
z^{k})^2}{4x^2} \right] +A \left(\bar z_k z^{k} -c  \right).
\end{equation}
Here $p=2\dot x$ is the canonical momentum for the coordinate $x$. Canonical
momentum for the field $A$ is vanishing, $p_A=0$.
This constraint and the fact that the field $A$ appears in the
action~(\ref{bose0}) linearly, suggest to treat $A$ as the Lagrange multiplier
for the constraint
\begin{equation}\label{D0-con1-s}
D^0-c := \bar z_k z^{k} -c \approx 0\,.
\end{equation}
The expressions for the canonical momenta $p_i$ and $\bar p^i$ of the
$z$-variables,
$[z^i, p_j]_{{}_P}= \delta^i_j$, $[\bar z_i, p^j]_{{}_P}= \delta^j_i\,$, follow
from the second-class
constraints
\begin{equation}\label{G-const-s}
G_k := p_k-{\textstyle\frac{i}{2}}\,\bar z_k\approx 0\,,\qquad
\bar G^k := \bar p^k+{\textstyle\frac{i}{2}}\, z^k\approx 0\,,\qquad
[G_k, \bar G^l]_{{}_P}=  -i\delta^l_k.
\end{equation}
Using their Dirac brackets
\begin{equation}
[A, B]_{{}_D}=[A, B]_{{}_P}+i[A, G_k]_{{}_P}[\bar G^k, B]_{{}_P}-i[A, \bar
G^k]_{{}_P}[G_k, B]_{{}_P}\,,
\end{equation}
we eliminate the spinor momenta $p_i$ and $\bar p^i$.
Dirac brackets for the residual variables are
\begin{equation}\label{DB-s}
[x, p]_{{}_D}= 1, \qquad [z^i, \bar z_j]_{{}_D}= -i\delta^i_j.
\end{equation}

To finish with the classical description, we point out that the spinor variables
describe a two-sphere.
Namely, using the first Hopf map we introduce three U(1) gauge invariant variables
\begin{equation}\label{cl-y0-s}
y_a= {\textstyle\frac{1}{2}}\,\bar z_i(\sigma_a)^i{}_j z^j\,,
\end{equation}
where $\sigma_a$, $a=1,2,3$ are Pauli matrices. The constraint~(\ref{D0-con1-s})
suggests that these variables parameterize a two-sphere with the radius $c/2$:
\begin{equation}\label{cl-sp-s}
y_a y_a=(z^{k}\bar z_k)^2/4\approx c^2/4\,.
\end{equation}
The group of motion of this 2-sphere is of course the ${\rm SU}(2)$ group
acting on the doublet indices $i, k$ and triplet indices $a$. In terms of the new
variables~(\ref{cl-y0-s}) the Hamiltonian~(\ref{bose-H0-s}), up to terms vanishing on the constraints, takes the form
\begin{equation}\label{bose-Hy-s}
H =  \frac{1}{4} \left[ p^2 + \frac{\alpha^2 y_a y_a}{x^2} \right] \approx \frac{1}{4} \left[ p^2 + \frac{\alpha^2 c^2}{4x^2} \right].
\end{equation}

At the quantum level, the algebra of the canonical operators obtained from
the algebra of Dirac brackets is
\begin{equation}\label{bose-cB-s}
[\hat x, \hat p] = i\,, \qquad [\hat z^i, \hat{\bar z}_j] = \delta^i_j \,.
\end{equation}
Then it is easy to check that the quantum counterparts of the
variables~(\ref{cl-y0-s})
\begin{equation}\label{qu-y-s}
\hat{y}_a= {\textstyle\frac{1}{2}}\,\hat{\bar z}_i(\sigma_a)^i{}_j \hat{z}^j
\end{equation}
form the $su(2)$ algebra
\begin{equation}\label{Y-cB-s}
[\hat{y}_a, \hat{y}_b] = i\,\epsilon_{abc} \hat{y}_c\,.
\end{equation}
Notice that no ordering ambiguity is present in the definition \p{qu-y-s}.

Moreover, the direct calculation yields
\begin{equation}\label{Y2-s}
\hat{y}_a \hat{y}_a = {\textstyle\frac{1}{2}}\,\hat{\bar z}_k \hat{z}^{k}
\left({\textstyle\frac{1}{2}}\,\hat{\bar z}_k \hat{z}^{k}+1\right)
\end{equation}
and, due to the constraints \p{D0-con1-s} (for definiteness, we adopt $\hat{\bar z}_k \hat{z}^{k}$-ordering
in it), one gets
\begin{equation}\label{Y2-1-s}
\hat{y}_a \hat{y}_a = \frac{c}{2} \left(\frac{c}{2}+1\right).
\end{equation}
But the relations~(\ref{Y-cB-s}) and~(\ref{Y2-1-s}) are the definition of the
{\it fuzzy sphere} coordinates.
Thus the angular variables, described, at the classical level, by spinor variables $z^i$ or vector variables $y_a$,
after quantization acquire a nice interpretation of the fuzzy sphere coordinates.
Comparing the expressions~(\ref{cl-sp-s}) and~(\ref{Y2-1-s}), we observe that
upon quantization the radius of the sphere changes as  $\frac{c^2}{4}\to \frac{c}{2} \left(\frac{c}{2}+1\right)$.

As suggested by the relation~(\ref{Y-cB-s}), the fuzzy sphere coordinates $\hat{y}_a$ are the generators
of $su(2)$ algebra and the relation~(\ref{Y2-1-s}) fixes the value of its Casimir
operator, with $c$ being the relevant ${\rm SU}(2)$ spin (``fuzzyness''). Then it follows that $c$ is quantized,
$c \in \mathbb{Z}$.

The wave functions inherit this internal symmetry through a  dependence on
additional ${\rm SU}(2)$ spinor degrees of freedom. Let us consider the
following realization for the operators $Z^i$ and $\bar Z_i$
\begin{equation}\label{bo-re-Z-s}
\hat{\bar z}_i=v^+_i, \qquad \hat{z}^i=  \partial/\partial v^+_i\,,
\end{equation}
where $v^+_i$ is a commuting complex SU(2) spinor.
Then the constraint~(\ref{D0-con1-s}) imposed on the wave function $\Phi(x,v^+_i)\,$,
\begin{equation}\label{q-con1-s}
D^0 \Phi=\hat{\bar z}_i \hat{z}^i \Phi=v^+_i\frac{\partial}{\partial v^+_i}\,\Phi=c\,\Phi\,,
\end{equation}
leads to the polynomial dependence of it on $v^+_i$:
\begin{equation}\label{wf-rep-s}
\Phi(x,v^+_i) = \phi_{k_1\ldots k_{c}}(x)v^{+k_1}\ldots v^{+k_{c}} \,.
\end{equation}
Thus, as opposed to the model of ref.~\cite{AFF}, in our case the
$x$-dependent wave function carries an irreducible spin $c/2$ representation of the group SU(2), being
an SU(2) spinor of the rank $c$.

Using~(\ref{bose-Hy-s}) and~(\ref{Y2-1-s}) we see that on physical states the
quantum Hamiltonian is
\begin{equation}\label{H-qu-bo-s}
{H} =\frac{1}{4}\,\left(\hat{p}^2  +\frac{g}{\hat{x}^2} \right)
\qquad\textrm{with}\qquad
g = \alpha^2 \,\frac{c}{2} \left(\frac{c}{2}+1\right)\,.
\end{equation}
It is easy to show that the SU(1,1) Casimir operator takes the value
\begin{equation}\label{Cas-qu-bo-s}
T^2 ={\textstyle\frac{1}{4}}\, g - {\textstyle\frac{3}{16}}\,,
\end{equation}
and the Hamiltonian~(\ref{H-qu-bo-s}) can be rewritten in the form~(\ref{charges-qu-H}).
Thus, like in~\cite{AFF}, on the fields $\phi_{k_1\ldots k_{c}}(x)$
the unitary irreducible representations of the group SU(1,1) are realized,
despite the fact that the wave function
is now multi-component, with $(c + 1)$ independent components.
Requiring the wave function $\Phi(v^+)$ to be single-valued once again
leads to the condition that $c\in \mathbb{Z}$.
This quantization of parameter $c$ could be important for the possible black hole
interpretation of the considered variant of conformal mechanics.

Note that the action \p{bose01} can be extended by adding the conformally and SU(2) invariant sigma-model kinetic term
for the SU(2)$/$U(1) variables $\gamma$ and $\beta$:
\be
S_b \;\Rightarrow \; S'_b = S_b + g' \int dt\, x^2\left[(\dot{\gamma})^2 + (\dot{\beta})^2 \sin^2\gamma \right],\lb{MoD}
\ee
where $g'$ is a renormalization constant. In such an extended model, the SU(2) fields loose their status of ``semi-dynamical'' isospin variables;
they become the full-fledged physical degrees of freedom and make a contribution to the corresponding hamiltonian.
The coefficient $c$ before the WZ term is still quantized on the topological grounds. This sort of the conformally invariant mechanics model
with three physical degrees of freedom (the radial variable $x$ and the angular variables $\beta$ and $\gamma$) and WZ term
related to the strength of the conformal potential as in \p{bose01} appears
as a bosonic part of the ${\cal N}{=}\,4$ superconformal mechanics based on the ${\cal N}{=}\,4, d{=}1$ off-shell supermultiplet
({\bf 3,4,1})~\cite{IKLech}.

\subsection{Conformally invariant multi-particle systems}

Above, we presented a model that has only one dynamic degree of freedom.
There are many models which possess conformal invariance and describe many dynamical degrees of freedom
that can be interpreted as degrees of freedom of different particles.
An example of such systems is provided by the well-known many-body Calogero model \cite{C1,C2,Per-b}, which describes
$n$ identical particles interacting pairwise through an inverse-square potential.
Here we present a formulation of the $n$-particle Calogero model as a matrix model with gauge $\textrm{U}(n)$ symmetry
\cite{Poly0,GorskyNekr1,GorskyNekr2,Poly1,Poly2,Poly3,FIL1}.

In this formulation the $n$-particle Calogero model is described by the Hermitian $(n\times n)$-matrix field
$X_a{}^b(t)$, $(\overline{X_a{}^b}) =X_b{}^a$,
and complex $\textrm{U}(n)$-spinor field
$Z_a(t)$, $\bar Z^a = (\overline{Z_a})$,
$a,b=1,\ldots ,n$, and involves $n^2$ gauge fields
$A_a{}^b(t)$, $(\overline{A_a{}^b}) =A_b{}^a$.
The action has the following form
\begin{equation}\label{b-Cal}
S_C =
\int dt  \,\Bigg[\, {\rm tr}\left(\nabla\! X \nabla\! X \right) + {\textstyle\frac{i}{2}}\, (\bar Z \nabla\!
Z -
\nabla\! \bar Z Z) + c\,{\rm tr} A  \,\Bigg]\, ,
\end{equation}
where the covariant derivatives are
\begin{equation}\label{cov-der-b}
\nabla\! X = \dot X +i [A, X]\,, \qquad \nabla\! Z = \dot Z + iAZ\,, \qquad \nabla\!
\bar Z
= \dot{\bar Z} -i\bar Z A\,.
\end{equation}
The last term (Fayet-Iliopoulos term) includes only $\textrm{U}(1)$ gauge field, $c$ is a
real constant.

The action \p{b-Cal} is invariant under the $d{=}1$ conformal $\textrm{SO}(2,1)$ transformations
\begin{equation}\lb{so12}
\begin{array}{ccc}
&& \delta t = a(t)\,, \quad \delta\partial_t = -\dot{a}\,\partial_t\, \\[7pt]
&& \delta X_a^b = \frac{1}{2} \dot{a}\, X_a^b\,, \quad \delta Z_a = 0\,, \quad  \delta A_a^b =  -\dot{a}\,A_a^b\,,
\end{array}
\end{equation}
where $a(t)$ obeys the constraint
\be
\partial_t^3 a = 0\,.
\ee
This is in agreement with the well-known fact that the Calogero model is conformal. Any gauge-invariant potential term of the field $X_a^b$
evidently breaks the conformal invariance. Non-conformal, though still integrable deformations of the Calogero model correspond to adding
some specific gauge invariant monomials of the matrix $X^a_b$ to the action \p{b-Cal}. In particular, the combination
\be
n\,{\rm tr}\, X^2 -({\rm tr}\,X)^2
\ee
in the gauge \p{X-fix} (see below)  yields the additional term in the final action
\be
\sum_{a<b} (x_a - x_b)^2\,,
\ee
which corresponds to passing to the Calogero-Moser model.

The action~(\ref{b-Cal}) is invariant with respect to the local $\textrm{U}(n)$
transformations, $g(\tau
)\in \textrm{U}(n)$,
\begin{equation}\label{Un-tran}
X \rightarrow \, g X g^+ \,, \qquad  Z \rightarrow \, g Z \,, \qquad \bar Z
\rightarrow \, \bar Z g^+\,, \qquad
A \rightarrow \, g A g^+ +i \dot g g^+\,.
\end{equation}

Using the gauge transformations~(\ref{Un-tran}) we can impose a (partial) gauge fixing
\begin{equation}\label{X-fix}
X_a{}^b =0\,,\qquad a\neq b.
\end{equation}
In this gauge the matrix variable $X$ takes the form
\begin{equation}\label{X-fix-com}
X_a{}^b = x_a \delta_a{}^b =
\left(
\begin{array}{cccccccc}
            x_1 & 0 & 0 & \cdots & 0 & 0 & 0 \\
            0 & x_2 & 0 & \cdots & 0 & 0 & 0 \\
            0 & 0 & x_3 & \cdots & 0 & 0 & 0 \\
            \vdots & \vdots & \vdots & \ddots  & \vdots & \vdots & \vdots  \\
            0 & 0 & 0 & \cdots & x_{n-2} & 0 & 0 \\
            0  & 0 & 0 & \cdots & 0 & x_{n-1} & 0 \\
            0  & 0 & 0 & \cdots & 0 & 0 & x_n \\
\end{array}
\right)\,.
\end{equation}
Then
\begin{equation}\label{rel-XA}
[X, A]_a{}^b = (x_a - x_b)A_a{}^b\,, \qquad
\end{equation}
and, therefore,
\begin{equation}\label{rel-0}
{\rm tr}[X, A] = 0\,, \qquad {\rm tr}(\dot X [X, A]) = 0\,, \qquad {\rm tr}([X,
A][X, A]) =
-\sum_{a,b}(x_a - x_b)^2 A_a{}^b A_b{}^a\,.
\end{equation}
As a result of this, the action~(\ref{b-Cal})
takes the form
\begin{equation}\label{b-Cal2}
S_C = \int dt  \sum_{a,b} \,\Big[\, \dot x_a \dot x_a + {\textstyle\frac{i}{2}}\, (\bar Z^a \dot Z_a -
\dot {\bar Z}{}^a Z_a) + (x_a - x_b)^2 A_a{}^b A_b{}^a -\bar Z^a A_a{}^b Z_b  + c\,
A_a{}^a\,\Big]\,.
\end{equation}

In the third term in the action~(\ref{b-Cal2}) there remain only non-diagonal
elements of the
matrix $A$, $A_a{}^b$ with $a\neq b$. Therefore the action~(\ref{b-Cal2}) has the
residual invariance under the gauge Abelian $[\textrm{U}(1)]^n$ symmetry with local
parameters $\varphi_a(t)$:
\begin{equation}\label{b-Ab}
Z_a \rightarrow \, e^{i\varphi_a} Z_a \,, \qquad \bar Z^a  \rightarrow \,
e^{-i\varphi_a}
\bar Z^a\,, \qquad A_a{}^a \rightarrow \, A_a{}^a - \dot \varphi_a \quad (\mbox{no
sum with
respect to } a)\,.
\end{equation}
Making use of this invariance, we can impose the further gauge
\begin{equation}\label{g-Z}
\bar Z^a = Z_a\,.
\end{equation}
In this gauge, the second term in the action~(\ref{b-Cal2}) vanishes and the action~(\ref{b-Cal2})
takes the form
\begin{equation}\label{b-Cal3}
S_C = \int dt  \sum_{a,b} \,\Big[\, \dot x_a \dot x_a
 + (x_a - x_b)^2 A_a{}^b A_b{}^a - Z_a Z_b A_a{}^b  + c\,
A_a{}^a\,\Big]\,.
\end{equation}

Let us verify that under passage from the action~(\ref{b-Cal2}) to the action~(\ref{b-Cal3})
the equations of motion are preserved and we do not obtain additional equations/constraints.

The equation of motion for $Z$, following from the action~(\ref{b-Cal2}),
\begin{equation}\label{eq1-z}
\nabla Z_a =  \dot{Z}_a -i \sum_b A_a{}^b Z_b =0\,, \qquad
\nabla \bar Z^a =  \dot{\bar{Z}}{}^a -i \sum_b \bar Z^b A_b{}^a  =0\,,
\end{equation}
\begin{equation}\label{eq1-ab}
A_a{}^b = \frac{Z_a \bar Z^b }{2(x_a - x_b)^2}  \qquad \mbox{for } a\neq b\,,
\end{equation}
\begin{equation}\label{eq1-aa}
\bar Z^a Z_a =c \qquad \forall \, a  \quad (\mbox{no sum with respect to }
a)
\end{equation}
in the gauge~(\ref{g-Z}) become the equations
\begin{equation}\label{eqZ1}
{\rm a)}\;\; 2\dot Z_a -i \sum_b (A_a{}^b - A_b{}^a)Z_b =0\,, \qquad\qquad
{\rm b)}\;\; \sum_b (A_a{}^b + A_b{}^a)Z_b =0 \,,
\end{equation}
\begin{equation}\label{eqA}
A_a{}^b =\frac{Z_a \bar Z^b }{2(x_a - x_b)^2}  \qquad \mbox{for } a\neq b\,,
\end{equation}
\begin{equation}\label{eqZ}
(Z_a)^2 =c \qquad \forall \, a  \quad (\mbox{no sum with respect to }
a)\,.
\end{equation}
The equations~(\ref{eqZ1}b), (\ref{eqA}), (\ref{eqZ}) are exactly the equations of motion,
following from the action~(\ref{b-Cal3}). And it is important that
the equations~(\ref{eqZ1}a) are corollary of the equations~(\ref{eqA}) and (\ref{eqZ})
(the equations~(\ref{eqZ}) imply $\dot Z_a =0$).

The equations~(\ref{eqZ}) imply $c>0$. Inserting~(\ref{eqZ1}b)
(which give in fact the expressions for diagonal $A_a{}^a$),
(\ref{eqA}) and (\ref{eqZ}) in the action~(\ref{b-Cal2}), we
finally obtain the standard Calogero action
\begin{equation}\label{st-Cal}
S_C = {\textstyle\frac{1}{2}}\,\int dt  \,\Big[\, \sum_{a} \dot x_a \dot x_a - \sum_{a\neq b} \frac{c^2}{(x_a -
x_b)^2}\,\Big]\,.
\end{equation}

Let us consider Hamiltonian formulation of this matrix model.
Expressions of the momenta, obtained from the action~(\ref{b-Cal}), are
\begin{equation}\label{P-Cal}
P_{\!\scriptscriptstyle{X}} = 2\nabla\! X\,,\qquad\qquad
P_{\!\scriptscriptstyle{Z}} = i
\, \bar Z \,,\qquad \bar P_{\!\scriptscriptstyle{Z}} = -i\, Z\,, \qquad\qquad
P_{\!\scriptscriptstyle{A}} = 0\,.
\end{equation}
Thus, there are the constraints
\begin{equation}\label{con-Z}
G \equiv P_{\!\scriptscriptstyle{Z}} - i \, \bar Z \approx 0\,,\qquad \bar G \equiv
\bar
P_{\!\scriptscriptstyle{Z}} + i\, Z \approx 0\,,
\end{equation}
\begin{equation}\label{con-A}
P_{\!\scriptscriptstyle{A}} \approx 0\,.
\end{equation}

Hamiltonian of the system has the following form
\begin{equation}\label{H-Cal}
H = {\textstyle\frac{1}{4}} {\rm tr}\left(
P_{\!\scriptscriptstyle{X}}P_{\!\scriptscriptstyle{X}} \right) + {\rm tr}\left( A\, T
\right),
\end{equation}
where
\begin{equation}\label{con-T}
T \equiv \, i [X, P_{\!\scriptscriptstyle{X}} ] \,-\, 2 Z \!\cdot\! \bar Z \,-\, c
I_n
\end{equation}
and $I_n$ is the $(n\times n)$ unity matrix.

The preservation of the constraints~(\ref{con-A}) leads to the secondary constraints
\begin{equation}\label{con-T1}
T \approx 0 \,.
\end{equation}
The fields $A$ are Lagrange multipliers for these constraints.

Using canonical Poisson brackets
\begin{equation}\label{CPB-Cal}
[X_a{}^b, P_{{\!\scriptscriptstyle{X}}\,c}{}^d ]_{{}_P} =\delta_a^d \delta_c^b \,,
\qquad
[Z_a, P_{\!\scriptscriptstyle{Z}}^b ]_{{}_P} =\delta_a^b  \,, \qquad [{\bar Z}^a, \bar
P_{{\!\scriptscriptstyle{Z}}\,b} ]_{{}_P} =\delta^a_b\,,
\end{equation}
we compute Poisson brackets of the constraints~(\ref{con-Z})
\begin{equation}\label{PB-G}
[G^a, \bar G_b ]_{{}_P} =-2i\delta^a_b\,.
\end{equation}
Next, we introduce Dirac brackets
\begin{equation}\label{DB-G}
[A, B ]_{{}_D} =[A, B ]_{{}_P} + {\textstyle\frac{i}{2}} [A, G^a]_{{}_P} [\bar G_a , B
]_{{}_P} - {\textstyle\frac{i}{2}} [A, \bar G_a ]_{{}_P} [ G^a , B ]_{{}_P}
\end{equation}
and treat the constraints~(\ref{con-Z}) in the strong sense,  eliminating
$P_{\!\scriptscriptstyle{Z}}$, $\bar P_{\!\scriptscriptstyle{Z}}$.

The remaining variables have the following  Dirac brackets
\begin{equation}\label{CDB-Cal}
[X_a{}^b, P_{{\!\scriptscriptstyle{X}}\,c}{}^d ]_{{}_D} =\delta_a^d \delta_c^b \,,
\qquad
[Z_a, {\bar Z}^b ]_{{}_D} =-{\textstyle\frac{i}{2}}\,\delta_a^b\,.
\end{equation}
The remaining constraints~(\ref{con-T}) form $(n\times n)$ hermitian matrix,
\begin{equation}
T = T^+\,,
\end{equation}
which generates $u(n)$ algebra with respect to the Dirac brackets
\begin{equation}\label{DB-T}
[T_a{}^b, T_c{}^d ]_{{}_D} =i(\delta_a{}^d T_c{}^b - \delta_c{}^b T_a{}^d )\,.
\end{equation}
They generate U(n) gauge transformations of $X_a{}^b$,
$P_{{\!\scriptscriptstyle{X}}\,a}{}^b$, $Z_a$, ${\bar Z}^a$. Let us
fix the gauges with respect to these transformations.

In the notations
\begin{equation}\label{not-Cal1}
x_a \equiv X_a{}^a\,, \qquad p_a \equiv
P_{{\!\scriptscriptstyle{X}}\,a}{}^a    \qquad (\mbox{no summation over }
a)\,,
\end{equation}
\begin{equation}\label{not-Cal2}
x_a{}^b \equiv X_a{}^b\,, \qquad p_a{}^b \equiv P_{{\!\scriptscriptstyle{X}}\,a}{}^b
  \qquad \mbox{for } a\neq b
\end{equation}
the constraints~(\ref{con-T}) take the form
\begin{equation}\label{T-ab}
T_a{}^b =i(x_a -x_b) p_a{}^b - i(p_a -p_b) x_a{}^b + i\sum_{c}(x_a{}^c p_c{}^b -
p_a{}^c
x_c{}^b) - 2 Z_a {\bar Z}^b \approx 0  \qquad \mbox{for } a\neq b \,,
\end{equation}
\begin{equation}\label{T-a}
T_a{}^a =i\sum_{c}(x_a{}^c p_c{}^a - p_a{}^c x_c{}^a) - 2 Z_a {\bar Z}^a - c \approx 0
\qquad (\mbox{no summation over } a) \,.
\end{equation}

The constraints~(\ref{T-ab}) generate the transformations
\begin{equation}
\delta x_a{}^b = [x_a{}^b, \epsilon_b{}^a T_b{}^a ]_{{}_D} \sim i(x_a
-x_b)\epsilon_b{}^a\,.
\end{equation}
When $x_a \neq x_b$ we can impose the gauge
\begin{equation}\label{gau-ab}
x_a{}^b \approx 0\,.
\end{equation}
For the constraints~(\ref{T-ab}), (\ref{gau-ab}) we introduce Dirac
brackets. As a result, we eliminate $x_a{}^b$, $p_a{}^b$ by these
constraints:
\begin{equation}\label{res-ab}
x_a{}^b = 0\,, \qquad p_a{}^b = -\frac{2i}{(x_a -x_b)}\, Z_a {\bar Z}^b\,.
\end{equation}
Thus
\begin{equation}\label{P-X-fix}
P_{{\!\scriptscriptstyle{X}}\,a}{}^b = \Big(p_a\delta_a{}^b \quad \mbox{for } a= b \,;\qquad
-\frac{2i}{(x_a -x_b)}\, Z_a {\bar Z}^b \quad \mbox{for } a\neq b \Big)\,.
\end{equation}
After this gauge fixing the constraint~(\ref{T-a}) becomes
\begin{equation}\label{T-a-fix}
- 2 Z_a {\bar Z}^a - c \approx 0 \qquad (\mbox{no summation over } a)
\end{equation}
and it generates local phase transformations of $Z_a$. We can impose
the gauge
\begin{equation}\label{Z-fix}
Z_a - {\bar Z}^a \approx 0 \,.
\end{equation}
As a result, we completely eliminated $Z_a$.

Finally, using the expressions~(\ref{P-X-fix}) and the
conditions~(\ref{Z-fix}), we obtain the following expression for the
Hamiltonian~(\ref{H-Cal})
\begin{equation}\label{H-Cal-fix}
H = {\textstyle\frac{1}{4}} {\rm tr}\left(
P_{\!\scriptscriptstyle{X}}P_{\!\scriptscriptstyle{X}} \right) =
\frac{1}{4}\left(\sum_{a}
(p_a )^2 + \sum_{a\neq b} \frac{c^2}{(x_a - x_b)^2}\right),
\end{equation}
i.e.~the standard Calogero Hamiltonian \cite{C1,C2}.

The Calogero model is an example of solvable multi-particle system.
It can be obtained by applying the reduction method to a matrix system corresponding
to the $A_n$ root system of Lie algebras \cite{Per-b}.

The quantization of the Calogero model was analyzed in the initial papers by Calogero \cite{C1,C2} where
the ground-state wave function and energy were found. The wave functions of some higher excitation states
were found in \cite{Per}. The progress in obtaining physical wave functions for the Calogero model
can be traced back to \cite{Vas1} where it was suggested to use Dunkl operators in the quantization procedure.

\setcounter{equation}{0}

\section{${\cal N}{=}\,1$ and ${\cal N}{=}\,2$ superconformal mechanics}

\subsection{${\cal N}{=}\,1$ \&\ ${\cal N}{=}\,2$ superconformal algebras and their representations}

\noindent
${\bf {\cal N}{=}\,1 }$ {\bf case. \quad}
The ${\cal N}{=}\,1$ superconformal algebra  is constituted by
the following set of generators ($\alpha,\beta=1,2$):
\begin{equation}\label{osp12-gen}
G^{(1)} =\left(  {Q}_{\alpha};  {T}_{\alpha\beta}\right)\,,\qquad
({Q}_{\alpha}){}^\dag = {Q}_{\alpha}\,.
\qquad
({T}_{\alpha\beta})^\dag = {T}_{\alpha\beta}= {T}_{\beta\alpha}\,.
\end{equation}
They form the real $osp(1|2)$ superalgebra which thus defines graded symmetries of ${\cal N}{=}\,1$ superconformal
mechanics \cite{AIPT,Britto}. The nonvanishing (anti)commutators are  (\ref{o21-spin}) and
\begin{equation}\label{osp12-QQ}
\{{Q}_{\alpha}, {Q}_{\beta} \}=
2\, {T}_{\alpha\beta}\,,
\qquad
[{T}_{\alpha\beta}, {Q}_{\gamma}]=-i\,\epsilon_{\gamma(\alpha}{Q}_{\beta)}\,.
\end{equation}
The fermionic $\textrm{O}(2,1)$ spinor supercharges ${Q}_{\alpha}$ encompass the standard supercharges
${Q}={Q}_{1}$ and the generators of superconformal boosts
${S}={Q}_{2}$.
The second-order Casimir operator of the supergroup $\textrm{OSp}\,(1|2)$ is given by the following
expression
\begin{equation}\label{qu-Cas1}
{C}^{({\cal N}{=}\,1)}_2={T}^2 +
{\textstyle\frac{i}{4}}\, {Q}_{\alpha}{Q}^{\alpha}\,.
\end{equation}
\vspace{0.3cm}

\noindent
${\bf {\cal N}{=}\,2 }$ {\bf case. \quad}
The ${\cal N}{=}\,2$ superconformal algebra involves the following set of the generators
\begin{equation}\label{osp22-gen}
G^{(2)}=\left( {Q}_{\alpha}, \bar{Q}_{\alpha}; {T}_{\alpha\beta}, {J}, {C}\right)\,,
\end{equation}
which define the $su(1,1|1)\,{\cong}\,osp\,(2|2)$ superalgebra with one central charge (see, e.g., \cite{AP,FR,IKL2,AIPT})
\begin{equation}\label{osp22-QQ}
\{{Q}_{\alpha}, \bar{Q}_{\beta} \}=
2\, {T}_{\alpha\beta}+ i\epsilon_{\alpha\beta} {J}+i \epsilon_{\alpha\beta} {C}\,,
\end{equation}
\begin{equation} \label{osp22-BQ1}
[{T}_{\alpha\beta}, {Q}_{\gamma}]=-i\,\epsilon_{\gamma(\alpha}{Q}_{\beta)}\,,\qquad
[{T}_{\alpha\beta}, \bar{Q}_{\gamma}]=-i\,\epsilon_{\gamma(\alpha}\bar{Q}_{\beta)}\,,
\end{equation}
\begin{equation} \label{osp22-BQ2}
[{J}, {Q}_{\alpha}]= {Q}_{\alpha}\,,\qquad
[{J}, \bar{Q}_{\alpha}]=- \bar{Q}_{\alpha}\,,
\end{equation}
where also the $\textrm{SO}(1,2)$ relations (\ref{o21-spin})  should be added. All other (anti)commutators vanish.
The hermiticity properties of the  $\textrm{SU}(1,1|1)$ generators are as follows
\begin{equation}\label{osp22-Her-Q}
({Q}_{\alpha}){}^\dag = \bar{Q}_{\alpha}\,,
\qquad
({T}_{\alpha\beta})^\dag = {T}_{\alpha\beta}= {T}_{\beta\alpha}\,,\qquad
({J})^\dag = {J}\,,\qquad
({C})^\dag = {C}\,.
\end{equation}
The generators ${T}_{\alpha\beta}$ form $so(1,2)$ algebra, whereas ${J}$
is the $\textrm{O}(2)$ generator. The generator ${C}$ is the central charge one.
Here we used the realization of the $osp(2|2)$ algebra
as in \cite{IKL2}, with fermionic supercharges being complex $\textrm{O}(2)$ spinors.
In particular, the standard supercharges
${Q}={Q}_{1}$, $\bar{Q}=\bar{Q}_{1}$ and the generators of conformal supertranslations
${S}={Q}_{2}$, $\bar{S}=\bar{Q}_{2}$ are combined into a single complex $\textrm{O}(2,1)$ doublet supercharge ${Q}_{\alpha}$.

The second-order Casimir operator of the supergroup ${\rm OSp}\,(2|2)$ is given by the following
expression
\begin{equation}\label{qu-Cas2}
{C}^{({\cal N}{=}\,2)}_2={T}^2 +
{\textstyle\frac{i}{4}}\, [{Q}_{\alpha},\bar{Q}^{\alpha}]-{\textstyle\frac{1}{4}}\,J^2-{\textstyle\frac{1}{4}}\,\{J,C\}\,.
\end{equation}
Unitary irreducible representations act on the eigenstates of the second order Casimir  ${C}_2$ with a fixed value.
Such  representations are decomposed into an infinite tower of the representations of compact supergroup
including compact generator $T_0$ (\ref{T-vec-def}) of the conformal group, similarly to bosonic case.
In ${\cal N}{=}\,2$ case this compact sub-superalgebra  of $osp(2|2)$ superalgebra
is spanned by the generators
\begin{equation}\label{comp-subalg2}
T_0\,,\quad J\,,\quad C\,,\quad \Gamma\equiv m^{1/2}S+im^{-1/2}Q\,,\quad \bar\Gamma\equiv m^{1/2}\bar S-im^{-1/2}\bar Q\,,
\end{equation}
with the following nonvanishing (anti)commutators
\begin{equation}\label{acom-subalg2}
\{\Gamma,\bar\Gamma \}=4\,T_0-2\,C-2\,J\,,
\end{equation}
\begin{equation}
[T_0,\Gamma ]={\textstyle\frac{1}{2}}\,\Gamma\,,\quad [T_0,\bar\Gamma ]=-{\textstyle\frac{1}{2}}\,\bar\Gamma\,,\qquad
[J,\Gamma ]= \Gamma\,,\quad [J,\bar\Gamma ]=- \bar\Gamma\,.
\end{equation}

\subsection{Single-particle mechanics: nonlinear realizations and actions}

We shall consider ${\cal N}{=}\,1$ superspace parameterized by $(t,\theta)$ where $t$ is even coordinate and
$\theta$ is a real Grassmann coordinate, $\theta^2=0$, $(\overline{\theta})=\theta$. The covariant spinor derivative is
\begin{equation}
D = \partial_{\theta} +i\theta\partial_{t}\,, \qquad \{D, D \}
= 2i\,
\partial_{t}\,.
\end{equation}
Coordinates of ${\cal N}{=}\,2$ superspace are
$(t,\theta, \bar\theta)$ where $t$ is even coordinate again and
$\theta$, $\bar\theta$ are two Grassmanian coordinates, $\theta^2=\bar\theta^2= 0$, $(\overline{\theta})=\bar\theta$.
${\cal N}{=}\,2$ covariant spinor derivatives are
\begin{equation}
D = \partial_{\theta} -i\bar\theta\partial_{t}\,, \qquad \bar D = -\partial_{\bar\theta}
+i\theta\partial_{t}\,, \qquad \{D, \bar D \} = 2i \partial_{t}\,.
\end{equation}

As given in \cite{CDKKTP,AIPT}, the ${\cal N}{=}\,1$ superconformal one-particle model is described by free particle action.
It follows from the superfield action
\begin{equation}\label{N1-free1}
S^{(N=1)}_{n=1} = -i\int dt d \theta \,\dot \Phi D
\Phi \,,
\end{equation}
where $\Phi(t,\theta)$ is a real superfield  with the following off-shell $({\bf 1,1,0})$ component contents
\begin{equation}\label{N1-comp-sf1}
\Phi(t,\theta) =  x(t) + i \theta \psi(t) \,.
\end{equation}
The action (\ref{N1-free1}) yields free actions for boson $x$ and fermion $\psi$, i.e. provides a supersymmetrization
of the AFF mechanics with the vanishing conformal potential.

The ${\cal N}{=}\,2$ superconformal mechanics \cite{AP,FR} can be described by a real ${\cal N}{=}\,2$ superfield
with the off-shell contents $({\bf 1,2,1})$
\begin{equation}\label{N2-comp-sf1}
\Phi(t,\theta,\bar\theta) =  x(t) + \theta \psi(t) -\bar\theta \bar\psi(t) +\theta \bar\theta y(t)\,,
\end{equation}
and with the action
\begin{equation}\label{N2-act-1}
S^{(N=2)}_{n=1} = \int dt d^2\theta \,\Big[\, \bar{D}
\Phi\,  {D} \Phi -  c\,\ln \Phi \,\Big] \,.
\end{equation}
The superpotential
\begin{equation}\label{N2-spot-1}
W =   c\,\ln \Phi
\end{equation}
produces the conformal potential for bosonic component field. In components, the action (\ref{N2-act-1}) takes the form
\begin{equation}\label{N2-act-1a}
S^{(N=2)}_{n=1} = \int dt  \,\left[\, \dot x^2
- i\left(\dot{\bar\psi}\psi -{\bar\psi}\dot\psi\right) +y^2 -\frac{c\, y}{x} +\frac{c\psi{\bar\psi}}{x^2}\,\right].
\end{equation}
After eliminating the auxiliary field $y$ we obtain
\begin{equation}\label{N2-act-1b}
S^{(N=2)}_{n=1} = \int dt  \,\left[\, \dot x^2
- i\left(\dot{\bar\psi}\psi -{\bar\psi}\dot\psi\right) - \frac{c(c-4 {\bar\psi}\psi)}{4x^2}\,\right],
\end{equation}
which is just ${\cal N}{=}\,2$ supersymmetric generalization of the AFF mechanics considered
firstly in \cite{FR} and in \cite{AP}.

Similarly to the bosonic case, the formulation (\ref{N2-act-1}) has a nice geometric interpretation within
the nonlinear realizations method \cite{IKL2}. One starts from the exponential parametrization of the coset
$\textrm{SU}(1,1|1)/\textrm{U}(1)$
\begin{equation}\label{N2-exp-param}
G = e^{itH}e^{i(\theta Q-\bar\theta\bar Q)}e^{izK}e^{i(\xi S-\bar\xi\bar S)}e^{iu D}\,.
\end{equation}
The coset parameters associated with the generators $K$, $S$, $\bar S$, $D$ are treated as Goldstone superfields,
$z=z(t,\theta,\bar\theta)$, $\xi=\xi(t,\theta,\bar\theta)$, $\bar\xi=\bar\xi(t,\theta,\bar\theta)$, $u=u(t,\theta,\bar\theta)$.
The generator $J$ corresponds to the vacuum stability subgroup U(1).
Imposing the inverse Higgs constraints on the left-covariant MC forms, one can eliminate a part of Goldstone superfields
and obtain the equations of motion for the residual fields.
The correct set of such constraints was derived in  \cite{IKL2} (see also \cite{ACP}). It corresponds to vanishing of all MC forms
except for those associated with the generators of the sub-superalgebra (\ref{comp-subalg2}). As a result, the appropriate
geodesic submanifold is singled out. The coset parameters $\tau$, $\eta$, $\bar\eta$, corresponding to the generators $T_0$, $\Gamma$,
$\bar\Gamma$, parametrize this geodesic supermanifold. As a result of imposing the constraints on the MC forms,
including the kinematical inverse Higgs ones, the only independent superfield is the dilaton superfield $u(t,\theta,\bar\theta)$.
The superfield  (\ref{N2-comp-sf1}) is expressed in terms of the dilaton as $\Phi=\exp(u/2)$.

The quantum generators of the Poincare superalgebra are calculated to be
\begin{equation}\label{N2-H-1}
H = \frac{1}{4}\left[  \hat p{}^2+\,\left(\frac{d W}{d \hat x}\right)^2\,\right]
+\frac12\,\frac{d^2 W}{d \hat x^2}\left(\hat \psi\hat {\bar\psi} -\hat {\bar\psi}\hat \psi\right),
\end{equation}
\begin{equation}\label{N2-Q-1}
Q = \hat \psi\left(\hat p-i\,\frac{d W}{d \hat x}\,\right),\qquad \bar Q = \hat {\bar\psi}\left(\hat p+i\,\frac{d W}{d \hat x}\,\right),
\end{equation}
whereas the superconformal boost generators are
\begin{equation}\label{N2-S-1}
S = -2\hat \psi\,\hat x\,,\qquad \bar S = -2\hat {\bar\psi}\,\hat x\,.
\end{equation}
Here $W(\hat x) = c \ln \hat x\,$. Using the algebra of the basic
quantum operators
\begin{equation}\label{N2-qu-br-1}
[ \hat x, \hat p]=i\,,\qquad \{ \hat\psi, \hat{\bar\psi}\}= {\textstyle\frac{1}{2}}\,,
\end{equation}
we find the non-vanishing anticommutators of the odd generators (\ref{N2-Q-1}) and
(\ref{N2-S-1}):
\begin{equation}\label{N2-QQ-SS-1}
\{ Q, \bar Q\}  =2H\,,\qquad \{ S, \bar S\} =2K\,,
\end{equation}
\begin{equation}\label{N2-QS-1}
\{ Q, \bar S\} =2D+iJ+iC \,,\qquad \{ \bar Q, S\} =2D-iJ-iC \,.
\end{equation}
Here, the generators
\begin{equation}\label{N2-KD-1}
K=\hat x^2\,,\qquad D=-{\textstyle\frac{1}{4}}\,(\hat x\hat p + \hat p \,\hat x)
\end{equation}
together with $H\,$ form the $su(1,1)$ algebra, the quantity
\begin{equation}\label{N2-J}
J= \hat \psi\hat {\bar\psi}-\hat {\bar\psi}\hat \psi
\end{equation}
is the U(1) generator, and
\begin{equation}\label{N2-C-1}
C=\hat x\,\frac{d W}{d \hat x}=c
\end{equation}
is the central charge. The remaining commutation relations in which the generators (\ref{N2-Q-1}), (\ref{N2-S-1}) and
(\ref{N2-KD-1})-(\ref{N2-C-1}) appear coincide with those present in the $su(1,1|1)$ superalgebra (\ref{osp22-gen})-(\ref{osp22-BQ2}).
A convenient realization of the quantum fermionic operators is through Pauli matrices,
\be
\hat \psi=\frac{1}{2\sqrt{2}}\left(\sigma_1 +i\sigma_2 \right), \quad \hat {\bar\psi}=\frac{1}{2\sqrt{2}}\left(\sigma_1 -i\sigma_2 \right),
\quad \hat \psi\hat {\bar\psi}-\hat {\bar\psi}\hat \psi = {\textstyle\frac{1}{2}}\,\sigma_3\,. \lb{Pauli5}
\ee

We see that, as opposed to the non-supersymmetric case, the energy spectrum in the one-particle ${\cal N}{=}\,2$ superconformal
model is doubly degenerate and the quantum Hamiltonian (\ref{N2-H-1}) takes the following form
\begin{equation}\label{N2-H-1-a}
H = \frac{1}{4}\left[  \hat p{}^2+\frac{c(c-\sigma_3)}{\hat x^2}\,\right].
\end{equation}
Since the second term in the brackets can be rewritten as $l(l+1)/x^2$, where $l=\pm c$ for the $\sigma_3$ entries $\mp 1$,
the strength $|c|$ was identified in \cite{AIPT} with the orbital angular momentum of
a particle near the horizon (i.e. in the large mass limit) of the extreme Reissner-Nordstr\"om black hole.\footnote{This limiting
case corresponds to what is called Bertotti-Robinson black hole.}

The quantum spectrum (with spontaneously broken superconformal symmetry) can be found by the same token
as in the bosonic case, using the group-theoretical information sketched in the previous subsection.
As was noticed in \cite{FR}, the ``naive'' application of the ``standard'' quantization scheme to this system
does not lead to the desired result. Indeed, defining the wave function $\Psi_0$ of the ground state by the conditions
\begin{equation}\label{Psi-Q0}
Q\,\Psi_0=0\,,\qquad \bar Q\,\Psi_0=0\,,
\end{equation}
which amount to the single matrix equation
\begin{equation}
\left(\partial_x-\frac{c\sigma_3}{x}\, \right)\Psi_0=0\,,
\end{equation}
we immediately find that the two-component wave function $\Psi_0$ has the following entries
\begin{equation}\label{Psi0-0n}
\Psi_0=\left(
\begin{array}{c}
A_+\,x^{c} \\
A_-\,x^{-c} \\
\end{array}
\right),
\end{equation}
with $A_\pm$ being some constants. Obviously, neither component of the ground state wave function (\ref{Psi0-0n}) is normalizable,
while, at the same time, the non-zero energy states are described by the plane-wave normalizable
Bessel wave functions (see \cite{AFF,FR} for details).

The basic step in the correct quantization of the ${\cal N}{=}\,2$ superconformal mechanics is to single out the compact
sub-superalgebra  (\ref{comp-subalg2}) in the full $osp(2|2)$ superalgebra of conserved charges.
Then, to define the ground state, we impose, instead of (\ref{Psi-Q0}), the following conditions~\cite{AP,FR}
\begin{equation}\label{Psi-Gamma}
\Gamma\,\Psi_0=0\,,\qquad \bar \Gamma\,\Psi_0=0\,,
\end{equation}
where $\Gamma$, $\bar\Gamma$ are defined in (\ref{comp-subalg2}). Explicitly,  these supercharges are as follows
(cf. (\ref{N2-Q-1}))
\begin{equation}\label{Gamma-2-Q-1}
m^{1/2}\Gamma = i\hat \psi\left(\hat p-i\,\frac{d \tilde W}{d \hat x}\,\right),\qquad
\bar \Gamma = -i\hat {\bar\psi}\left(\hat p+i\,\frac{d \tilde W}{d \hat x}\,\right),
\end{equation}
where
$$
\tilde W:=W-mx^2=c\ln x-mx^2\,.
$$
Eqs. (\ref{Psi-Gamma}) amount to the condition
\begin{equation}\label{N2-vac-eq}
\left(\partial_x-\sigma_3\, \frac{d \tilde W}{d \hat x}\, \right)\Psi_0=0\,,
\end{equation}
which is solved by
$$
\Psi_0=\left(
\begin{array}{c}
\Psi_0^{(+)} \\
\Psi_0^{(-)} \\
\end{array}
\right)=
\left(
\begin{array}{c}
A_+ x^{c}e^{- mx^2} \\
A_- x^{- c}e^{mx^2} \\
\end{array}
\right).
$$
Assuming, without loss of generality, that the ``mass'' parameter $m$ is positive
we can choose the wave function $\Psi_0^{(+)}$ as the true normalizable ground state wave function.
For definiteness, we can also choose this vacuum state to be bosonic \cite{FR}.
Taking into account the relation (\ref{acom-subalg2}), we conclude that
$$
\left(T_0-{\textstyle\frac{1}{2}}\,C-{\textstyle\frac{1}{2}}\,J\right)\Psi_0^{(+)} = 0\,.
$$
Since  $C=c$ and $J=\frac{1}{2}\,$ on the vacuum wave function,  the vacuum eigenvalue of the conformal
operator $T_0$ is
\begin{equation}\label{N2-vac-r0}
r_0={\textstyle\frac{1}{2}}(c+{\textstyle\frac{1}{2}})\,.
\end{equation}
Note that the relations \p{Psi-Gamma} can be interpreted as expressing the property that the whole OSp(2$|$2) superconformal
symmetry is spontaneously broken down to the compact supergroup SU(1$|$1) $\propto (\Gamma, \bar\Gamma,
T_0-{\textstyle\frac{1}{2}}\,C-{\textstyle\frac{1}{2}}\,J)\,$.

The full spectrum of the ${\cal N}{=}2$ model considered in this
subsection was described in details in \cite{AP,FR}. It should be
emphasized that the choice of the ``true'' spontaneously broken
phase can be treated as passing to the effective theory with the
redefined time and dynamical variables \cite{AP}, quite similarly to
the analogous phenomenon in the bosonic conformal mechanics
explained in Sect.\,2.2. Redefining, in the action
(\ref{N2-act-1b}), the evolution parameter and bosonic variable as
in (\ref{new-tau-q}),  and making an additional redefinition of the
fermionic variable as  $\psi(t)\rightarrow
\tilde{\psi}(\tau){=}\psi(t)$, we obtain a system which on the quantum
level is described  by the new Hamiltonian
\begin{equation}\label{tHamFF-T}
\tilde H=2m T_0=H+m^2 K=\frac{1}{4}\left[  \hat p{}^2+\frac{c(c-2J)}{\hat x^2}\,\right] +m^2 \hat x^2 \,.
\end{equation}
It basically coincides with the compact operator $T_0$ and so has the discrete spectrum.

Finally, we remark that on shell the ${\cal N}{=}\,2$ superconformal mechanics associated with the multiplet $({\bf 1,2,1})$
has an equivalent description in terms of the chiral multiplet $({\bf 2,2,0})\,$ \cite{IKL2}.

\subsection{Multi-particle mechanics}

\subsubsection{The Freedman--Mende model}

Direct construction of multiparticle system possessed ${\cal N}{=}\,2$ superconformal symmetry
is achieved by using $n$ real superfields $\Phi_a$, $a=1,2,\ldots,n$ (\ref{N2-comp-sf1})
describing $n$  $({\bf 1,2,1})$ multiplets. Superfield action is the following
\begin{equation}\label{N2-act-n-gen}
S^{(N=2)}_{n} = \int dt d^2\theta \,\Big[\, \sum_{a=1}^{n}\bar{D}
\Phi_a\,  {D} \Phi_a -  W (\Phi_a )\,\Big] \,,
\end{equation}
where the superpotential $W (\Phi_a )$ is included. The component action of this model is
\begin{equation}\label{N2-act-n-comp-gen}
S^{(N=2)}_{n} = \int dt  \,\left[\, \sum_{a=1}^{n}\left(\dot x_a^2
- i\dot{\bar\psi}_a\psi_a +i{\bar\psi}_a\dot\psi_a - {\textstyle\frac{1}{4}}\,\partial_a W \partial_a W\right) +
\sum_{a, b} {\bar\psi}_a \psi_b\partial_a \partial_b W\,\right],
\end{equation}
where $\partial_a :=\partial/\partial x_a$.
Of course, the action (\ref{N2-act-n-comp-gen}) possesses ${\cal N}{=}\,2$ Poincare supersymmetry, which fermionic charges are
\begin{equation}\label{N2-Q-n}
Q = \sum_{a}\psi_a\Big(\, p_a-i\,\partial_a W \Big),\qquad
\bar Q = \sum_{a}\bar\psi_a\Big(\, p_a+i\,\partial_a W \Big).
\end{equation}
The requirement of ${\cal N}{=}\,2$ superconformal symmetry imposes strong constraints on the superpotential.
If we consider the superconformal boost transformation of the $({\bf 1,2,1})$ multiplets
\cite{IKL2}
\begin{equation}  \label{2N-n-Phi}
\delta^\prime \Phi_a= -i(\eta\bar\theta+\bar\eta\theta)\,\Phi_a
\end{equation}
and take into account the invariance of the measure $\delta^\prime (dtd^2\theta)=0$,
we find that the action (\ref{N2-act-n-gen}) is invariant under superconformal boosts only if the superpotential
satisfies the condition
\begin{equation}  \label{2N-cond-s-pot}
x_a \partial_a\, W(x)= C\,,
\end{equation}
where $C$ is a constant. As shown in \cite{GL} by detail  calculations,
this constant $C$ coincides with the central charge in general ${\cal N}{=}\,2$ superconformal algebra
(\ref{osp22-QQ})-(\ref{osp22-BQ2}). The full set of the generators of the ${\cal N}{=}\,2$ superconformal algebra,
including quantum case, was presented in \cite{GL}. In particular, besides the Poincar\'e supersymmetry
generators (\ref{N2-Q-n}) there are additional  fermionic generators which correspond to the superconformal boosts
and are given by the expressions
\begin{equation}\label{N2-S-n}
S = t Q -2 \sum_{a}\psi_a\,x_a\,,\qquad \bar S = t\bar Q -2 \sum_{a}\bar\psi_a\,x_a\,.
\end{equation}
The closure of the generators  (\ref{N2-Q-n}), (\ref{N2-S-n}) reproduces the whole ${\cal N}{=}\,2$ superconformal algebra.
As shown in \cite{GL}, in the case when quantum Hamiltonian contains only boson-fermion couplings without boson-boson interaction,
the superpotential $W$ is defined by a harmonic homogeneous function of $x_a$ and the central charge $C$
satisfies some ``quantization'' conditions.

An important particular case of the superfield action (\ref{N2-act-n-gen}) corresponds to the superpotential of the form \cite{BGK}
\begin{equation}\label{N2-spot-n}
W =   c\,\sum_{a\neq b}\ln \left(\Phi_a -\Phi_b\right),
\end{equation}
which is a generalization of the one-particle superpotential (\ref{N2-spot-1}). Then, the component action of this model is
\begin{equation}\label{N2-act-n-comp}
S^{(N=2)}_{n} = \int dt  \,\left[\, \sum_{a=1}^{n}\left(\dot x_a^2
- i\dot{\bar\psi}_a\psi_a +i{\bar\psi}_a\dot\psi_a\right) -
\sum_{a\neq b}\frac{c^2+4c({\bar\psi}_a-{\bar\psi}_b)(\psi_a-\psi_b)}{4(x_a-x_b)^2}\,\right].
\end{equation}
The action (\ref{N2-act-n-comp}) provides just ${\cal N}{=}\,2$ superconformal generalization of the Calogero model proposed
by Freedman and Mende \cite{FM}. Thus, in the Freedman--Mende model the strength $c$ of the conformal potential
links with the central charge $C$ in  the ${\cal N}{=}\,2$ superconformal algebra by
\begin{equation}\label{N2-C-g}
C=n(n{-}1)c\,,
\end{equation}
which directly follows from the condition \p{2N-cond-s-pot}.

A different interesting new case of $n$-particle system with ${\cal N}{=}\,2$ superconformal symmetry was obtained in
\cite{GLP1} via nontrivial nonlinear transformation from the free ${\cal N}{=}\,2$ superconformal $n$-particle
system.\footnote{The geometric meaning of such transformations for simple (super)conformal systems was
explained in~\cite{HakNer}.}
This new interacting system is described by the superpotential
\begin{equation}\label{N2-spot-n0}
W (x)=   \nu\,\ln \Big(\sum_{a}x_a x_a\Big)\,,
\end{equation}
where $\nu$ is a constant. It presents a ${\cal N}{=}\,2$ superconformal generalization of the motion of the $n$-particle center
of mass.

\subsubsection{Gauged models}

Using the $d{=}1$ gauging method applied earlier for deriving the Calogero system,
one can construct new many-body systems with ${\cal N}{=}\,1$ and ${\cal N}{=}\,2$ superconformal symmetry.
In ${\cal N}{=}\,2$ case these new systems and the Freedman--Mende system differ in  their fermionic sectors.
This type of superconformal extensions of the Calogero model was proposed and briefly described in \cite{FIL1}.
Here we present them in more detail.\\

\noindent
{\bf ${\cal N}{=}\,1$ multiparticle mechanics. \quad}
We start with the model which uses the hermitian ${\cal N}{=}\,1$ $(n\times n)$-matrix superfield
$\mathscr{X}_a{}^b(t, \theta)$, $(\mathscr{X})^+
=\mathscr{X}$,
and  ${\cal N}{=}\,1$ U($n$)-spinor superfield
$\mathcal{Z}_a (t, \theta)$, $\bar \mathcal{Z}^a (t, \theta)
= (\mathcal{Z}_a)^+$
$a,b=1,\ldots ,n$.
Gauge superfields in the present case are the anti-hermitian odd
$(n\times n)$-matrix superfield
$\mathscr{A}_a{}^b(t, \theta )$, $\mathscr{A}_a{}^b \equiv
-\overline{(\mathscr{A}_b{}^a)}$, $(\mathscr{A} \equiv
-\mathscr{A}^+)$,
which are spinor connections covariantizing the spinor derivatives.
The even gauge potential superfield is
$
\mathscr{A}_t = -iD\mathscr{A} - \mathscr{A}\mathscr{A}
$.
This composite superfield is the gauge connection covariantizing
the derivative $\partial_t$.

The gauge invariant action has the following form
\begin{equation}\label{N1-Cal}
S^{(N=1)}_{} = -i\int dt d\theta \,\Bigg[\, {\rm tr}
\left( \nabla_t \mathscr{X} \,  \mathscr{D} \mathscr{X}\, \right) +
{\textstyle\frac{i}{2}}\,(\bar \mathcal{Z}\, \mathscr{D} \mathcal{Z} - \mathscr{D}\bar
\mathcal{Z}\, \mathcal{Z} )+ c\,{\rm tr} \mathscr{A} \,\Bigg] \,.
\end{equation}
Here the covariant derivatives are defined as
\begin{equation}\label{cov1-der-X}
\mathscr{D} \mathscr{X} =  D \mathscr{X} +i [ \mathscr{A} ,
\mathscr{X}] \,, \qquad \nabla_t \mathscr{X} =
\partial_{t}\mathscr{X} +i [ \mathscr{A}_t , \mathscr{X}]\,.
\end{equation}
\begin{equation}\label{cov1-der-Z}
\mathscr{D} \mathcal{Z} =  D \mathcal{Z} +i \mathscr{A} \mathcal{Z}
\,, \qquad \mathscr{D} \bar\mathcal{Z} =  D \bar\mathcal{Z}-i
\bar\mathcal{Z} \mathscr{A} \,.
\end{equation}
Being composed only of the gauge covariant objects, the action~(\ref{N1-Cal}) is invariant with respect to the local
U(n) transformations:
\begin{equation}\label{tran1-X}
\mathscr{X}^{\,\prime} =  e^{i\tau}\, \mathscr{X}\, e^{-i\tau} \,,
\end{equation}
\begin{equation}\label{tran1-Z}
\mathcal{Z}^{\prime} =  e^{i\tau} \mathcal{Z} \,, \qquad \bar
\mathcal{Z}^{\prime} =  \bar \mathcal{Z}\, e^{-i\tau}\,,
\end{equation}
\begin{equation}\label{tran1-A}
\mathscr{A}^{\,\prime} =  e^{i\tau}\, \mathscr{A}\, e^{-i\tau} - i\,
e^{i\tau} (D e^{-i\tau})\,, \qquad \mathscr{A}_t^{\,\prime} =
e^{i\tau}\, \mathscr{A}_t\, e^{-i\tau} - i\, e^{i\tau} (\partial_t
e^{-i\tau})\,,
\end{equation}
where $ \tau_a{}^b(t, \theta) \in u(n) $ is the hermitian $(n\times
n)$-matrix parameter, $(\tau)^+ =\tau$.

Let us check superconformal invariance of the action~(\ref{N1-Cal}).
We shall consider only superconformal boosts since all superconformal transformations
are contained in the closure of the latter and Poincar\'e supersymmetry which is manifest in the superfield description.
The superconformal boost transformations of the coordinates are
\begin{equation}  \label{1N-sc-c}
\delta^\prime t= -i\,\eta\theta t\,,\qquad
\delta^\prime \theta = \eta t\,,\qquad
\delta^\prime (dtd\theta)=(dtd\theta)(-i\,\eta\theta)\,,\qquad
\delta^\prime D = i\,\eta\theta\,D\,,
\end{equation}
whereas the corresponding transformations of the superfields read
\begin{equation}  \label{1N-sc-X}
\delta^\prime \mathscr{X}= -i\,\eta\theta\,\mathscr{X}\,;\qquad \delta^\prime \mathscr{A} = i\,\eta\theta\,\mathscr{A}\,;
\qquad\,\delta^\prime \mathcal{Z} = 0\,,\qquad \delta^\prime \bar \mathcal{Z} = 0.
\end{equation}
Note that
\begin{equation}  \label{1N-sc-cDX}
\delta^\prime \left(\mathscr{D}\mathscr{X}\right) =
i\,\eta\,\mathscr{X} \,,\qquad
\delta^\prime \left(\nabla_t\mathscr{X}\right) =
i\,\eta\theta\left(\nabla_t\mathscr{X}\right)
-\eta \left(\mathscr{D}\mathscr{X}\right) \,.
\end{equation}
Using the expressions~(\ref{1N-sc-c})-(\ref{1N-sc-cDX}) we find that the variation of the action~(\ref{N1-Cal})
is a total derivative. For example,
\begin{equation}
\delta^\prime \int dt d\theta \,{\rm tr} \left( \nabla_t \mathscr{X} \,  \mathscr{D} \mathscr{X}\, \right)=
-i\eta\int dt d\theta \,{\rm tr} \left( \mathscr{X} \nabla_t \mathscr{X} \right)=
-{\textstyle\frac{i}{2}}\,\eta\int dt d\theta \,\partial_t{\rm tr} \left( \mathscr{X}^{\,2}\right).
\end{equation}

The component contents of the ${\cal N}{=}\,1$ superfields are
\begin{equation}\label{com-sf}
\mathscr{X} =  X + i \theta \Psi \,, \qquad \mathcal{Z} = Z+
\theta\Upsilon\,,\quad \bar\mathcal{Z} = \bar Z- \theta\bar\Upsilon
\,, \qquad \mathscr{A} = i(\Phi + \theta A)\,.
\end{equation}
Due to the gauge invariance~(\ref{tran1-A}) we can choose WZ gauge
\begin{equation}\label{1N-WZ}
\mathscr{A} = i\theta A(t)\,.
\end{equation}
Substituting this expression in the action~(\ref{N1-Cal}), integrating over $\theta$ ($\int d\theta\,\theta =1$) and
eliminating the auxiliary fields $\Upsilon$ and $\bar\Upsilon$ by
their equations of motion, $\Upsilon=0$ and $\bar\Upsilon=0$, we obtain the physical component action
\begin{equation}\label{1N-Cal-WZ}
S^{(N=1)}_{WZ} =
\int dt  \,\Bigg[\, {\rm tr}\left(\nabla\! X \nabla\! X \right)-
i\,{\rm tr}\left(\Psi \nabla \Psi \right) +
{\textstyle\frac{i}{2}}\, (\bar Z \nabla\!Z - \nabla\! \bar Z Z) + c\,{\rm tr} A  \,\Bigg]\, ,
\end{equation}
where $\nabla \Psi=\Psi+i[A,\Psi]$ is ${\rm U}(n)$-covariant derivative of matrix Grassmannian od field $\Psi$.
First, third and fourth terms in (\ref{1N-Cal-WZ}) include only bosonic fields and precisely
yield the action~(\ref{b-Cal}) which was the starting point of deriving the Calogero model by the $d{=}1$
gauging approach. The second term with fermionic fields
ensures ${\cal N}{=}\,1$  supersymmetrization of the Calogero model.
Note that the system involves $n^2$ (real) fermionic degrees of freedom comprised by the matrix fermionic field $\Psi(t)$.
The bosonic and fermonic terms involve the same $d{=}1$ gauge field $A(t)$, which, being integrated out, produces non-trivial
interaction of the bosonic fields of  Calogero model with the fermions.

Equally, we can analyze the model in a supersymmetric gauge.
Using the gauge $\tau$ transformations (\ref{tran1-X}), we can
impose a (partial) gauge fixing
\begin{equation}\label{s-X-fix1}
\mathscr{X}_a{}^b =0\,,\qquad a\neq b\,.
\end{equation}
In this gauge the action~(\ref{N1-Cal}) takes the form
\begin{eqnarray}\label{N1-Cal-fix}
S^{(N=1)}_{} &= -i\int dt d \theta \,\Bigg[&\!\! \sum_a \dot
\mathscr{X}_a D \mathscr{X}_a + {\textstyle\frac{i}{2}}\, \sum_a (\bar \mathcal{Z}^a\, D
\mathcal{Z}_a - D\bar \mathcal{Z}^a\,
\mathcal{Z}_a ) \\
&& - i\sum_{a,b} (\mathscr{X}_a -\mathscr{X}_b)^2 D\mathscr{A}_a{}^b
\mathscr{A}_b{}^a - \sum_{a,b} (\mathscr{X}_a -\mathscr{X}_b)^2
(\mathscr{A}\mathscr{A})_a{}^b \mathscr{A}_b{}^a  \nonumber \\
&& +  \sum_{a,b} \bar \mathcal{Z}^a \mathscr{A}_a{}^b \mathcal{Z}_b
+ c\,\sum_{a} \mathscr{A}_a{}^a\,\Bigg] \,. \nonumber
\end{eqnarray}

Let us consider the model~(\ref{N1-Cal-fix}) for small $n$.

In the $n=1$ case the generic action~(\ref{N1-Cal-fix}) yields the action $S^{(N=1)}_{n=1} =
-i\int dt d \theta \,\dot \mathscr{X} D \mathscr{X}$ of the free real ${\cal N}{=}\,1$ supermultiplet. No any potential
term is present in the component action.

The first non-trivial case is $n=2$.  After imposing the gauge condition
conditions
\begin{equation}\label{gf-1-Z}
\mathcal{Z}_1 = \bar\mathcal{Z}^{1} \,, \qquad \mathcal{Z}_{2} =
\bar \mathcal{Z}^2
\end{equation}
for the residual gauge symmetries and eliminating the auxiliary fields $\mathcal{Z}_1$, $\mathcal{Z}_2$,
we obtain
\begin{equation}\label{N1-Cal-fix55}
S^{(N=1)}_{n=2} = -i\int dt d \theta \,\Bigg[\,
{\textstyle\frac{1}{2}} \dot \mathscr{Y} D \mathscr{Y} -
{\textstyle\frac{i}{2}} \mathscr{C} D \mathscr{C} +
{\textstyle\frac{1}{2}} \dot \mathscr{X} D \mathscr{X} -
{\textstyle\frac{i}{2}} \mathscr{B} D \mathscr{B}-
c\,\epsilon_1\epsilon_2 \frac{\mathscr{B}}{\mathscr{X}} \,\Bigg],
\nonumber
\end{equation}
where
\begin{equation}\label{n-X-Y}
\mathscr{X} \equiv \mathscr{X}_1 -\mathscr{X}_2 \,, \qquad
\mathscr{Y} \equiv \mathscr{X}_1 +\mathscr{X}_2 \,,\qquad
\mathscr{B} \equiv
\mathscr{X}\,(\mathscr{A}_1{}^2 + \mathscr{A}_2{}^1)\,, \qquad \mathscr{C}  \equiv
i\mathscr{X}\,(\mathscr{A}_1{}^2 - \mathscr{A}_2{}^1)\,.
\end{equation}
Here the constants $\epsilon_1=\pm 1$, $\epsilon_2=\pm 1$ arise from the constraint
$\mathcal{Z}_1
\mathcal{Z}_2=-{\textstyle\frac{c\,\epsilon_1\epsilon_2}{2}}$
which follows from the equations of motion for $\mathcal{Z}_1$, $\mathcal{Z}_2$. The superfields $\mathscr{X}$ and $\mathscr{Y}$
are bosonic, while $\mathscr{B}$ and $\mathscr{C}$ are fermionic

The action \p{N1-Cal-fix55} is a sum of the actions of two free ${\cal N}{=}\,1$
multiplets (superfields $\mathscr{Y}$ and $\mathscr{C}$) and two mutually interacting ${\cal N}{=}\,1$ multiplets
(superfields $\mathscr{X}$ and $\mathscr{B}$). It is easy
to see that it is the ${\cal N}{=}\,1$ superfield form of the off-shell action
of $N=2$ superconformal mechanics based on the ${\cal N}{=}\,2$ multiplet
$({\bf 1,2,1})$, supplemented by a free action  of an extra
multiplet $({\bf 1,2,1})$. The latter, from the viewpoint of the
two-particle Calogero model, is ${\cal N}{=}\,2$ superextension of the action
corresponding to the center-mass motion. The hidden ${\cal N}{=}\,1$
supersymmetry completing the manifest one to ${\cal N}{=}\,2$ and leaving the
above action invariant (up to a total derivative under the integral)
is realized as
\be
\delta\mathscr{Y} = \varepsilon \, \mathscr{C}\,,
\quad \delta \mathscr{C} = i  \varepsilon \,\dot  \mathscr{Y}\,,
\qquad \delta\mathscr{X} = \varepsilon\, \mathscr{B}\,, \quad \delta
\mathscr{B} = i  \varepsilon \,\dot  \mathscr{X}\,,
\ee
where $\varepsilon = \bar\varepsilon $ is a Grassmann transformation
parameter.
The action \p{N1-Cal-fix55}  is also invariant (each of its two constituent terms separately)
under the ${\cal N}{=}\,1$ superconformal transformations:
\bea
&& \delta t = a(t) + i\theta \chi(t)\,, \quad \delta \theta  = \chi(t) + \frac{1}{2} \theta \dot{a}(t)\,, \quad
(\partial_t)^3{a} = (\partial_t)^2{\chi} = 0\,, \nn
&& \delta D = - A \,D\,, \delta (dtd\theta) = (dt d\theta)\, A\,, \quad A = \frac{1}{2}\dot{a} + i\theta\dot{\chi}\,, \lb{N1sc1} \\
&& \delta {\mathscr{X}} = A\,{\mathscr{X}}\,, \quad \delta {\mathscr{Y}} = A\,{\mathscr{Y}}\,, \quad
\delta {\mathscr{C}} = \delta {\mathscr{B}} = 0\,. \lb{N1sc2}
\eea
Together with the hidden ${\cal N}{=}\,1$ supersymmetry, these transformations close on the ${\cal N}{=}\,2, d{=}1$ superconformal symmetry.
The component form of the action \p{N1-Cal-fix55} contains a conformal potential for the physical bosonic field
$x = \mathscr{X}|_{\theta = 0}\,$.

In the $n=3$ case, passing through the same steps as before finally
yields the following action
\begin{eqnarray}
S^{(N=1)}_{n=3} &=& -i\int dt d \theta
\,\Bigg[\sum_{a=1}^3\dot{\mathscr{X}}_a D\mathscr{X}_a
-\frac{i}{2}\sum_{a=1}^3 \left({\mathscr{B}}_a D{\mathscr{B}}_a +
{\mathscr{C}}_a D{\mathscr{C}}_a \right) \nn &&\,
+\frac{i}{2}\, \left(\frac{1}{\mathscr{X}_1 -
\mathscr{X}_2} + \frac{1}{\mathscr{X}_2 -
\mathscr{X}_3} + \frac{1}{\mathscr{X}_3 -
\mathscr{X}_1}  \right) \left({\mathscr{C}}_1
{\mathscr{B}}_2{\mathscr{B}}_3  + {\mathscr{B}}_1
{\mathscr{C}}_2{\mathscr{B}}_3 - {\mathscr{B}}_1
{\mathscr{B}}_2{\mathscr{C}}_3 + {\mathscr{C}}_1
{\mathscr{C}}_2{\mathscr{C}}_3  \right) \nn && -c
\,\frac{{\mathscr{B}}_1}{(\mathscr{X}_1
-\mathscr{X}_2)}\epsilon_1\epsilon_2 \left[1 -
\frac{i}{2c}\left({\mathscr{B}}_2{\mathscr{C}}_2 +
{\mathscr{B}}_3{\mathscr{C}}_3\right)
-\frac{1}{4c^2}({\mathscr{B}}_2{\mathscr{C}}_2)(
{\mathscr{B}}_3{\mathscr{C}}_3)\right] \nn && -c
\,\frac{{\mathscr{B}}_2}{(\mathscr{X}_2
-\mathscr{X}_3)}\epsilon_2\epsilon_3 \left[1 +
\frac{i}{2c}\left({\mathscr{B}}_1{\mathscr{C}}_1 +
{\mathscr{B}}_3{\mathscr{C}}_3\right)
-\frac{1}{4c^2}({\mathscr{B}}_1{\mathscr{C}}_1)(
{\mathscr{B}}_3{\mathscr{C}}_3)\right] \nn && -c
\,\frac{{\mathscr{B}}_3}{(\mathscr{X}_1
-\mathscr{X}_3)}\epsilon_1\epsilon_3 \left[1 -
\frac{i}{2c}\left({\mathscr{B}}_1{\mathscr{C}}_1 -
{\mathscr{B}}_2{\mathscr{C}}_2\right)
+\frac{1}{4c^2}({\mathscr{B}}_1{\mathscr{C}}_1)(
{\mathscr{B}}_2{\mathscr{C}}_2)\right]\Bigg], \lb{Actn33}
\end{eqnarray}
where
\bea
&& {\mathscr{B}}_1 = (\mathscr{X}_1 -
\mathscr{X}_2)\left(\mathscr{A}_1^2 + \mathscr{A}_2^1\right), \quad
{\mathscr{C}}_1 = i(\mathscr{X}_1 -
\mathscr{X}_2)\left(\mathscr{A}_1^2 - \mathscr{A}_2^1\right), \nn
&&{\mathscr{B}}_2 = (\mathscr{X}_2 -
\mathscr{X}_3)\left(\mathscr{A}_2^3 + \mathscr{A}_3^2\right), \quad
{\mathscr{C}}_2 = i(\mathscr{X}_2 -
\mathscr{X}_3)\left(\mathscr{A}_2^3 - \mathscr{A}_3^2\right), \nn &&
{\mathscr{B}}_3 = (\mathscr{X}_1 -
\mathscr{X}_3)\left(\mathscr{A}_1^3 + \mathscr{A}_3^1\right), \quad
{\mathscr{C}}_3 = i(\mathscr{X}_1 -
\mathscr{X}_3)\left(\mathscr{A}_1^3 - \mathscr{A}_3^1\right).
\lb{n3fermdef}
\eea
The superfield action \p{Actn33} produces a new ${\cal N}{=}\,1$ superconformal invariant system, which
reveals no clear links with the known ${\cal N}{=}\,2$ or ${\cal N}{=}\,3$ superconformally invariant systems,
contrary to the two-particle ($n=2$) case. In components, in the limit when all fermionic fields are omitted,
\p{Actn33} yields of course the 3-particle Calogero model for the fields $x_a = \mathscr{X}_a|_{\theta =0}\,$.\\

\noindent
{\bf ${\cal N}{=}\,2$ supersymmetric extension. \quad}
Here we present ${\cal N}{=}\,2$ superconformal gauged matrix model, which produces ${\cal N}{=}\,2$ supersymmetric extension
of Calogero model. Such a system is described by the hermitian $(n\times n)$-matrix superfield
$\mathscr{X}_a{}^b(t, \theta,\bar\theta)$, $(\mathscr{X})^+ =\mathscr{X}$, $a,b=1,\ldots ,n$,
and chiral U(n)-spinor superfield
$\mathcal{Z}_a (t_{\!\scriptscriptstyle{R}}, \theta) $, $\bar \mathcal{Z}^a
(t_{\!\scriptscriptstyle{L}}, \bar\theta) = (\mathcal{Z}_a)^+$,
$t_{\!\scriptscriptstyle{L,R}}=t\mp i\theta\bar\theta$,
$a,b=1,\ldots ,n$, subjected to the constraints
\begin{equation}\label{N2-chir-Z}
D \mathcal{Z}_a=0 \,, \qquad \bar D \bar \mathcal{Z}^a=0.
\end{equation}
An important ingredient of the gauging procedure is gauge superfields. Similar to the ${\cal N}{=}\,2,$ $d{=}4$ supersymmetric theories,
they can be described either by  complex $(n\times n)$ matrix bridge superfields
\begin{equation}
b_a{}^b(t, \theta,\bar\theta)\,, \qquad \bar b_a{}^b \equiv \overline{(b_b{}^a)}\,
\qquad (\bar b \equiv b^+)\,,
\end{equation}
or by the prepotential $V$ defined by
\begin{equation}\label{prepot}
e^{2V} = e^{-i\bar b} \, e^{ib} \,.
\end{equation}
The covariant derivatives of the superfield $\mathscr{X}$ are
\begin{equation}\label{cov-der-v}
\mathscr{D} \mathscr{X} =  D \mathscr{X} -i [ \mathscr{A} , \mathscr{X}] \,, \qquad
\bar\mathscr{D} \mathscr{X} = \bar D \mathscr{X} -i [ \bar\mathscr{A} , \mathscr{X}]\,,
\end{equation}
where the potentials are constructed from the bridges in the standard way
\begin{equation}\label{pot}
\mathscr{A} =  -i \, e^{i\bar b} (D e^{-i\bar b}) \,, \qquad
\bar\mathscr{A} = -i \, e^{ib} (\bar D e^{-ib})\,.
\end{equation}

The action we propose is
\begin{equation}\label{N2-Cal-v}
S^{(N=2)}_{sC} = \int dt d^2\theta \,\Bigg[\, {\rm tr} \left( \bar\mathscr{D}
\mathscr{X}
\,  \mathscr{D} \mathscr{X}\, \right) + {\textstyle\frac{1}{2}}\,\bar \mathcal{Z}\, e^{2V}\!
\mathcal{Z} - c\,{\rm tr} V \,\Bigg] \,.
\end{equation}
It is invariant with respect to the local U(n) transformations
\begin{equation}\label{tran-b}
e^{ib^\prime} = e^{i\tau} \, e^{ib} e^{-i\lambda}\,, \qquad
e^{i\bar b^\prime} = e^{i\tau} \, e^{i\bar b} e^{-i\bar\lambda}\,,
\qquad e^{2V^\prime} = e^{i\bar\lambda} \, e^{2V} e^{-i\lambda}\,,
\end{equation}
\begin{equation}\label{tran-X}
\mathscr{X}^{\,\prime} =  e^{i\tau}\, \mathscr{X}\, e^{-i\tau} \,,
\qquad
\mathcal{Z}^{\prime} =  e^{i\lambda} \mathcal{Z} \,, \qquad
\bar \mathcal{Z}^{\prime} =  \bar \mathcal{Z}\, e^{-i\bar\lambda}\,,
\end{equation}
where $\tau$ is the Hermitian $(n\times n)$-matrix parameter,
$\tau_a{}^b(t, \theta,\bar\theta) \in u(n)$, $(\tau)^+ =\tau$, and
$\lambda$ are $(n\times n)$ complex gauge parameters, $\lambda=
(\lambda_a{}^b)$, which are (anti)chiral superfields
$\lambda (t_{\!\scriptscriptstyle{L}}, \theta) \in u(n)$,
$\bar\lambda(t_{\!\scriptscriptstyle{R}}, \theta) = (\lambda)^+  \in u(n)$.

Global invariances of the action~(\ref{N2-Cal-v}) form the ${\cal N}{=}\,2$ superconformal group SU(1,1$|$1).
The transformations of the ${\cal N}{=}\,2$ superspace coordinates under the superconformal boosts
of SU(1,1$|$1) were found in \cite{IKL2}:
\begin{equation}  \label{2N-sc-c}
\delta^\prime t= -i(\eta\bar\theta+\bar\eta\theta)t\,,\qquad
\delta^\prime \theta = -\eta(t-i\theta\bar\theta)\,,\qquad
\delta^\prime \bar\theta = -\eta(t+i\theta\bar\theta)\,,
\qquad
\delta^\prime (dtd^2\theta)=0\,,
\end{equation}
\begin{equation}  \label{2N-sc-X}
\delta^\prime \mathscr{X}= -i(\eta\bar\theta+\bar\eta\theta)\,\mathscr{X}\,,
\qquad \delta^\prime \mathcal{Z} = 0\,,\qquad
\delta^\prime b = 0\,,\qquad
\delta^\prime V = 0\,.
\end{equation}

Similarly to ${\cal N}{=}\,2$ $d{=}4$ supersymmetric theories,
there is a possibility to deal only with the $\lambda$-group covariant objects.
After passing to the new Hermitian $(n\times n)$-matrix superfield
\begin{equation}\label{X-tX}
\mathcal{X} = e^{-ib} \,\mathscr{X}\, e^{i\bar b}\,,\qquad
\mathcal{X}^{\,\prime} =  e^{i\lambda}\, \mathcal{X}\, e^{-i\bar\lambda} \,,
\end{equation}
the action~(\ref{N2-Cal-v}) takes the more economical form
\begin{equation}\label{N2-Cal}
S^{(N=2)}_{sC} = \int dt d^2\theta \,\Bigg[\, {\rm tr} \left( \bar\mathcal{D}
\mathcal{X}
\, e^{2V} \mathcal{D} \mathcal{X}\, e^{2V} \right) + {\textstyle\frac{1}{2}}\,\bar \mathcal{Z}\, e^{2V}\!
\mathcal{Z} - c\,{\rm tr} V \,\Bigg],
\end{equation}
where the covariant derivatives of the superfield $\mathcal{X}$ are
\begin{equation}\label{cov-der-s2}
\mathcal{D} \mathcal{X} =  D \mathcal{X} + e^{-2V} (D e^{2V}) \, \mathcal{X} \,, \qquad
\bar\mathcal{D} \mathcal{X} = \bar D \mathcal{X} - \mathcal{X} \, e^{2V} (\bar D
e^{-2V})\,.
\end{equation}

In the abelian ($n=1$) case, similarly to the  $\mathcal{N}=0,1$ cases, the
action~(\ref{N2-Cal}) describes the free  $\mathcal{N}=2$ real supermultiplet. Nontrivial superconformal models start from $n\geq 2\,$.
Let us find dynamical component contents of this generic $n\geq 2\,$ model.

The component expansions of the involved superfields are
\begin{equation}\label{com-V}
V = v + \theta \Phi - \bar\theta \bar\Phi + \theta\bar\theta A \,,
\end{equation}
\begin{equation}\label{com-X}
\mathcal{X} = X + \theta \Psi - \bar\theta \bar\Psi + \theta\bar\theta Y \,,
\qquad
\mathcal{Z} = Z + 2i\theta \Upsilon + i\theta\bar\theta \dot Z \,, \qquad
\bar\mathcal{Z} =
\bar Z + 2i\bar\theta \bar\Upsilon - i\theta\bar\theta \dot{\bar Z}\,,
\end{equation}
where $\Psi_a{}^b$, $\bar\Psi_a{}^b=(\overline{\Psi_b{}^a})$
($\bar\Psi = \Psi^+$), $\Phi_a{}^b$,
$\bar\Phi_a{}^b=(\overline{\Phi_b{}^a})$ ($\bar\Phi = \Phi^+$) and
$\Upsilon_a$, $\bar\Upsilon^a=(\overline{\Upsilon_a})$ are fermionic component
fields. Let us consider the action~(\ref{N2-Cal}) in the Wess--Zumino gauge
\begin{equation}\label{WZ-2}
V (t, \theta,\bar\theta) = \theta\bar\theta A (t)\,.
\end{equation}
Eliminating the
auxiliary fields $\Upsilon$, $\bar\Upsilon$ in this gauge by their equations of motion,
we find the component form of the action~(\ref{N2-Cal}) ($\int d^2\theta\, (\theta\bar\theta) =1 $):
\begin{equation}\label{N2Cal-com}
S^{\scriptscriptstyle{WZ}}_{sC} =  {\displaystyle\int} dt \,\Bigg[\, {\rm
Tr}\Big(\nabla\!
X \nabla\! X \Big) + {\textstyle\frac{i}{2}}\, (\bar Z \nabla\! Z - \nabla\! \bar Z Z) - c\,{\rm tr} A  \, +
\,i\,{\rm tr} \Big( \bar\Psi \nabla \Psi - \nabla \bar\Psi \Psi \Big)\,\Bigg].
\end{equation}
The covariant derivatives $\nabla\! X$, $\nabla\! Z$, $\nabla\! \bar Z$, $\nabla
\Psi$,
$\nabla \bar\Psi$ were defined in~(\ref{cov-der-b}) and
\begin{equation}\label{cov-der-Psi}
\nabla \Psi = \dot \Psi +i [\Psi, A]\,, \qquad \nabla \bar\Psi = \dot {\bar\Psi} +i
[\bar\Psi, A]\,.
\end{equation}
We see that the bosonic part of the model~(\ref{N2Cal-com}) is exactly the Calogero
system~(\ref{b-Cal}).

The action~(\ref{N2Cal-com}) is invariant with respect to residual local
U($n$) transformations,
$g(\tau )\in U(n)$, defined by~(\ref{Un-tran}) and
\begin{equation}\label{Un-tran-Psi}
\Psi \rightarrow \, g \Psi g^+ \,, \qquad  \bar\Psi \rightarrow \, g \bar\Psi g^+\,.
\end{equation}
Exactly as in the pure bosonic case, this local U(n) invariance
eliminates the nondiagonal fields $X_a^b$,
$a{\neq}b$, and all spinor fields $Z_a$. Thus, the physical fields
in our ${\cal N}{=}\,2$ supersymmetric generalization of the Calogero system are $n$ bosons
$x_a=X_a^a$ and $2n^2$ fermions $\Psi_a^b$. These fields present on-shell content
of $n$ multiplets ({\bf 1,2,1}) and $n^2{-}n $
multiplets ({\bf 0,2,2}). These multiplets are produced from the original $n^2$ multiplets ({\bf 1,2,1})
by gauging procedure~\cite{DI1}. We can depict this multiplet structure on the
plot:
\begin{center}
$\underbrace{\mathscr{X}_a^a=({X}_a^a, \Psi_a^a)}_{{{\bf (1,2,1)}\,multiplets}}$\qquad\quad
\qquad $\underbrace{\mathscr{X}_a^b=({{X}_a^b}, \Psi_a^b),
{\scriptstyle a\neq b}}_{{\bf (1,2,1)}\,multiplets}$\\
$\,$\\
$\,$\\
{$\Downarrow$ \qquad{gauging}\qquad $\Downarrow$}\quad$\,$\\
$\,$\\
$\,$\\
$\underbrace{\mathscr{X}_a^a=({X}_a^a, \Psi_a^a)}_{{\bf (1,2,1)}\,multiplets}$\,\,\quad
{interact}\quad $\underbrace{\Omega_a^b=(\Psi_a^b, {B_a^b}), {\scriptstyle a\neq b}}_{{\bf
(0,2,2)}\,multiplets}\,$.
\end{center}
\noindent Here, the bosonic fields $B_a^b$ are auxiliary components of the multiplets ({\bf 0,2,2}), which are not present in the action~(\ref{N2Cal-com}), being eliminated by their equations of motion.
Thus, we obtained some new ${\cal
N}{=}\,2$ extensions of the $n$-particle Calogero models with {$n$} bosons and {$2n^2$}
fermions,  as opposite to the standard ${\cal N}{=}\,2$ super Calogero system of Freedman and Mende which involves only
{$2n$} fermions~\cite{FM}.

It is easy to explicitly express the considered system in terms of physical variables only.
This can be done either by going to the Hamiltonian formalism, like we did in the purely bosonic case,
or by eliminating the auxiliary variables after  the gauge fixing.
Using the latter method, i.e. by effecting the gauge-fixing conditions (\ref{X-fix}), (\ref{g-Z})  and eliminating the auxiliary fields  $A_a{}^b$,
we obtain the physical action of  the considered ${\cal N}{=}\,2$ multiparticle superconformal model of Calogero type
\begin{equation}\label{st-Cal-N2}
S^{(N=2)}_{sC} = \int dt  \,\Big[\, \sum_{a} \dot x_a \dot x_a + \,i\,( \bar\Psi_a{}^b
\dot\Psi_b{}^a - \dot{\bar\Psi}_a{}^b\Psi_b{}^a ) - H\Big]\,.
\end{equation}
Here, the Hamiltonian has the form
\begin{equation}\label{H-Cal-fix-s}
H = \frac{1}{4}\sum_{a} (p_a )^2 + \sum_{a\neq b} \frac{4}{(x_a - x_b)^2}\Big( Z_a {\bar
Z}^a Z_b {\bar Z}^b +  2 {\bar Z}^a \{ \Psi, \bar\Psi\}_a{}^b Z_b + \{ \Psi,
\bar\Psi\}_a{}^b\{ \Psi, \bar\Psi\}_b{}^a \Big)\,.
\end{equation}
In the expression (\ref{H-Cal-fix-s}), the variables $Z_a$ (which are real after gauge fixing (\ref{g-Z}), $\bar Z^a = Z_a$)
are found from the equations of motion for diagonal components of $A_a{}^b$
\begin{equation}\label{eqZ-N2-a}
(Z_a)^2 = c - R_a\,,
\end{equation}
where
\begin{equation}\label{R}
R_a \equiv \{ \Psi , \bar\Psi \}_a{}^a = \sum_b ( \Psi_a{}^b \bar\Psi_b{}^a +
{\bar\Psi}_a{}^b\Psi_b{}^a ) \qquad \quad (\mbox{no summation over } a)\,.
\end{equation}
The quantities $R_a$ contains $\Psi_a{}^b$ and $\bar\Psi_a{}^b$ with $a\neq b$ only
and has $2(n-1)$ terms, so that $(R_a)^{2n-1} \equiv 0$.
Therefore, the solutions of eqs.~(\ref{eqZ-N2-a}) are
\begin{equation}\label{sol-eqZ-N2}
Z_a = \epsilon_a \sqrt{c} \,\sum_{k=0}^{2(n-1)}
\frac{\alpha_k}{c^k}(R_a)^{k}\,,
\end{equation}
where $\epsilon_a =\pm 1$, independently for each $a$, and $\alpha_k$ are some constants.
The first constants from this set are $\alpha_0 = \alpha_1 = 1$, $\alpha_2 = -\alpha_3 = -\frac{1}{2}$,
$\alpha_3 = -\frac{5}{8}$.
Let us consider the $n=2$ case as an example.
In this case the action~(\ref{st-Cal-N2}) has the form
\begin{eqnarray}  \label{n2Cal-com-3}
S^{(n=2)}_{sC} =\!\! & \!\!{\displaystyle\int} dt \!\! &\!\!\Bigg\{
\dot x_0 \dot x_0 + \,i\,( \bar\psi_0 \dot\psi_0 - \dot{\bar\psi}_0\psi_0 )
+ \,i\,( \bar\chi \dot\chi - \dot{\bar\chi}\chi )
\\
&& + \dot \rho \dot \rho + \,i\,( \bar\psi \dot\psi - \dot{\bar\psi}\psi )
+ \,i\,( \bar\varphi \dot\varphi - \dot{\bar\varphi}\varphi )
\nonumber \\
&& -\frac{1}{\rho^2}\Bigg[\frac{c^2}{4} + c\epsilon_1\epsilon_2
(\psi\bar\varphi + \varphi\bar\psi ) +
(\psi\bar\psi + \bar\varphi\varphi +
\bar\chi\chi)^2 \Bigg] \Bigg\} \,,\nonumber
\end{eqnarray}
where we used the notations
\begin{equation}
x_0\equiv {\textstyle\frac{1}{\sqrt{2}}}\, (x_1 + x_2) \,, \qquad
\rho\equiv {\textstyle\frac{1}{\sqrt{2}}}\, (x_1 - x_2) \,,
\end{equation}
\begin{equation}
\psi_0\equiv {\textstyle\frac{1}{\sqrt{2}}}\, (\Psi_1{}^1 + \Psi_2{}^2) \,, \qquad
\psi\equiv {\textstyle\frac{1}{\sqrt{2}}}\, (\Psi_1{}^1 - \Psi_2{}^2) \,,
\qquad
\chi \equiv {\textstyle\frac{1}{\sqrt{2}}}\, (\Psi_1{}^2 + \Psi_2{}^1) \,, \qquad
\varphi \equiv {\textstyle\frac{1}{\sqrt{2}}}\, (\Psi_1{}^2 - \Psi_2{}^1) \,,
\end{equation}
After introducing new fermionic fields via the nonlinear transformations
\begin{equation}
\phi_0\equiv \chi\Big(1+ {\textstyle\frac{1}{2m}}\,
(\psi\bar\varphi - \varphi\bar\psi ) +
{\textstyle\frac{1}{4m^2}\,\psi\bar\psi\varphi\bar\varphi}\, \Big) \,, \qquad
\phi_1\equiv \psi + {\textstyle\frac{1}{2m}}\, (\bar\chi\chi)\,\varphi \,, \qquad
\phi_2 \equiv \bar\varphi - {\textstyle\frac{1}{2m}}\, (\bar\chi\chi)\,\bar\psi \,,
\end{equation}
where $m=-\frac{c\epsilon_1\epsilon_2}{2}$,
the action~(\ref{n2Cal-com-3}) becomes
\begin{eqnarray}  \label{n2Cal-com-4}
S^{(n=2)}_{sC} =\!\! & \!\!{\displaystyle\int} dt \!\! &\!\!\Bigg\{
\dot x_0 \dot x_0 + \,i\,( \bar\psi_0 \dot\psi_0 - \dot{\bar\psi}_0\psi_0 )
+ \,i\,( \bar\phi_0 \dot\phi_0 - \dot{\bar\phi_0}\phi_0 )
\\
&& + \,\dot \rho \dot \rho + \,i\,( \bar\phi_1 \dot\phi_1 - \dot{\bar\phi_1}\phi_1 )
+ \,i\,( \bar\phi_2 \dot\phi_2 - \dot{\bar\phi_2}\phi_2 )
\nonumber \\
&& -\,\frac{1}{\rho^2}\Bigg[m^2 -2m \phi_1\phi_2 +2m \bar\phi_1\bar\phi_2 +
(\bar\phi_1\phi_1 +\bar\phi_2\phi_2)^2 \Bigg] \Bigg\} \,.\nonumber
\end{eqnarray}
This action is none other than a sum of the action of free ${\cal N}{=}\,4$ ({\bf 1,4,3}) supermultiplet $\Omega$
with the physical components ($x_0$; $\psi_0$, $\phi_0$) and the ${\cal N}{=}\,4$ superconformal mechanics action \cite{IKL2}
for ${\cal N}{=}\,4$ ({\bf 1,4,3}) supermultiplet $Y$ with physical components ($\rho$; $\phi_1$, $\phi_2$). It can be recovered
from the following ${\cal N}{=}\,4$ superfield action
\begin{equation}
S^{(n=2)}_{sC} =-\int dt d^4 \!\theta \,\Big({\textstyle\frac{1}{2}}\,\Omega^2+Y \ln Y\Big)\,,
\end{equation}
where $\Omega$ and $Y$ are some constrained ${\cal N}{=}\,4$ superfields representing two independent off-shell ({\bf 1,4,3})
supermultiplets (see next Section). This appearance of hidden higher-rank superconformal symmetry is quite similar to
what happens in the ${\cal N}{=}\,1$, $n{=}2$ case. No such an intriguing feature arises in ${\cal N}{=}\,2$ models with  $n{>}2$.

The ${\cal N}{=}\,1$ and ${\cal N}{=}\,2$ superextensions of the conformal Calogero model considered
in this section (and sketched in~\cite{FIL1}) are new and so need a more detailed analysis,
including the study of their possible integrability (e.g., along the line of~\cite{Vas2,Vas1,Brink}).
Though they contain non-minimal sets of fermionic fields as compared to the Freedman and Mende models,
these sets are necessarily implied by the supersymmetric gauge procedure which is a well-defined generalization of
the bosonic $d{=}1$ gauge procedure yielding the ordinary Calogero model. It is an open question whether it is possible to somehow reduce the
number of fermions - either by imposing some extra covariant conditions on the original superfields
or through the proper enlargement of the underlying gauge invariance. In Section 4.3 we shall show that in the ${\cal N}{=}\,4$ case the
similar gauging procedure naturally leads to ${\cal N}{=}\,4$ superconformal extension of the so called spin Calogero
model~\cite{FIL1}.

\setcounter{equation}{0}

\section{${\cal N}{=}\,4$ superconformal mechanics}

\subsection{A bestiary of ${\cal N}{=}\,4$, $d{=}1$ superconformal algebras}

The full family of ${\cal N}{=}\,4$, $d{=}1$ superconformal algebras is spanned by the following generators in the
spinorial basis
\begin{equation}\label{D21-gen}
G^{(4)}=\left( {Q}_{\alpha i i^\prime}; T_{\alpha\beta}, {J}_{ij}, {I}_{i^\prime j^\prime}\right).
\end{equation}
In general they form the superalgebra $D(2,1;\alpha)$ (for more details see, e.g., \cite{Sorba,BILS,IKLech})
\begin{equation}\label{D21-QQ}
\{ {Q}_{\alpha i i^\prime},  {Q}_{\beta j j^\prime}\}=
2\,\Big(\epsilon_{ij}\epsilon_{i^\prime j^\prime} {T}_{\alpha\beta}+
\alpha \,\epsilon_{\alpha\beta}\epsilon_{i^\prime j^\prime} {J}_{ij}-
(1{+}\alpha)\,\epsilon_{\alpha\beta}\epsilon_{ij} {I}_{i^\prime j^\prime}\Big)\,,
\end{equation}
\begin{eqnarray}\label{D21-BB}
[{T}_{\alpha\beta}, {T}_{\gamma\delta}] &=&
i\left(\epsilon_{\alpha\gamma}{T}_{\beta\delta} +\epsilon_{\beta\delta}{T}_{\alpha\gamma}\right)\,,\\
{}[{J}_{ij}, {J}_{kl}] &=&
i\left(\epsilon_{ik}{J}_{jl} +\epsilon_{jl}{J}_{ik}\right)\,,\label{D21-BB-b}\\
{}[{I}_{i^\prime j^\prime }, {I}_{k^\prime l^\prime }] &=&
i\big(\epsilon_{i^\prime k^\prime }{I}_{j^\prime l^\prime } +
\epsilon_{j^\prime l^\prime }{I}_{i^\prime k^\prime }\big)\,,\label{D21-BB-c}
\end{eqnarray}
\begin{eqnarray}\label{D21-BQ}
[{T}_{\alpha\beta}, {Q}_{\gamma i i^\prime}] &=&
-i\,\epsilon_{\gamma(\alpha}{Q}_{\beta) i i^\prime}\,, \\
{}[{J}_{ij}, {Q}_{\alpha k i^\prime}] &=&
-i\,\epsilon_{k(i}{Q}_{\alpha j) i^\prime}\,,\label{D21-BQ-b}\\
{}[{I}_{i^\prime j^\prime}, {Q}_{\alpha i k^\prime}] &=&
-i\,\epsilon_{k^\prime (i^\prime}{Q}_{\alpha i j^\prime)}\,,\label{D21-BQ-c}
\end{eqnarray}
with other commutators vanishing.
All  the $D(2,1;\alpha)$ generators are assumed to be Hermitian, i.e. they satisfy the relations
\begin{equation}\label{Her-Q}
({Q}_{\alpha i i^\prime}){}^\dag = \epsilon^{ij}\epsilon^{i^\prime j^\prime}{Q}_{\alpha jj^\prime}\,,
\end{equation}
\begin{equation}\label{Her-B}
({T}_{\alpha\beta})^\dag = {T}_{\alpha\beta}\,,\qquad
({J}_{ij})^\dag = \epsilon^{ik}\epsilon^{jl}{J}_{kl}\,,\qquad
({I}_{i^\prime j^\prime}){}^\dag = \epsilon^{i^\prime k^\prime}\epsilon^{j^\prime l^\prime}{I}_{k^\prime l^\prime}\,.
\end{equation}
The generators ${T}_{\alpha\beta}$ form the $o(2,1)$ algebra
(${T}_{11}={H}$, ${T}_{22}={K}$, ${T}_{12}={D}$), whereas ${J}_{ij}$ and ${I}_{i^\prime j^\prime}$
constitute two mutually commuting $o(3)$ algebras forming $o(4)$ algebra.
The indices $\alpha,\beta,\gamma=1,2$ are spinor $o(2,1)$ indices and  $i,j,k=1,2$;  $i^\prime,j^\prime,k^\prime=1,2$ are doublet
indices of two $o(3)$ algebras.
Everywhere in this paper we take $\epsilon_{12}=\epsilon^{21}=1$.
The fermionic $o(2,1)$ spinor supercharges ${Q}_{\alpha i i^\prime}$ unify the standard $d{=}1$ supercharges
${Q}_{i i^\prime}={Q}_{1 i i^\prime}$ and the generators of superconformal boosts
${S}_{i i^\prime}={Q}_{2 i i^\prime}$. In the complex notation, with only one $O(3)$ symmetry being manifest,
the supercharges are rewritten as
\begin{equation}\label{not-Qq}
{Q}^{i 1^\prime}=-{Q}^{i}\,,\qquad {Q}^{i2^\prime}=-\bar{Q}^{i}\,,\qquad ({Q}^{i}){}^\dag=\bar{Q}_{i}\,,
\end{equation}
\begin{equation}\label{not-Qq-c}
{S}^{i1^\prime}={S}^{i}\,,\qquad {S}^{i2^\prime}=\bar{S}^{i}\,,\qquad ({S}^{i}){}^\dag=\bar{S}_{i}\,.
\end{equation}

The $D(2,1;\alpha)$ superalgebra, parametrized by a real parameter $\alpha$, in fact encompasses all
${\cal N}{=}\,4$ superconformal algebras:
\begin{equation}\label{D21a}
\begin{array}{lcl}
\alpha=0,\quad \alpha=-1&:\qquad &D(2,1;\alpha)\cong su(1,1|2)\,\,{+\!\!\!\!\!\!\supset} \,\,su(2)\,,\\[7pt]
\alpha=1,\quad \alpha=-2&:\qquad &D(2,1;\alpha)\cong osp(4^\ast|\,2)\,,\\[7pt]
\alpha=-1/2&:\qquad &D(2,1;\alpha)\cong osp(4|\,2)\,.
\end{array}
\end{equation}
The discrete transformation $\alpha\leftrightarrow-(1{+}\alpha)$
switches the roles of two $su(2)$ algebras presented by the generators ${J}_{ij}$ and ${I}_{i^\prime j^\prime}$ in
(\ref{D21-QQ}). The discrete transformation $\alpha\leftrightarrow\alpha^{-1}$ exchanges the generators  ${T}_{\alpha\beta}$
of noncompact $sl(2,\mathbb{R})$ algebra and compact ${J}_{ij}$ of $su(2)$ algebra and so it is ill defined in case of real
$D(2,1;\alpha)$ superalgebra considered here.

In the degenerate cases $\alpha=0$ and $\alpha=-1$ we may retain only eight fermionic generators ${Q}_{\alpha i i^\prime}$
and six bosonic generators ${T}_{\alpha\beta}$, ${J}_{ij}$ forming $su(1,1|2)$ superalgebra and
do not require  $su(2)$ symmetry which is produced by the generators ${I}_{i^\prime j^\prime}$ and acts on primed indices.
In this case it is possible to extend this $su(1,1|2)$ superalgebra by the addition of central charges.
As result of it, the anticommutators (\ref{D21-QQ}) turn in the following form
\begin{equation}\label{D21-QQ-C}
\{ {Q}_{\alpha i i^\prime},  {Q}_{\beta j j^\prime}\}=
2\,\Big(\epsilon_{ij}\epsilon_{i^\prime j^\prime} {T}_{\alpha\beta}
-\epsilon_{\alpha\beta}\epsilon_{i^\prime j^\prime} {J}_{ij}-
\epsilon_{\alpha\beta}\epsilon_{ij} {C}_{i^\prime j^\prime}\Big)\,,
\end{equation}
where central charges ${C}_{i^\prime j^\prime}$ commute with all other generators
\begin{equation}
[{Q}_{\alpha i i^\prime}, {C}_{k^\prime j^\prime}]=
[ {T}_{\alpha\beta}, {C}_{k^\prime j^\prime}]=
[ {J}_{ij}, {C}_{k^\prime j^\prime}]=0\,.
\end{equation}
By using ${\rm SU}(2)$ transformations acting on indices $i^\prime, j^\prime$
we can leave only one nonvanishing central charge, for example,  its third component:
\begin{equation}
C:={C}_{1^\prime 2^\prime}\neq 0\,,\qquad {C}_{1^\prime 1^\prime}={C}_{2^\prime 2^\prime}=0\,.
\end{equation}

The second-order Casimir operator of the whole supergroup $D(2,1;\alpha)$ is given by the following
expression~\cite{Je}
\begin{equation}\label{qu-Cas}
{C}^{({\cal N}{=}\,4)}_2={T}^2 +\alpha\, {J}^2- (1{+}\alpha)\,{I}^2 +
{\textstyle\frac{i}{4}}\, {Q}^{\alpha ii^\prime}{Q}_{\alpha ii^\prime}\,,
\end{equation}
where
\begin{equation}\label{q-Cas-3}
{T}^2:= {\textstyle\frac{1}{2}}\,
{T}^{\alpha\beta}{T}_{\alpha\beta}\,,
\qquad
{J}^2:= {\textstyle\frac{1}{2}}\,
{J}^{ik}{J}_{ik} \,,
\qquad
{I}^2:= {\textstyle\frac{1}{2}}\, {I}^{i^\prime k^\prime}{I}_{i^\prime k^\prime }
\end{equation}
and
\begin{equation}\label{QQ-quCas}
{\textstyle\frac{i}{4}}\, {Q}^{\alpha i i^\prime}{Q}_{\alpha i i^\prime} =
{\textstyle\frac{i}{4}}\, [{Q}^{i},\bar {S}_{i}] + {\textstyle\frac{i}{4}}\,
[\bar {Q}_{i}, {S}^{i}] \,.
\end{equation}

\subsection{Superconformal models from standard ${\cal N}{=}\,4$ superspace}

A natural arena for ${\cal N}{=}\,4, d{=}1$ supersymmetric theories is the
${\cal N}{=}\,4, d{=}1$ superspace~\cite{IKL2}
\be
(t,\theta_i, \bar\theta^i)\,, \qquad \bar\theta^i=(\overline{\theta_i})\,,
\quad (i
= 1, 2)\,. \lb{Rss}
\ee
The corresponding spinor covariant derivatives have the form
\begin{equation}
D^i=\frac{\partial}{\partial\theta_i}-i\bar\theta^i \partial_t\,,\qquad \bar
D_i=\frac{\partial}{\partial\bar\theta^i}-i \theta_i \partial_t = -
\overline{(D^i)}\,.
\end{equation}
One of the two ${\rm SU}(2)$ factors
of the full R-symmetry (automorphism) group ${\rm SO}(4)_R\,$ acts on the doublet indices $i$ and will be denoted ${\rm
SU}(2)_R\,$. The second ${\rm SU}(2)$ mixes
$\theta_i$ with their complex conjugates and is not manifest in the
considered approach.

\subsubsection{Single-particle models}

The one-particle model is built on
the multiplet ({\bf 1,4,3}), which is described by the even real superfield $\mathscr{X}$
subjected to the constraints
\begin{equation}  \label{cons-X-g1}
D^iD_i \,\mathscr{X}=0\,,
\qquad
\bar D_i\bar D^i \,\mathscr{X}=0\,,\qquad
[D^i,\bar D_i]\,\mathscr{X}=0\,.
\end{equation}
The superconformal action of the ({\bf 1,4,3}) multiplet is given by
($\alpha\neq 0$)
\begin{equation}\label{4N-X1}
S_{\mathscr{X}} =-\textstyle{\frac{1}{4(1{+}\alpha)}}\,\displaystyle{\int} dt\,d^4\theta \,
\mathscr{X}^{\,-1/\alpha} \,.
\end{equation}

Note that the action~(\ref{4N-X1}) is in fact non-singular at $\alpha=-1\,$. Indeed, making use of the fact that
${\int} \mu_H \, \mathscr{X}$ is an integral of total derivative (in virtue of constraints (\ref{cons-X-g1})), we cast the
action~(\ref{4N-X1}) in the equivalent form
\begin{equation}
S_{\mathscr{X}} =-\textstyle{\frac{1}{4(1{+}\alpha)}}\,\displaystyle{\int} dt\,d^4\theta \,
\left(\mathscr{X}^{\,-1/\alpha} -\mathscr{X}\right).
\end{equation}
Thus in the limit $\alpha=-1$ we obtain the meaningful action
\begin{equation}\label{4N-X10}
S_{\mathscr{X}} \Big|_{\alpha=-1} =-\textstyle{\frac{1}{4}}\,\displaystyle{\int} dt\,d^4\theta \,
\mathscr{X} \,\ln\! \mathscr{X} \,.
\end{equation}

The action~(\ref{4N-X1}) is not defined at $\alpha{=}0$, and this special case needs a separate
analysis (see \cite{DI1, DI2}). In what follows we assume that $\alpha \neq 0\,$.

The action~(\ref{4N-X1}) is invariant with respects to the rigid
${\cal N}{=}\,4$ superconformal symmetry $D(2,1;\alpha)\,$.
All superconformal transformations are
contained in the closure of the supertranslations and superconformal boosts.

Invariance of the action~(\ref{4N-X1}) under the supertranslations
($\bar\varepsilon^i=\overline{(\varepsilon_i)}$)
\begin{equation}
\delta t =i(\varepsilon_k\bar\theta^k-\theta_k\bar\varepsilon^k),\qquad
\delta
\theta_k=\varepsilon_k, \qquad \delta \bar\theta^k=\bar\varepsilon^k
\end{equation}
is automatic because we work in the ${\cal N}{=}\,4$ superfield approach.

The coordinate realization of the superconformal boosts of $D(2,1;\alpha)$ \cite{IKLech,IKLecht,IL} is
as follows ($\bar\eta^i=\overline{(\eta_i)}$):
\begin{equation}  \label{sc-coor-c-b}
\delta^\prime t=i t(\theta_k \bar\eta^k + \bar\theta^k\eta_k) - (1{+}\alpha)\,\theta_i
\bar\theta^i (\theta_k \bar\eta^k + \bar\theta^k\eta_k)\,,
\end{equation}
\begin{equation}  \label{sc-coor-c-f1}
\delta^\prime \theta_i= -\eta_i t -2i\alpha \,\theta_i(\theta_k \bar\eta^k) +
2i(1{+}\alpha)\,\theta_i (\bar\theta^k\eta_k)- i(1{+}2\alpha)\,\eta_i (\theta_k
\bar\theta^k)\,,
\end{equation}
\begin{equation}  \label{sc-coor-c-f2}
\delta^\prime \bar\theta^i= -\bar\eta^i t -2i\alpha \,\bar\theta^i(\bar\theta^k\eta_k) +
2i(1{+}\alpha)\,\bar\theta^i (\theta_k \bar\eta^k)+ i(1{+}2\alpha)\,\bar\eta^i (\theta_k
\bar\theta^k)\,,
\end{equation}
\begin{equation}  \label{sc-me1}
\delta^\prime (dtd^4\theta)=-\alpha^{-1}\,(dtd^4\theta)\,\Lambda_0\,,
\end{equation}
where
\begin{equation}  \label{def-La2}
\Lambda_0 = 2i\alpha (\theta_k \bar\eta^k +
\bar\theta^k\eta_k)\,.
\end{equation}
Taking the superfield transformations in the form (here we use the ``passive''
interpretation of them)
\begin{equation}  \label{sc-1n}
\delta^\prime \mathscr{X}= -\Lambda_0\,\mathscr{X}\,,
\end{equation}
it is easy to check the invariance of the action~(\ref{4N-X1}).

The solution of the constraint~(\ref{cons-X-g1}) is as
follows \begin{equation}  \label{sing-X1}
\mathscr{X}(t,\theta_i,\bar\theta^i)= x + \theta_i\psi^i +
\bar\psi_i\bar\theta^i +
\theta^i\bar\theta^k
N_{ik}+{\textstyle\frac{i}{2}}(\theta)^2\dot{\psi}_i\bar\theta^i
+{\textstyle\frac{i}{2}}(\bar\theta)^2\theta_i\dot{\bar\psi}{}^i +
{\textstyle\frac{1}{4}}(\theta)^2(\bar\theta)^2 \ddot{x}\,,
\end{equation}
where $(\theta)^2\equiv \theta_i\theta^i$,
$(\bar\theta)^2\equiv
\bar\theta^i\bar\theta_i$.
Inserting~(\ref{sing-X1}) in~(\ref{4N-X1}) and integrating there over the Grassmann
variables, we obtain
\begin{eqnarray}\label{4N-X1a}
S_{\mathscr{X}} &=& \textstyle{\frac{1}{4\alpha^2}}\,\displaystyle{\int} dt\,
x^{-\frac{1}{\alpha}-2}\,\left[ \dot x\dot x +i \left(\bar\psi_k \dot\psi^k
-\dot{\bar\psi}_k \psi^k \right) -{\textstyle\frac{1}{2}}N^{ik}N_{ik}\, \right] \\
&&  -\, \textstyle{\frac{1}{4\alpha^2}}\, (\textstyle{\frac{1}{\alpha}}+2)
\,\displaystyle{\int} dt\, x^{-\frac{1}{\alpha}-3}\, N^{ik} \psi_{(i}\bar\psi_{k)}    - \,
\textstyle{\frac{1}{12\alpha^2}}\,
(\textstyle{\frac{1}{\alpha}}+2)(\textstyle{\frac{1}{\alpha}}+3)\, \displaystyle{\int} dt\,
x^{-\frac{1}{\alpha}-4}\, \psi^{i}\bar\psi^{k} \psi_{(i}\bar\psi_{k)} \,.\nonumber
\end{eqnarray}

Eliminating the auxiliary fields $N^{ik}$ by their algebraic equations of motion,
\begin{equation}\label{4N-eq-N}
N_{ik} = -
(\textstyle{\frac{1}{\alpha}}+2)\, x^{-1}\, \psi_{(i}\bar\psi_{k)} \,,
\end{equation}
we obtain the on-shell form of the action~(\ref{4N-X1a})
\begin{equation}\label{4N-1ph}
S = \int dt \,\Big[\dot x\dot x  +i  \left( \bar\psi_k \dot\psi^k -\dot{\bar\psi}_k \psi^k \right) \Big]  +
\textstyle{\frac{2}{3}}\,(1{+}2\alpha) \!\displaystyle{\int}\! dt\,
\frac{\psi^{i}\bar\psi^{k} \psi_{(i}\bar\psi_{k)}}{x^2}  \,.
\end{equation}
This is the action of one-particle ${\cal N}{=}\,4$ superconformal mechanics. It contains no pure conformal potential at the classical level.
It can appear from the last term with fermions upon quantization.

In Sect.\,4.3.2 below we shall consider the system in which one-particle model
(\ref{4N-X1}) appears as a subsystem. Therefore,
using formulas of Sect.\,4.3.2 (for example, (\ref{Q-qu})--(\ref{I-qu}))
and making there the appropriate truncations, we can recover the generators of the ${\cal N}{=}\,4$ superconformal $D(2,1;\alpha)$
symmetry in the considered one-particle superconformal system (\ref{4N-X1})
(see also \cite{W,GLP2,GLP3} for such a realization of the $su(1,1|2)$ superconformal algebra
as a particular case of $D(2,1;\alpha)$).

When only $su(1,1|2)$ symmetry is required,
while the $su(2)_L$ symmetry appearing in the semi-direct sum (see (\ref{D21a})) is allowed to be broken, the
constraints~(\ref{cons-X-g1}) for the even real superfield
$\mathscr{X}$ can be weakened~\cite{IKL2} by adding nonzero constants in their right-hand sides.
These constants can always be rotated between the equations~(\ref{cons-X-g1}),
for example to
\begin{equation}  \label{cons-X-rel1}
{\rm (a)} \quad D^iD_i \,\mathscr{X}=0\,, \;\; \bar D_i\bar D^i \,\mathscr{X}=0\,; \qquad  {\rm (b)} \quad
[D^i,\bar D_i]\, \mathscr{X}=m\,,
\end{equation}
where the constant $m$ provides a central charge to the $su(1,1|2)$ algebra.
The solution to~(\ref{cons-X-rel1}) is a sum
of~(\ref{sing-X1}) and an additional term $-\frac{1}{4}\theta\bar\theta m$.
Then the
action~(\ref{4N-X1}) (with $\alpha{=}{-}1$) gives rise to additional contributions to the physical
component Lagrangian~(\ref{4N-1ph}) which now becomes~\cite{IKL2}
\begin{equation}\label{4N-1ph-a}
\tilde S = \int dt \,\Big[\dot x\dot x +i  \left( \bar\psi_k \dot\psi^k -\dot{\bar\psi}_k \psi^k \right) -
\frac{\left(m+\bar\psi_{k} \psi^{k}\right)^2}{x^2}\Big]  \,.
\end{equation}
The additional terms,  proportional
to $m^2/x^2$ and $m\psi\bar\psi /x^2$, also appear in the Hamiltonian, and they are induced by the appropriate new terms
in the Noether supercharges, which correspond to
the $su(1,1|2)$ algebra with a central charge proportional to $m$.
Such a central-charge deformation is possible only for $\alpha{=}{-}1$ (and $\alpha{=}0$).
Thus in this case the conformal potential
comes out at the classical level, and its strength is the square of the central charge $m\,$.
Note, that the action~(\ref{4N-1ph-a}) and the action~(\ref{n2Cal-com-4}) for the superfield $Y$ are equivalent to each other.
Instead of the constraints~(\ref{cons-X-rel1}), the superfield $Y$ is subject to the constraints
\begin{equation}  \label{cons-Y-rel1}
{\rm (a)} \quad D^iD_i \,Y=m\,, \;\; \bar D_i\bar D^i \,Y=m\,; \qquad  {\rm (b)} \quad
[D^i,\bar D_i]\, Y=0\,.
\end{equation}
The constraints (\ref{cons-X-rel1}) and (\ref{cons-Y-rel1}), as well as the actions (\ref{4N-1ph-a}) and (\ref{n2Cal-com-4}),  are related
to each other by the (broken) $\textrm{SU}(2)$ rotations which mix  $\theta$ with $\bar\theta$ (see details in~\cite{IKL2}).

\subsubsection{Multi-particle models with $D(2,1;\alpha)$ symmetry}

Despite the physical importance, multiparticle systems with ${\cal N}{=}\,4$
superconformal symmetry are poorly understood to date. Unlike multi-particle
${\cal N}{=}\,2$ superconformal systems, direct generalizations
of ${\cal N}{=}\,4$ one-particle superconformal systems do not yield
${\cal N}{=}\,4$ superconformal systems of Calogero type.
For this reason, until now, studies of the ${\cal N}{=}\,4$ multi-particle
superconformal systems were performed not only
in ordinary superspace, but also at the component level.
Furthermore, it turns out that the realization of $D(2,1;\alpha)$ superconformal
symmetry on the multi-particle phase space for $\alpha\neq-1$ or~0 requires at
least one pair of (bosonic) isospin variables
$\{u^i,\bar{u}_i|\,i{=}1,2\}$ parametrizing an internal two-sphere.
This subsection mainly summarizes the results of~\cite{kl}.

We consider $n$ particles on a real line,
with coordinates and momenta $\{x^a,p_a|\,a{=}1,\ldots,n\}$
as well as associated complex pairs of fermionic variables
$\{\psi_i^a,\bpsi^{ai}|\,a{=}1,\ldots,n,\,i{=}1,2\}$.\footnote{
Viewed as a one-particle system, the bosonic target is $\R^n$.
Its metric~$(\delta_{ab})$ allows us to pull down all particle indices.
Spinor indices are raised and lowered with the invariant tensor
$\eps^{ik}$ and its inverse $\eps_{ki}$, respectively.}
The basic nonvanishing Poisson brackets read
\be\label{PB}
\bigl\{x^a, p_b\bigr\}=\delta^a_b\ ,\qquad
\bigl\{\psi^a_i, \bpsi^{bk} \bigr\}= -\sfrac\im 2\,\delta_i^{\ k}\delta^{ab}\ ,\qquad
\bigl\{u^i,\bar{u}_k\bigr\}=-\im\,\delta^i_{\ k}\ .
\ee
We would like to realize the ${\cal N}{=}\,4$ superconformal algebra
$D(2,1;\alpha)$ on the (classical) phase space of this mechanical system,
thereby severely restricting the particle interactions.
It is convenient to start with an ansatz for the supercharges
$Q_a$ and $\bar Q^a$.
An important novelty of the ${\cal N}{=}\,4$ supercharges
compared the ${\cal N}{=}\,2$ ones is the necessity of a term cubic
in the fermions besides the standard linear one in (\ref{N2-Q-n})
and~(\ref{N2-S-n}). Therefore, the ansatz should take the form
\be\label{ansQ}
Q_i \=p_a\psi^a_i+\im U_a(x)_i^{\ k}\,\psi^a_k
+ \im F_{abc}(x)_{i\ell}^{jk}\,\psi^a_j\psi^b_k\bpsi^{c\ell}\ ,
\ee
where $U_a(x)_i^{\ k}$ and $F_{abc}(x)_{i\ell}^{jk}$ are homogeneous
functions of degree~$-1$ in $\{x^1,\ldots,x^n\}$, to be determined.

For the $F$ functions we adopt the simplest possibility,
\be
F_{abc}(x)_{i\ell}^{jk}=F_{abc}(x)\,\e_{i\ell}\e^{jk}\ ,
\ee
while the analogous option $\ U_a(x)_i^{\ k}=W_a(x)\,\delta_i^{\ k}\ $
for the $U$~functions turns out to work only for the case of $\alpha{=}-1$
or~0 (see below). To open the way for a realization of the generic
$D(2,1;\alpha)$ superalgebra, one must generalize to
\be\label{J}
\im U_a(x)_i^{\ k}\=U_a(x)\,{\cal J}_i^{\ k}
\qquad\textrm{with}\qquad
{\cal J}_{ik}=\sfrac{\im}{2}\bigl(u_i\bar{u}_k+u_k\bar{u}_i\bigr)\ ,
\ee
utilizing the isospin $su(2)$ current. Therefore, the ansatz in~\cite{kl} reads
\be\label{Q}
Q_i \=p_a\,\psi_i^a\ +\ U_a(x)\,{\cal J}_{ik}\,\psi^{ak}
\ +\ \im F_{abc}(x)\,\psi^{ak}\psi^b_k\bpsi_i^c\ .
\ee
The spin variables just serve to produce these currents and do not appear
by themselves. In the quantum case, the cubic terms should be Weyl ordered.

Following~\cite{kl}, let us try to build the $D(2,1;\alpha)$ algebra based on \p{Q}.
Firstly, the ${\cal N}{=}\,4$ super-Poincar\'e subalgebra
\be\label{Poincare}
\left\{ Q_i, Q_k\right\} =0 \und
\left\{ Q_i, {\bar Q}^k \right\} =2\im\,\delta_i^{\ k} H
\ee
defines a Hamiltonian~$H$ and enforces the following conditions on
our functions $U_a$ and $F_{abc}$,
\bea
&& \pa_a U_b-\pa_b U_a =0\ ,\quad \pa_a F_{bcd}-\pa_b F_{acd}=0\ ,
\label{integr} \\[4pt]
&& F_{cae}F_{ebd} -F_{cbe}F_{ead}=0\ ,
\label{wdvv} \\[4pt]
&& -\partial_a U_b +U_a U_b +F_{abc}U_c=0\ .
\label{eq1}
\eea
The integrability conditions \p{integr} are solved by
\be\label{integr1}
U_a =\pa_a U \und F_{abc} = \pa_a \pa_b \pa_c F
\ee
with two scalar prepotentials $F(x)$ and $U(x)$,
and hence we read subscripts on $U$ and $F$ as derivatives.\footnote{
Note that $U(x)$ and $F(x)$ are defined only up to polynomials of
degree zero and two, respectively.}
Thus, the other two conditions become nonlinear differential equations
for $F(x)$ and $U(x)$, whose solutions define the various possible models.
In particular, \p{wdvv} is the celebrated WDVV equation~\cite{WDVV1,WDVV2},
and \p{eq1} describes the (logarithm of) so-called twisted periods
related to~$F$~\cite{Dub,FeiS}.
With the above conditions fulfilled, the Hamiltonian acquires the form
\be\label{Ham}
H\=\sfrac14 p_a p_a +\sfrac18 {\cal J}^{ik}{\cal J}_{ik}\,U_a U_a(x)
-\im U_{ab}(x)\,{\cal J}_{ik}\,\psi^{ai}\bpsi^{bk}
-\sfrac12 F_{abcd}(x)\,\psi^{ai}\psi^b_i\,\bpsi^c_k\bpsi^{dk}\ .
\ee
One may check that $[H,{\cal J}^{ik}{\cal J}_{ik}]=0$, and thus the Casimir
${\cal J}^{ik}{\cal J}_{ik}=:g^2$ appears as a coupling constant
in the bosonic potential
\be
V \= \sfrac{g^2}{8}\,U_aU_a\ .
\ee

Secondly, for the full $D(1,2;\alpha)$ superconformal invariance one has
to realize the additional generators. This can be done via
\be\label{Dalpha}
D= -\sfrac12 x^a p_a\ ,\qquad K= x^a x^a\ ,\qquad
S_i = - 2 x^a \psi_i^a\ ,\qquad {\bar S}^i = - 2 x^a \bpsi^{ia}\ ,
\ee
together with two sets of composite $su(2)$ currents,
\be\label{su2}
J_{ik} = {\cal J}_{ik} +2\im \psi^a_{(i}\bpsi^a_{k)} \und
I_{11}=\im \psi_i^a\psi^{ia}\ ,\quad
I_{22}=-\im\bpsi^{ia}\bpsi_i^a\ ,\quad
I_{12}=\im \psi_i^a \bpsi^{ia}\ .
\ee
Now, dilatation invariance requires homogeneity,
\be\label{dil}
(x^a\pa_a+1)\,U_b=\pa_b(x^a U_a)=0 \und
(x^a\pa_a+1)\,F_{bcd} =\pa_b(x^a F_{acd})=0\ .
\ee
Thirdly, the remaining superalgebra commutators only fix the integration
constants to
\be\label{conD}
x^a U_a =2\alpha \und x^a F_{abc}=-(1{+}2\alpha)\,\delta_{bc}
\quad\Rightarrow\quad (x^a\pa_a-2)\,F=-\sfrac12(1{+}2\alpha)\,x^ax^a\ .
\ee
Clearly, $F$ cannot vanish unless $\alpha{=}-\sfrac12$, the $osp(4|2)$ case.

It is instructive to introduce the exponential of the prepotential~$U$,
since this linearizes~\p{eq1},
\be \label{Weq}
W = \ep^{-U} \qquad\Rightarrow\qquad
W_{ab}-F_{abc}W_c=0 \und
(x^a\pa_a+2\alpha)\,W = 0\ .
\ee
For $\alpha{=}{-}1$ or~0, the presence of a central charge $m$
perturbs the homogeneity of $U$ and $W$, and it is better to work
with $W$ in these cases (see below).
Clearly, the WDVV prepotential takes the form
\be
F(x)\= -\sfrac14\,(1{+}2\alpha)\,x^2\ln x^2\ +\ F_0(x)\ ,
\ee
while (for $m{=}0$) the other prepotential reads
\be
U(x)\= \alpha\,\ln x^2\ +\ U_0(x) \qquad\Leftrightarrow\qquad
W(x)\= x^{-2\alpha}\,W_0(x)\ ,
\ee
where $F_0$, $U_0$ and $W_0$ are homogeneous of degree~2, 0 and~0,
respectively, and $x^2$ is any expression quadratic in the coordinates.
Obviously, the $F$~prepotentials for any two values of~$\alpha$ are related
by a mere rescaling as long as $\alpha{\neq}{-}\sfrac12$.
The mathematical literature usually does not introduce a euclidean metric
$\delta_{bc}$ but defines an induced metric $G_{bc}=-x_aF_{abc}$
which is constant and nondegenerate.
Hence, for any $\alpha{\neq}{-}\sfrac12$ we can import all known
WDVV solutions~\cite{margra,ves,chaves,feives1,feives2,lp,LST} up to constant coordinate
transformations. The special case
of $D(2,1;{-}\sfrac12)\simeq D(2,1;1)\simeq osp(4|2)$ only appears
as a singular limit, where $F$ can no longer be `normalized' via~\p{conD}
and the induced metric degenerates.

The system \p{integr}--\p{eq1} is form-invariant under SO($n$) rotations.
It can happen that such a rotation reduces the system \p{integr}--\p{eq1}
to two decoupled subsystems. For example, in the presence of global translation
invariance, $x^a\to x^a+\xi$, the center of mass, $X=\sum_a x^a$, can be
decoupled by passing to $n{-}1$ appropriate relative coordinates. Furthermore,
in any coordinates the contractions of \p{wdvv} and \p{eq1} with $x^a$ are
already a consequence of~\p{conD}, which fixes the `radial' dependence.
This effectively reduces the dimensionality further to $n{-}2$ `angular'
coordinates. The number of independent equations in \p{wdvv} and \p{eq1} is
$\sfrac1{12}(n{-}1)(n{-}2)^2(n{-}3)$ and $\sfrac12(n{-}1)(n{-}2)$,
respectively, so that for up to three particles the WDVV~equation~\p{wdvv}
is empty, and \p{eq1} reduces to at most one angular condition, which can
always be solved. Hence, the construction of irreducible multi-particle models
becomes nontrivial with four particles.

All known WDVV solutions are of the form~\footnote{
We disregard here the possibility of `radial' terms, where
$z=\sqrt{\sum_a x_a^2}$ \ or \
$z=\sqrt{\sum_{a<b}(x_a{-}x_b)^2}$~\cite{GLP2,lp}.}
\be \label{Fansatz}
F(x) \= \sum_\beta f_\beta\,K(\beta{\cdot}x) \qquad\textrm{with}\qquad
f_\beta\in\R \und \beta{\cdot}x=\beta(x)=\beta^a x_a\ ,
\ee
where the sum runs over a collection $\{\beta\}$ of $p$ non-parallel
covectors comprising a deformed (super) Lie-algebra root system,
and the function $K$ is universal up to a quadratic polynomial,
\be \label{K}
K'''(z)\=-\frac1z \qquad\Rightarrow\qquad
K(z)\=-\sfrac14 z^2\ln z^2\ .
\ee
Similarly, all known explicit solutions for~$W$ are of the form~\cite{FeiS}
\be
W(x) =  \prod_\ell P_\ell(x)^{-u_\ell}
\qquad\Leftrightarrow\qquad
U(x) = \sum_\ell u_\ell\,\ln P_\ell(x)
\qquad\textrm{with}\qquad \sum_\ell n_\ell u_\ell=2\alpha\ ,
\ee
where $P_\ell$ is a homogeneous polynomial of degree~$n_\ell$.

However, it is a deep and unsolved mathematical problem to find all
solutions~$F$ to the WDVV equation, and only for a part of the known
solutions~\p{Fansatz} the twisted periods~$W$ have been constructed.
Here, we present only the special case of polynomials which completely
factorize into linear factors (with the notation of \p{Fansatz}),\footnote{
For more general solutions involving symmetric polynomials of higher degree,
see~\cite{FeiS}.}
\be \label{Uansatz}
P_\ell(x) = \prod_{\beta_\ell} \beta_\ell{\cdot}x
\qquad\Leftrightarrow\qquad
W(x) = \prod_\beta (\beta{\cdot}x)^{-u_\beta}
\qquad\Leftrightarrow\qquad
U(x) = \sum_\beta u_\beta\,\ln\beta{\cdot}x\ ,
\ee
where we choose a collection of covectors~$\beta$ common to $U$ and~$F$.
Note however that not all covectors from $\{\beta\}$ need to appear in
$F$ or~$U$, because some $f_\beta$ or $u_\beta$ may vanish.
The normalizations~(\ref{conD}) imposed by conformal invariance translate into
simple conditions for the coefficients $u_\beta$ and $f_\beta$,
\be \label{norm}
\sum_\beta u_\beta = 2\alpha \und
\sum_\beta f_\beta\;\beta_b\beta_c = (1{+}2\alpha)\,\delta_{bc}
\qquad\Rightarrow\qquad
\sum_\beta \beta{\cdot}\beta\,f_\beta = (1{+}2\alpha)\,n
\ee
with $\beta{\cdot}\gamma\equiv\delta^{ab}\beta_a\gamma_b$,
and the bosonic potential becomes
\be
V(x) \= \frac{g^2}{8}\sum_{\beta,\gamma} u_\beta u_\gamma\,
\frac{\beta\cdot\gamma}{\beta{\cdot}x\ \gamma{\cdot}x}\ .
\ee
One can actually employ the twisted-period equation~(\ref{eq1}) to solve for
$u_\beta$ in terms of~$f_\beta$.
When inserting the forms (\ref{Fansatz}) and~(\ref{Uansatz}) into~(\ref{eq1}),
the vanishing of each double pole $(\beta{\cdot}x)^{-2}$ yields
\be \label{doublepole}
u_\beta(u_\beta{+}1)\=\beta{\cdot}\beta\,f_\beta\,u_\beta
\qquad\Rightarrow\qquad
u_\beta=0 \qquad\textrm{or}\qquad
u_\beta=\beta{\cdot}\beta\,f_\beta-1
\ee
for each covector~$\beta$.
Inserting this into the `sum rule' (\ref{norm}) for~$u_\beta$, one obtains
a second necessary condition for $\{f_\beta\}$, namely
\be \label{norm2}
\sum_\beta\delta_\beta\,(\beta^2 f_\beta-1) \= 2\alpha
\qquad\textrm{with}\qquad \delta_\beta\in\{0,1\}\ .
\ee
It restricts the $F$~solutions to those which may admit a $U$~solution as well.
However, by no means it guarantees that the single-pole terms in~(\ref{eq1})
work out as well.
More concretely, in a moduli space of WDVV~solutions $F_{\vec t}$, where
$\vec{t}=(t_1,t_2,\ldots)$ represents continuous moduli parameters,
the necessary condition~\p{norm2} for a twisted period~$W$ yields
a hypersurface of co-dimension one,
\be
h_\alpha(\vec t)\=0\ ,
\qquad\textrm{for every yes/no collection $\{\delta_\beta\}$}\ .
\ee
This means that $D(2,1;\alpha)$ symmetric models can exist only on such
hypersurfaces in moduli space, or it can be turned around to fix the value
of~$\alpha$ for any given WDVV solution~$F_{\vec t}$. In such a way,
families of solutions~$(F,U)$ were found for deformed root systems
of types $A_n$, $BCD_n$ and $EF_n$ as well as for (a reduction of) the
super root system~$AB(1,3)$~\cite{kl}.

\subsubsection{Multi-particle models with $osp(4|2)$ symmetry}

For the value $\alpha=-\sfrac12$, the superconformal algebra
$D(2,1;\alpha)$ reduces to $osp(4|2)$. This is a special case, since some
formulae of the preceding subsection become singular for $2\alpha{+}1=0$.
In particular, $F$ becomes homogeneous of degree~2, implying that \
$\sum_\beta f_\beta\,\beta{\otimes}\beta=0$.
Thus the induced metric degenerates,
but the scale of $F$ is determined via~\p{norm2}.~\footnote{
It is now possible to put $F\equiv0$, but it yields the trivial rank-one
solution $U=-\ln\beta{\cdot}x$ for a single covector~$\beta$.}
It also means that the
covector collection~$\{\beta\}$ degenerates to rank~$n{-}1$. Among the known
deformed root systems which solve the WDVV~equation, there exists a degenerate
limit in the moduli space of deformed $A_n$ roots~\cite{kl}: fix the positive roots
of an $A_{n-1}$ subalgebra spanning an $\R^{n-1}$ subspace and project the
remaining $n$~positive roots onto this subspace. There, they form the
fundamental weights of the $A_{n-1}$ subalgebra. Embedding this system again
into $\R^n$, one arrives at the translation and permutation invariant
collection
\be
\bigl\{\beta{\cdot}x\bigr\}\=
\bigl\{ x^a{-}x^b\,, \,x^a{-}\sfrac1n X\ \big|\ 1{\le}a{<}b{\le}n \bigr\}\ .
\ee
The corresponding prepotentials read~\cite{kl}
\bea
F(x) &=& \frac{1}{4n}\sum_{a<b}(x^a{-}x^b)^2\ln(x^a{-}x^b)^2 \ -\
\frac{1}{4n^2}\sum_a(nx^a{-}X)^2\ln(nx^a{-}X)^2 \ ,\\[4pt]
U(x) &=& -\frac{1}{2n} \sum_a \ln(nx^a{-}X)^2 \ .
\eea
The bosonic potential then becomes
\be
V(x)\=\frac{g^2}8\,\Bigl\{ \sum_a \frac1{(nx^a{-}X)^2}\ -\
\frac{1}{n} \Bigl( \sum_a \frac1{(nx^a{-}X)} \Bigr)^2 \Bigr\}
\=\frac{g^2}{8n}
\sum_{a<b} \Bigl(\frac{1}{nx^a{-}X}\ -\ \frac{1}{nx^b{-}X}\Bigr)^2\ ,
\ee
where only the fundamental weights of~$A_{n-1}$ appear.
Of course, the center-of-mass motion can still be decoupled.

The simplest nontrivial example occurs for $n{=}4$,
featuring the 6 positive roots and 4 fundamental weights of~$A_3$.
Due to the isometry $A_3\simeq D_3$, which maps the fundamental $A_3$
weights to the fundamental spinor weights of~$D_3$, there exists a nice
three-dimensional coordinate system after elimination of the center of mass
(note that $a,b=1,2,3$ only)~\cite{kl}:
\bea
F(x) &=& \frac{1}{16}\sum_{a<b,\pm}(x^a{\pm}x^b)^2\ln(x^a{\pm}x^b)^2 \ -\
\frac{1}{16}\sum_{\pm,\pm}(x^1{\pm}x^2{\pm}x^3)^2\ln(x^1{\pm}x^2{\pm}x^3)^2
\\[4pt]
U(x) &=& -\frac18 \sum_{\pm,\pm} \ln(x^1{\pm}x^2{\pm}x^3)^2 \ ,
\eea
where the sums run over all sign choices indicated.
Abbreviating $(x^1,x^2,x^3)=(x,y,z)$ we can write the bosonic potential as
\be
V \= \frac{g^2}{512}\ \frac{
x^4(x^2{-}y^2{-}z^2)+y^4(y^2{-}z^2{-}x^2)+z^4(z^2{-}x^2{-}y^2)+6x^2y^2z^2}
{(x{+}y{+}z)^2\ (x{+}y{-}z)^2\ (x{-}y{+}z)^2(x{-}y{-}z)^2}\ .
\ee
The singular loci of this rational function are best visualized by projecting
onto the unit two-sphere $x^2{+}y^2{+}z^2=1$. On this sphere, the singular
lines form the edges of (the celestial projection of) a regular cuboctahedron
\cite{hny}.

\subsubsection{Multi-particle models with $su(1,1|2)$ symmetry}

The other special case occurs at $\alpha=-1$ or $0$, where
$D(2,1;\alpha)$ becomes a semidirect sum of $su(1,1|2)$ with $su(2)$.
Note that these two special values are related by the discrete flip
$\alpha\leftrightarrow-(1{+}\alpha)$, which flips the sign of $2\alpha{+}1$
and therefore the sign of the WDVV solution~$F$ (see~\p{conD}).
In this subsection, we follow~\cite{KLP} and discuss only $\alpha{=}0$.\footnote{
The case $\alpha{=}{-}1$ is obtained via a `duality', interchanging
$u^A$ and $w_A$ in~\cite{KLP} and giving $W=R^2(W_0-m\ln R)$.}
The possibility of a central charge~$m$ gives us a new option:
It perturbs the homogeneity of~$W$ to
\be
x^a\partial_a W \= -m \qquad\Rightarrow\qquad
W(x)\= -m\ln R\ +\ W_0(x)
\qquad\textrm{with}\quad R^2:={\textstyle\sum_a}(x^a)^2
\ee
and a degree-0 homogeneous function~$W_0$.
Suddenly, the naive choice $\ U_a(x)_i^{\ k}=W_a(x)\,\delta_i^{\ k}\ $
works in~\p{ansQ}, and we can dispense with the isospin degrees of freedom.
As a result one obtains
\bea
Q_i &=& p_a\,\psi_i^a\ +\ \im W_a(x)\,\psi^a_i
\ +\ \im F_{abc}(x)\,\psi^{ak}\psi^b_k\bpsi_i^c \ ,\\[4pt]
H &=& \sfrac14 p_a p_a +\sfrac18 \,W_a W_a(x)
- W_{ab}(x)\,\psi^a_i\bpsi^{bi}
-\sfrac12 F_{abcd}(x)\,\psi^{ai}\psi^b_i\,\bpsi^c_k\bpsi^{dk}\ ,
\eea
and the bosonic potential $V=\sfrac18W_aW_a$ depends on $m$ and possibly
other coupling constants.

In the $su(1,1|2)$ situation,
the standard ${\cal N}{=}\,4$ superspace description applied to the one-particle
case above generalizes to the multi-particle case. With
$n$ copies $\mathscr{X}^a$, $a=1,\ldots,n$, of the ({\bf 1,4,3}) multiplet
subject to the constraints
\be
{\rm (a)} \quad D^iD_i \,\mathscr{X}^a=0\,, \;\;
\bar D_i\bar D^i \,\mathscr{X}^a=0\,; \qquad
{\rm (b)} \quad [D^i,\bar D_i]\, \mathscr{X}^a=2m^a\,,
\ee
one can formulate the action
\be
S_n \=-\displaystyle{\int} dt\,d^4\theta \  G(\mathscr{X}^a)\ ,
\ee
where the superpotential $G(x)$ is subject to the conformality condition
\be\label{supconf}
x^a G_a - G \= c_a x^a
\ee
for some constants $c_a$.
The bosonic part of this action takes the sigma-model form
\be
S_B \= \sfrac12 \displaystyle{\int} dt\, \Bigl\{
G_{ab}(x)\,\dot{x}^a\dot{x}^b\ -\ G_{ab}(x)\,m^am^b \Bigr\}\ .
\ee
To serve as an action for $n$ particles on a line, the target-space
metric $G_{ab}$ must be flat. This yields an integrability condition,
namely the vanishing of the Riemann tensor,
\be\label{Riemann}
G_{abe}G^{ef}G_{fcd}-G_{ace}G^{ef}G_{fbd}\=0
\qquad\textrm{with}\qquad G^{ef}G_{fg}=\delta^e_{\ g}\ .
\ee
If this is met, there exists an `inertial' coordinate system in which the
kinetic term takes the standard form
$\ \sfrac12\delta_{ab}\dot{x}^a\dot{x}^b$,
and the superpotential produces both prepotentials $F$ and $W$~\cite{KLP}.
Converting to the `inertial' coordinates,
the condition \p{Riemann} is in fact equivalent to
\be \label{structure}
F_{abe}F_{ecd} -F_{ace}F_{ebd}=0
\und W_{ab}-F_{abc}W_c=0\ .
\ee
In these coordinates, the conformality condition~\p{supconf} becomes
\be
x^a G_a - 2G \= -\sfrac12 x^a x^a\ ,
\ee
which yields the homogeneity relations
\be \label{homogeneity}
x^a W_a =-m \und x^a F_{abc}=-\delta_{bc}
\quad\Rightarrow\quad (x^a\pa_a-2)\,F=-\sfrac12\,x^ax^a\ ,
\ee
and the central charge comes out as $\ m=-2c_a m^a$.

Unfortunately, it is very difficult to solve \p{structure} together with
\p{homogeneity} for an arbitrary number~$n$ of particles.\footnote{
One cannot simply map the $(F,U)$ solutions of subsection~4.2.2
to $(F,\ep^{-U})$ since in $Q$ and $H$ there appears $W$ instead of~$U$.
Hence, the $su(1,1|2)$ system without isospin variables differs essentially
from the $\alpha\to0$ or $\alpha\to-1$ limit of the isospin system of
subsection~4.2.2.}
At $n{=}3$, the general solution depends on one free function
and can be given. For $n{=}4$, only sporadic solutions, mostly
involving hypergeometric functions, have been found~\cite{KLP}.

\subsection{Superconformal models from harmonic ${\cal N}{=}\,4$ superspace}

In the previous subsections we dealt with the formulations of the one- and many-body ${\cal N}{=}\,4$ superconformal Calogero
models in the ordinary ${\cal N}{=}\,4$ superspace and in the component approach. Here we discuss which new possibilities
for construction of such models are provided by harmonic ${\cal N}{=}\,4, d{=}1$ superspace \cite{IL}.\footnote{
For a description of ${\cal N}{=}\,4, d{=}1$ superconformal models in bi-harmonic superspace
see~\cite{IN}.}
Sometimes we shall use
the harmonic superfields in parallel with the ordinary ${\cal N}{=}\,4$ ones, hoping that this will not give rise to misunderstanding.

\subsubsection{Harmonic description of the ${\cal N}{=}\,4$ representations}

Off-shell ${\cal N}{=}\,4, d{=}1$ supermultiplets admit a concise
formulation in
the harmonic superspace (HSS)~\cite{IL}, an extension of \p{Rss} by
the harmonic
coordinates $u^\pm_i$:
\be
(t,\theta^\pm, \bar\theta^\pm, u_i^\pm)\,, \qquad \theta^\pm=\theta^i
u_i^\pm \,,
\qquad
\bar\theta^\pm=\bar\theta^i u_i^\pm\,,\qquad u^{+i}u_i^-=1\,. \lb{HSS}
\ee
The commuting ${\rm{SU}}(2)$ spinors $u_i^\pm$ parametrize the 2-sphere
$S^2 \sim {\rm SU}(2)_R/{\rm{U}}(1)_R$. The salient property of HSS is the
presence of
an important subspace in it, the harmonic analytic superspace (ASS) with
half of
Grassmann co-ordinates
as compared to \p{Rss} or \p{HSS}:
\be
(\zeta,u)=(t_A,\theta^+, \bar\theta^+, u_i^\pm)\,, \qquad t_A=t+i(\theta^+
\bar\theta^-
+\theta^-\bar\theta^+)\,. \lb{Ass}
\ee
It is closed under the ${\cal N}{=}\,4$ supersymmetry transformations. Most
of the off-shell ${\cal N}{=}\,4, d{=}1$ multiplets are represented by the analytic superfields, i.e. those
``living'' on the subspace \p{Ass}.

Spinor covariant derivatives in the analytic basis of HSS, viz. $(\zeta, u,
\theta^-, \bar\theta{}^-)$, take the form
\begin{equation}\label{D-ferm}
D^+ =\frac{\partial}{\partial\theta^-}\,,\quad \bar D^+
=-\frac{\partial}{\partial\bar\theta^-}\,,\qquad D^-
=-\frac{\partial}{\partial\theta^+}-2i\bar\theta^-\partial_{t_A}\,,\quad
\bar D^-
=\frac{\partial}{\partial\bar\theta^+}- 2i\theta^-\partial_{t_A}\,.
\end{equation}
In the central basis \p{HSS}, the same derivatives are defined as the
projections
$D^\pm=D^i u_i^\pm$ and $\bar D^\pm=\bar D^i u_i^\pm$.
Harmonic covariant derivatives in the analytic basis read
\begin{equation}\label{D-harm}
D^{\pm\pm} =\partial^{\pm\pm} +2i\theta^\pm\bar\theta^\pm\partial_{t_A} +
\theta^\pm\frac{\partial}{\partial\theta^\mp}
+\bar\theta^\pm\frac{\partial}{\partial\bar\theta^\mp}\,.
\end{equation}
The integration measures are defined by
\begin{equation}
\mu_H =dudtd^4\theta =\mu^{(-2)}_A(D^+\bar D^+)\,, \qquad
\mu^{(-2)}_A=dud\zeta^{(-2)}
=dudt_A d\theta^+d\bar\theta^+ =dudt_A(D^-\bar D^-)\,.
\end{equation}

Let us briefly sketch the harmonic superspace description of some ${\cal N}{=}\,4, d{=}1$ supermultiplets
(for the details see \cite{IL,DI1,DI2}).

The ${\cal N}{=}\,4, d{=}1$ supermultiplets are described in harmonic superspace by the harmonic superfields
$\Phi^{(q)}(t,\theta^\pm, \bar\theta^\pm, u)$ with $U(1)$ charge $q$ reflecting local $U(1)$ symmetry of harmonic formulation
\begin{equation}\label{D0-harm}
D^{0} \Phi^{(q)}=q  \Phi^{(q)}\,,
\end{equation}
where
\begin{equation}\label{D-harm-equ}
D^{0} =\partial^{0}  +
\theta^+\frac{\partial}{\partial\theta^+}
+\bar\theta^+\frac{\partial}{\partial\bar\theta^+}-
\theta^-\frac{\partial}{\partial\theta^-}
-\bar\theta^-\frac{\partial}{\partial\bar\theta^-}\,.
\end{equation}
Analytic superfields $\Phi^{(q)}(\zeta, u)$ depending on the supercoordinates  \p{Ass} are
defined by the Grassmann Cauchy-Riemann constraints
\begin{equation}\label{D-anal}
D^{+} \Phi^{(q)}=\bar D^{+} \Phi^{(q)}=0\,.
\end{equation}
Analytic superfields can satisfy generalized reality conditions which yields usual reality conditions
for component fields. Generalized conjugation of the harmonic superfields,
which is the combination of complex conjugation and antipodal reflection on harmonic two-sphere,
is denoted by tilde: $\Phi^{(q)}\to\tilde\Phi^{(q)}$  (for details see \cite{GIOS,IL}).
Below we shortly present  the superconformal description of the basic ${\cal N}{=}\,4, d{=}1$ supermultiplets.\\

\noindent
{\bf The (1,4,3) supermultiplet. \quad}
The formulation of this supermultiplet in the standard ${\cal N}{=}\,4$ superspace was given in Sect.\,4.2.1.
In HSS it is described by the  even real  harmonic superfield ${\mathscr{X}}(t,\theta^\pm, \bar\theta^\pm, u)$
which has zero harmonic charge and is subjected to the constraints
\begin{equation}  \label{cons-X-g-V-1}
D^{++} \,\mathscr{X}=0\,,
\end{equation}
\begin{equation}  \label{cons-X-g-1}
D^{+}{D}^{-} \,\mathscr{X}=0\,,\qquad
    \bar D^{+}\bar D^{-}\, \mathscr{X}=0\,,\qquad
    (D^{+}\bar D^{-} +\bar D^{+}D^{-})\, \mathscr{X}=0\,.
\end{equation}
The set of conditions \p{cons-X-g-V-1} and (\ref{cons-X-g-1}) is equivalent to
the standard constraints  \p{cons-X-g1} in the central basis \p{HSS}: the equation  \p{cons-X-g-V-1}
implies independence of the superfield $\mathscr{X}$ on harmonic variables, i.e.
$\mathscr{X}=\mathscr{X}(t,\theta_i, \bar\theta^i)$, and then the equations (\ref{cons-X-g-1})
are reduced to the constraints  \p{cons-X-g1}.

There is another, equivalent description of the same ({\bf 1,4,3}) supermultiplet. It makes use of
real analytic gauge superfield $\mathcal{V}(\zeta,u)$ , ${D}^{+} \,\mathcal{V}=\bar{D}^{+}\, \mathcal{V}=0$,
which is defined up to the abelian gauge freedom
\be\lb{ga-V0}
\delta \mathcal{V} = D^{++}\lambda^{--}\,, \quad \lambda^{--} =
\lambda^{--}(\zeta,
u)\,.
\ee
The analytic superfield $\mathcal{V} = \mathcal{V}(\zeta, u)$ plays the role of the prepotential
for the ({\bf 1,4,3}) multiplet and is related to the superfield
${\mathscr{X}}(t,\theta_i,\bar\theta^i)$ by the harmonic integral
transform \cite{DI1}
\begin{equation} \label{X-V}
{\mathscr{X}}(t,\theta_i,\bar\theta^i)=\int du
\,\mathcal{V}(t_A,\theta^+,\bar\theta^+,u)\Big|_{\theta^\pm=\theta^i u^\pm_i,\,\,\,
\bar\theta^\pm=\bar\theta^i u^\pm_i} \,.
\end{equation}
The prepotential representation~(\ref{X-V}) automatically solves the basic
constraints  \p{cons-X-g1}.

The superconformal action for the ({\bf 1,4,3}) multiplet is given by the formula~(\ref{4N-X1}).
As shown below, the manifestly analytic prepotential formulation of this multiplet allows one to
construct new ${\cal N}{=}\,4$ superconformally-invariant models in which the  ({\bf 1,4,3}) multiplet
is coupled to other ${\cal N}{=}\,4$ supermultiplets.\\

\noindent
{\bf The (3,4,1) tensor supermultiplet. \quad}
This supermultiplet is described by the even analytic gauge superfield
$L^{++}(\zeta,u)$, which satisfies generalized reality condition, $\widetilde{L^{++}} = L^{++}$,  and is subjected to
the additional harmonic constraint \cite{IL}
\begin{equation}\label{con-lm}
{D}^{++} \,L^{++}=0\,.
\end{equation}
The constraints \p{con-lm} can be directly solved.
The off-shell component content of the tensor multiplet is formed by the fields $v_{ij} = v_{ji}$, $B$,
$\psi_i$ and $\bar\psi_i$. They enter the $\theta$ -expansion of the
superfield $L^{++}$ subjected to \p{con-lm} as \cite{IL}
\begin{equation}\label{sol-LV-com}
L^{++} = v^{++} +\theta^+ \psi^+ +\bar\theta^+ \bar\psi^+ + 2i \,\theta^+ \bar\theta^-
\left(\dot v^{+-} +B \right),
\end{equation}
where $v^{++}=v^{ij}u^+_iu^+_j\,, \;v^{+-}=v^{ij}u^+_iu^-_j$, $\psi^+=\psi^i u^+_i$ and $\bar\psi^+=\bar\psi^i u^+_i\,$,
and SU(2)-triplet $v^{ij}$ describes three bosonic physical degrees of freedom.

In central basis, the superfield $L^{++}$ is represented in the form
\begin{equation}
L^{++}=u^+_i u^+_k L^{ik}\,,
\end{equation}
where $L^{ik}(t,\theta,\bar\theta)$ is the usual superfield  subjected to the constraints
\begin{equation}
D^{(i}L^{kl)}=0\,,\qquad \bar D^{(i}L^{kl)}=0\,.
\end{equation}

Sigma-model type actions for $L^{++}$ are written as an integral over the whole harmonic superspace of the Lagrangian
which is a function of $L^{++}$, $L^{+-}=\frac12 \, D^{--}L^{++}$, $L^{--}=\frac12  \, D^{--}L^{+-}$ and harmonics
(see the details in \cite{IL}). The ${\cal N}{=}\,4$ superconformal subclass of these actions have the form~(\ref{4N-X1})
in which we must make the substitution (see details in \cite{IL,IKLech,IKLecht,DI1,DI2})
\begin{equation}\label{X-L}
\mathscr{X}\;\to\; L^{-1},\quad\mbox{where}\quad L:=\sqrt{L^{ik}L_{ik}}=\sqrt{2[L^{++}L^{--}-(L^{+-})^2]}\,.
\end{equation}

The WZ action, generating the superpotential term for $L^{++}$, is given by the following integral over the analytic superspace
\begin{equation}\label{A}
S_{WZ}= {\textstyle\frac{i}{2}} \,\gamma\displaystyle{\int} \mu^{(-2)}_A \,
\mathscr{L}^{(+2)} (L^{++},u)= \gamma \int dt \, \Big({\textstyle\frac{1}{2}}\,\mathscr{A}_{ik}\dot v^{ik} -
{\textstyle\frac{i}{2}}\, \,\mathscr{R}_{ik}\bar\psi^{(i}\psi^{k)} + \mathscr{U} B \Big),
\end{equation}
where
\begin{equation}\label{pot-lin}
\mathscr{A}_{ik}= 2\int du \,u^+_{(i}u^-_{k)}\, \frac{\partial \mathscr{L}^{++} }{\partial
v^{++}}\,,\qquad \mathscr{R}_{ik}= \int du \,u^+_{i}u^+_{k}\, \frac{\partial^2
\mathscr{L}^{++} }{\partial (v^{++})^2}\,,\qquad \mathscr{U}=\int du \, \frac{\partial
\mathscr{L}^{++} }{\partial v^{++}}\,.
\end{equation}
{}From the definition of these potentials follow the relations between them:
\begin{equation}\label{const-pot-lin1}
\triangle_{\mathbb{R}^3}\mathscr{U}=0\,,\qquad
\triangle_{\mathbb{R}^3}\mathscr{A}_{ik}=0\,,\qquad
\partial^{ik}\mathscr{A}_{ik}=0\,,
\end{equation}
\begin{equation}\label{const-pot-lin2}
\partial_{ij}\mathscr{A}_{kl}-\partial_{kl}\mathscr{A}_{ij}=
\left(\epsilon_{ik}\partial_{jl}+\epsilon_{jl}\partial_{ik}\right)\mathscr{U}\,,
\end{equation}
\begin{equation}\label{const-pot-lin3}
\mathscr{R}_{ik}=
\partial_{ik}\mathscr{U}\,.
\end{equation}
Here, $\partial_{ik}=\partial/\partial v^{ik}$ and $
\triangle_{\mathbb{R}^3}=\partial^{ik}\partial_{ik} $ is Laplace operator on
$\mathbb{R}^3$.
Eqs. (\ref{const-pot-lin1}), (\ref{const-pot-lin2}) are recognized as the equations
defining the monopole (static) solution for a self-dual Maxwell or gravitation fields in $\mathbb{R}^4$. The new striking
feature of the ${\cal N}{=}\,4$ mechanics models associated with the multiplet $({\bf 3,4,1})$ as compared with
those based on the multiplet $({\bf 1,4,3})$ is just the appearance of coupling of SU(2) vector physical bosonic component
to the external magnetic 3-potential $\mathscr{A}_{ik}$ in \p{A}.

The superconformal action of $L^{++}$ was constructed in \cite{IL} as the harmonic superspace reformulation of the model
previously constructed in \cite{IKLech} (which used the ordinary ${\cal N}=\,4$ superspace). It consists of the sigma model part
and WZ part, each being
superconformal separately (with respect to the most general ${\cal N}=\,4$ superconformal symmetry $D(2,1;\alpha)$).
The unique superconformal WZ term corresponds to the one-monopole potential $\mathscr{U}$ in (\ref{A}).
The explicit expression
for the superconformal superfield WZ term is given in the next subsection. In components, after elimination of the auxiliary field $B$
from both the sigma model and WZ parts, there emerges the conformal potential for $v^{ik}$ and SU(2)/U(1) WZ term,
with the strengths specified by  the coefficient before the superfield WZ term and so related to each other. Thus in this case a sort
of 3-dimensional superconformal mechanics arises. The full bosonic action, in the parametrization in which $v^{ik}$ is split into the
radial and angular parts, is given by the expression \p{MoD} with $g = (2\alpha)^{-2}$.\\

\noindent
{\bf The (4,4,0) ``root'' supermultiplet. \quad}
{}From this multiplet all others may be obtained via a reduction process on either the component-action \cite{BKMO} or
the superfield-action \cite{DI1,DI2,DI2a} levels.
It is described by the harmonic charge-one complex analytic superfields
${\mathcal{Z}}^+ $, $\bar{\mathcal{Z}}^+ $,
subjected to the constraint
\begin{equation}\label{con-hm}
{D}^{++} \,{\mathcal{Z}}^+=0\,.
\end{equation}
The solution of the constraint~(\ref{con-hm}) is the following
\begin{equation}  \label{Ph-WZ}
{\mathcal{Z}}^+ = z^{i}u_i^+ + \theta^+ \varphi + \bar\theta^+ \phi - 2i\,
\theta^+
\bar\theta^+\partial_{t_A}z^{i}u_i^-
\,,
\end{equation}
where SU(2)-dublet $z^{i}(t_A)$ describes four bosonic physical degrees of freedom.
We can combine the
superfields ${\mathcal{Z}}^+$ and $\tilde {\mathcal{Z}}^+$ into a doublet of
some extra (``Pauli-G\"ursey'') SU(2)$_{PG}$ group according to
\be \label{PG}
q^{+ a} := (\tilde {\mathcal{Z}}^+, {\mathcal{Z}}^+)\,, \; a =1,2\,.
\ee
In central basis, the ``root'' supermultiplet is represented in the form
\begin{equation}
q^{+a}=u^+_i q^{ia}\,,
\end{equation}
where $q^{ia}(t,\theta,\bar\theta)$ is the ordinary ${\cal N}{=}\,4$  superfield  subjected to the constraints
\begin{equation}
D^{(i}q^{k)a}=0\,,\qquad \bar D^{(i}q^{k)a}=0\,.
\end{equation}

Sigma-model action of the ``root'' supermultiplet can be written as an integral over
the whole harmonic superspace of a function of $q^{+ a}$, $q^{- a}={D}^{--} q^{+ a}\,$,
and harmonics. Taking into account that properties of the product $a_{(cb)}q^{+ c}q^{+ b}$,
where $a_{(cb)}$ are some constants, are similar to the tensor superfield $L^{++}$, it is natural
to construct the ${\cal N}{=}\,4$ superconformal sigma-action for the multiplet ({\bf 4,4,0}) in the same way as for
the ({\bf 3,4,1}) supermultiplet (\ref{X-L}). More precisely,
${\cal N}{=}\,4$ superconformal actions for the ({\bf 4,4,0}) multiplet have the form~(\ref{4N-X1})
in which one should make the substitution
\begin{equation}\label{X-q}
\mathscr{X}\;\to\; q^{-2},\quad\mbox{where}\quad q^2:=q^{ia}q_{ia}=2q^{- a}q^{+}_{ a}\,.
\end{equation}
More details on the structure of superconformal sigma-model actions for the ({\bf 4,4,0}) multiplet can be
found in \cite{IL,IKLecht}.

The superpotential term for the superfield $q^{+ a}$ is represented as
the integral over the analytic superspace of an analytic Lagrangian which can depend on the superfield $q^{+ a}$
and harmonics (see details in \cite{IL}). As demonstrated in  \cite{IL}, such a WZ term involving only
the ({\bf 4,4,0}) supermultiplet cannot be superconformally invariant. So, based only on the multiplet ({\bf 4,4,0}) one cannot
construct ${\cal N}{=}\,4$ superconformal mechanics model with the conformal potential for the physical bosonic fields, as distinct from the
case of the multiplet  ({\bf 3,4,1}). Some ways of gaining conformal potential for the multiplet ({\bf 4,4,0}),
through its superconformal couplings to some other ${\cal N}{=}\,4$ multiplets, are discussed in \cite{DI1,DI2}.
In the next section we shall consider a modification
of the ({\bf 4,4,0}) WZ term, such that it involves coupling to the ({\bf 1,4,3}) prepotential $\mathcal{V}(\zeta, u)$. This modification
will allow us to construct a new ${\cal N}{=}\,4$ superconformal mechanics with a new mechanism of generating  conformal potential for the field
$x(t)$.

Finally, we point out that the WZ potential terms for all ${\cal N}{=}\,4$ supermultiplets can be represented in manifestly
${\cal N}{=}\,4$ superfield form only in harmonic superspace but not in the usual superspace.\footnote{In the
formulations in ordinary ${\cal N}{=}\,4$ superspace such terms either include manifest
$\theta$s~\cite{IKLech} or are expressed through
appropriate unconstrained prepotentials with a complicated pregauge freedom.}
Such terms are important for the description of the isospin supermultiplets
recently introduced in the construction of new integrable supersymmetric systems,
including superconformal ones. Also note that for each ${\cal N}{=}\,4$ multiplet listed above
there exists a ``mirror'' (or ``twisted'') counterpart~\cite{IKLecht}, for which the roles of
the manifest R-symmetry group~${\rm SU}(2)_R$ (acting on the doublet indices $i,k$
of the component fields and Grassmann coordinates) and a hidden R-symmetry group SU(2)
(joining, e.g., $\theta_i$ and $\bar\theta_i$ into doublets) are interchanged. When these two types of ${\cal N}{=}\,4$ multiplets
are considered together, new models of supersymmetric (and superconformal) mechanics can be constructed (see, e.g. \cite{IN}).
Not too much is known to date about such models, so we will not discuss them here.

\subsubsection{Single-particle models with isospin degrees of freedom}

Here we present a non-trivial model of ${\cal N}{=}\,4$
superconformal mechanics in which the conformal potential is generated by the gauging method
with making use of the ``semi-dynamical'' ${\cal N}{=}\,4$ isospin supermultiplet. This model is the one-particle
case of the Calogero-like multi-particle model~\cite{FIL1} which will be considered in the next section.

The model is built on superfields corresponding to three off-shell ${\cal N}{=}\,4$
supermultiplets: ({\bf i}) the ``radial'' multiplet
({\bf 1,4,3}); ({\bf ii}) the Wess-Zumino (``isospin'') multiplet ({\bf
4,4,0}); and ({\bf iii}) the gauge (``topological'') multiplet ({\bf 0,0,0}). The
action is a sum of three terms
\begin{equation}\label{4N-gau-1}
S =S_{\mathscr{X}} + S_{FI} + S_{WZ}\,.
\end{equation}

First term in~(\ref{4N-gau-1}) is the standard free action (\ref{4N-X1}) of the ({\bf 1,4,3}) multiplet.

Second term in~(\ref{4N-gau-1}) is Fayet--Iliopoulos (FI) term
\begin{equation}\label{4N-FI-1}
S_{FI} =-{\textstyle\frac{i}{2}}\,\,c\int \mu^{(-2)}_A \,V^{++}
\end{equation}
for the gauge supermultiplet. The even analytic
gauge superfield $V^{++}(\zeta,u)$, ${D}^{+} \,V^{++}=0$, $ \bar{D}^{+}\,
V^{++}=0\,,$ is subjected to the gauge transformations
\be
V^{++}{}' = V^{++} - D^{++}\lambda, \quad \lambda = \lambda(\zeta,
u)\,,\label{tran4-V}
\ee
which are capable to gauge away, {\it locally}, all the components from
$V^{++}$. However, the latter contains a component which
    cannot be gauged away {\it globally}. This is the reason why this $d{=}1$
supermultiplet was called ``topological'' in \cite{DI1}.

Last term in~(\ref{4N-gau-1}) is Wess--Zumino (WZ) term
\begin{equation}\label{4N-WZ-1}
S_{WZ} = -{\textstyle\frac{1}{2}}\,\int \mu^{(-2)}_A \, \mathcal{V}\,
\tilde{\mathcal{Z}}{}^+\, {\mathcal{Z}}^+\, .
\end{equation}
Here, the complex analytic superfield ${\mathcal{Z}}^+,
\tilde{\mathcal{Z}}^+$ $(D^+
{\mathcal{Z}}^+ = \bar D^+ {\mathcal{Z}}^+ =
0)\,,$ is subjected to the harmonic constraints
\begin{equation}  \label{cons-Ph-g-1}
\mathscr{D}^{++} \,{\mathcal{Z}}^+\equiv (D^{++} + i\,V^{++})
\,{\mathcal{Z}}^+=0\,,\qquad \mathscr{D}^{++}
\,\tilde{\mathcal{Z}}{}^+\equiv (D^{++} - i\,V^{++}) \,\tilde{\mathcal{Z}}{}^+=0
\end{equation}
and represents a gauge-covariantized version of the ${\cal N}{=}\,4$
multiplet ({\bf 4,4,0}). The relevant gauge transformations are
\be
{\mathcal{Z}}^+{}' = e^{i\lambda} {\mathcal{Z}}^+,\qquad
\tilde{\mathcal{Z}}{}^+{}' =
e^{-i\lambda}\tilde{\mathcal{Z}}{}^+\,.\label{tran4-Phi}
\ee

The superfield $\mathcal{V}(\zeta,u)$ in \p{4N-WZ-1} is a prepotential
of the $({\bf 1,4,3})$ superfield $\mathscr{X}$.
The coupling to the multiplet $({\bf 1,4,3})$ in \p{4N-WZ-1} is introduced
to ensure ${\cal N}{=}\,4$ superconformal $D(2,1;\alpha)$ invariance.
The coordinate realization of the superconformal boosts of
$D(2,1;\alpha)$ in analytic subspace is given by \cite{IL,DI1}:
\begin{equation}  \label{sc-coor-bo}
\delta^\prime t_A=\alpha^{-1}\Lambda t_A\,,\qquad \delta^\prime \theta^+= -\eta^+ t_A +2i(1{+}\alpha)\eta^- \theta^+\bar\theta^+\,,\qquad
\delta^\prime u^+_i= \Lambda^{++}u^-_i\,,
\end{equation}
\begin{equation}  \label{sc-me1-h}
\delta^\prime
\mu_H= \mu_H\left(2\Lambda -\alpha^{-1}(1{+}\alpha)\Lambda_0\right)\,,\qquad \delta^\prime
\mu^{(-2)}_A= 0\,,
\end{equation}
where
\begin{equation}  \label{def-La1}
\Lambda = 2i\alpha (\bar\eta^-\theta^+ -\eta^-\bar\theta^+ )\,,\qquad
\Lambda^{++} = D^{++}\Lambda=2i\alpha (\bar\eta^+\theta^+ - \eta^+\bar\theta^+ )\,.
\end{equation}
Based on the results of ref. \cite{IL}, it is easy to find the appropriate transformation laws of all involved superfields
\begin{equation}  \label{sc-1n-h}
\delta^\prime \mathcal{V} =
-2\Lambda\,\mathcal{V}\,,\qquad \delta^\prime {\mathcal{Z}}^+ =
\Lambda\,{\mathcal{Z}}^+\,,\qquad\delta^\prime V^{++} = 0\,.
\end{equation}
The invariance of the action~(\ref{4N-gau-1}) and harmonic constraints under these transformations can be easily checked.
Note that the constraints~(\ref{cons-X-g-V-1}), (\ref{cons-X-g-1})
and~(\ref{cons-Ph-g-1}), as well as the actions~(\ref{4N-FI-1})
and~(\ref{4N-WZ-1}), are invariant with respect to the
$D(2,1;\alpha)$  transformations with an arbitrary $\alpha$. It is
worth to point out that the action~(\ref{4N-WZ-1}) is superconformally
invariant just due to the presence of the analytic prepotential
$\mathcal{V}\,$.

In order to clarify
the off-shell superfield content of our model,
it is instructive to fix the underlying U(1) gauge freedom by choosing a
gauge which preserves manifest ${\cal N}{=}\,4$ supersymmetry. A gauge suitable for
our purpose was used in \cite{DI2}.
To make contact with the consideration in \cite{DI2}, we introduce
SU(2)$_{PG}$ spinor field by \p{PG}
and rewrite the transformation law \p{tran4-Phi} and the constraints
\p{cons-Ph-g-1} as
\be
\delta q^{+ a} = \lambda c^a_{\;b} q^{+ b}\,, \quad D^{++}q^{+ a} + V^{++}
c^a_{\;b}q^{+ b} = 0\,. \label{qc}
\ee
Here, the traceless constant tensor $c^{a}_{\;b}$ breaks SU(2)$_{PG}$ down
to U(1) which is just the symmetry to be gauged. In the frame where the only non-zero entries
of $c^a_{\; b}$ are $c^1_{\;1} =-c^2_{\;2} = -i$,
we recover the transformation law \p{tran4-Phi} and the constraints
\p{cons-Ph-g-1}. It is easy to show that
\be
\tilde{\mathcal{Z}}^+ {\mathcal{Z}}^+ = -\frac{i}{2}\,q^{+a}\,c_{ab}\,q^{+
b}\,.
\ee

In \cite{DI1} (following \cite{GIOS}) an invertible equivalence
redefinition of $q^{+ a} \Rightarrow (\omega, L^{++})$
has been used, such that the ${\rm U}(1)$ gauge transformation in
\p{qc} is realized as
\be
\delta \omega = -2\lambda\,, \quad \delta L^{++} = 0\,.
\ee
One can fully fix
the ${\rm U}(1)$ gauge freedom by imposing the manifestly
${\cal N}{=}\,4$ supersymmetric gauge
\be
\omega = 0\,.
\ee
In this gauge, the harmonic constraint in \p{qc} amounts to the following relations
\bea
&& \mbox{(a)} \;q^{+a}\,c_{ab}\,q^{+ b} = 4(c^{++} + L^{++})\,, \quad
\mbox{(b)} \;
V^{++} = \frac{L^{++}}{(1 + \sqrt{1 + c^{--}L^{++}})\sqrt{1 +
c^{--}l^{++}}}\,,
\nonumber \\
&& \mbox{(c)} \; D^{++}(c^{++} + L^{++}) = D^{++}L^{++} =0\,, \lb{gc2}
\eea
where $c^{\pm\pm} = c^{(ab)}u^\pm_a u^\pm_b$. After substituting the
expressions
(\ref{gc2}a) and (\ref{gc2}b) into
\p{4N-WZ-1} and \p{4N-FI-1}, the total superfield action \p{4N-gau-1} takes the
form:
\be\lb{Sman}
S = -\textstyle{\frac{1}{4(1{+}\alpha)}}\,\displaystyle{\int} \mu_H \,
\mathscr{X}^{\,-1/\alpha}
+ i \int \mu_A^{(-2)}
\left[\mathcal{V}\,(c^{++} + L^{++}) - \frac{c}{2}\, \frac{L^{++}}{(1 + \sqrt{1 + c^{--}L^{++}})\sqrt{1 +
c^{--}L^{++}}}\right].
\ee

As we already mentioned, the superfield $L^{++}$ with the constraint (\ref{gc2}c)
accommodates an off-shell ${\cal N}{=}\,4$ multiplet (${\bf 3,4,1}$)
\cite{IL}. So, the action \p{Sman} describes a system of two
interacting off-shell ${\cal N}{=}\,4, d{=}1$ multiplets: (${\bf 1,4,3}$) described by the superfield $\mathscr{X}$
and (${\bf 3,4,1}$) described by the analytic superfield $L^{++}$. This is
the off-shell content of our $D(2,1;\alpha)$ model.

As distinct from the superconformal mechanics based on a single $({\bf 3,4,1}$)
multiplet the action of which is a sum of the sigma-model type
term and superconformal WZ term of $L^{++}$ \cite{IKLech,IL}, the action \p{Sman} involves
only superconformal superfield WZ term of this multiplet (the last term
in the square brackets). The interaction with the multiplet (${\bf 1,4,3}$) is introduced through a superconformal bilinear coupling
of both multiplets (the first term in the square
brackets).
Notice that, due to the absence of the kinetic term for $L^{++}$ in
\p{Sman}, the on-shell content of the model appears to be
drastically different from the off-shell one: the eventual component
action contains only three bosonic fields and four fermionic fields,
which are combined into some new on-shell (${\bf 3,4,1}$) multiplet
(see the next section).

Now we consider the model in WZ gauge. Using the ${\rm U}(1)$ gauge freedom~\p{tran4-V}, (\ref{tran4-Phi}) we can
choose WZ gauge
\begin{equation}  \label{WZ-4N-1}
V^{++} =2i\,\theta^{+}
    \bar\theta^{+}A(t_A)\,.
\end{equation}
Then,
using component expansion for the prepotential superfield $\mathcal{V}\,$,
\begin{equation}  \label{V0-WZ}
\mathcal{V} (t_A, \theta^+, \bar\theta^+, u^\pm) =x(t_A)- 2\,\theta^+
\psi^{i}(t_A)u^-_i  -
2\,\bar\theta^+ \bar\psi^{i}(t_A)u^-_i + 3\,\theta^+ \bar\theta^+
N^{ik}(t_A)u^-_i
u^-_k
\end{equation}
and the solution of the constraint~(\ref{cons-Ph-g-1}) in WZ gauge~(\ref{WZ-4N-1})
\begin{equation}  \label{Ph-WZ-m}
{\mathcal{Z}}^+ = z^{i}u_i^+ + \theta^+ \varphi + \bar\theta^+ \phi - 2i\,
\theta^+
\bar\theta^+\nabla_{t_A}z^{i}u_i^-
\,,\qquad \nabla z^k:=\dot z^k + i A \, z^k\,,
\end{equation}
as well as eliminating auxiliary fields and making the redefinition
\begin{equation}\label{4N-nZ}
x^{\prime} = x^{-\frac{1}{2\alpha}} \,, \qquad \psi_{k}^{\prime} =
-{\textstyle\frac{1}{2\alpha}}\,x^{-\frac{1}{2\alpha}-1} \psi_{k}\,, \qquad z^\prime{}^i =
x^{1/2}\,z^i\,,
\end{equation}
we arrive at the on-shell form of the action~(\ref{4N-gau-1}) in WZ gauge (we
omitted the primes on $x$, $\psi$ and $z$)~\footnote{
For obtaining such systems via Hamiltonian reduction see~\cite{KL}.}
\begin{equation}\label{4N-ph}
S =  \int dt \,\Big[p\,\dot x  +i
\left( \bar\psi_k \dot\psi^k -\dot{\bar\psi}_k \psi^k \right)
+ {\textstyle\frac{i}{2}}\, \left(\bar z_k \dot z^k -
\dot{\bar z}_k z^k\right)-H \Big] \,.
\end{equation}
The Hamiltonian $H$ is
\begin{equation}\label{ga-u1-H}
H ={\textstyle\frac{1}{4}}\,p^2  +\alpha^2\,\frac{(\bar z_k z^{k})^2}{4x^2} - 2\alpha  \,
\frac{\psi^{i}\bar\psi^{k} z_{(i} \bar z_{k)}}{x^2} - (1{+}2\alpha) \,
\frac{\psi_{i}\psi^{i}\,\bar\psi^{k} \bar\psi_{k}}{2x^2}\,.
\end{equation}
The field $A(t)$, playing the role  $d{=}1$ $U(1)$-connection,
is the Lagrange multiplier for the first-class constraint
\begin{equation}\label{D0-con}
D^0 -c\equiv \bar z_k z^{k} -c \approx 0\,,
\end{equation}
which should be imposed on the wave functions in quantum case.

Quantum operators of physical coordinates and momenta satisfy the quantum
brackets, obtained in the standard way from Dirac brackets. The variables of the model (\ref{4N-ph})
satisfy the following quantum algebra
\begin{equation}\label{cB}
[\hat x, \hat p] = i\,, \qquad [\hat z^i, \hat{\bar z}_j] = \delta^i_j \,, \qquad \{\hat\psi^i,
\hat{\bar\psi}_j\}=
{\textstyle\frac{1}{2}}\,\delta^i_j \,.
\end{equation}
Quantum supertranslation and superconformal boost generators defined by the corresponding classical
expressions are
\begin{equation}\label{Q-qu}
{Q}^i =\hat p \hat\psi^i+ 2i\alpha\frac{\hat z^{(i} \hat{\bar z}{}^{k)}\hat\psi_k}{\hat x}+
i(1{+}2\alpha)\frac{\langle\hat\psi_{k} \hat\psi^{k}\hat{\bar\psi}{}^i\rangle}{\hat x}\, ,
\end{equation}
\begin{equation}\label{Qb-qu}
\bar{Q}_i=\hat p\hat{\bar\psi}_i- 2i\alpha\frac{\hat z_{(i} \hat{\bar z}_{k)}\hat{\bar\psi}{}^k}{\hat x} +
i(1{+}2\alpha)\frac{\langle\hat{\bar\psi}{}^{k} \hat{\bar\psi}_{k}\hat\psi_i\rangle}{\hat x}\,,
\end{equation}
\begin{equation}\label{S-qu}
{S}^i =-2\,\hat x \hat\psi^i + t\,{Q}^i,\qquad \bar{S}_i=-2\,\hat x
\hat{\bar\psi}_i+ t\,\bar{Q}_i\,,
\end{equation}
where the symbol $\langle\ldots\rangle$ means Weyl ordering.
Evaluating the anticommutators of the odd generators (\ref{Q-qu}), (\ref{S-qu}), one
determines uniquely the full set of quantum generators of superconformal algebra
$D(2,1;\alpha)$. We obtain
\begin{eqnarray}
{H} &=&{\textstyle\frac{1}{4}}\,\hat p^2  +\alpha^2\frac{(\hat{\bar z}_k \hat z^{k})^2+2\hat{\bar z}_k
{\hat z}^{k}}{4\hat x^2} - 2\alpha    \frac{\hat z^{(i} \hat{\bar z}{}^{k)} \hat\psi_{(i}\hat{\bar\psi}_{k)}}{\hat x^2} \label{H-qu}\\
&& -\, (1{+}2\alpha)
\frac{\langle\hat\psi_{i}\hat\psi^{i}\,\hat{\bar\psi}{}^{k} \hat{\bar\psi}_{k}\rangle}{2\hat x^2}+
  \frac{(1{+}2\alpha)^2}{16\hat x^2}\,,\nonumber
\\ \label{K-qu}
{K} &=&\hat x^2  - t\,{\textstyle\frac{1}{2}}\,\{\hat x, \hat p\} +
    t^2\, {H}\,,
\\ \label{D-qu}
{D} &=&-{\textstyle\frac{1}{4}}\,\{\hat x, \hat p\} +
    t\, {H}\,,
\\ \label{T-qu}
{J}^{ik} &=& i\left[ \hat z^{(i} \hat{\bar z}{}^{k)}-
2\hat\psi^{(i}\hat{\bar\psi}{}^{k)}\right]\,,
\\ \label{I-qu}
{I}^{1^\prime 1^\prime} &=& i\hat\psi_k\hat\psi^k\,,\qquad
{I}^{2^\prime
2^\prime} =
-i\hat{\bar\psi}{}^k\hat{\bar\psi}_k\,,\qquad {I}^{1^\prime 2^\prime}
={\textstyle\frac{i}{2}}\,
[\hat\psi_k,\hat{\bar\psi}{}^k]\,.
\end{eqnarray}
It can be directly checked that the generators~(\ref{Q-qu})--(\ref{I-qu}) indeed obey the (anti)commutation relations of the
$D(2,1;\alpha)$ superalgebra (\ref{D21-QQ})--(\ref{D21-BQ}).

{}For the realization~(\ref{Q-qu})--(\ref{I-qu}) the second-order Casimir operator (\ref{qu-Cas})
of $D(2,1;\alpha)$ is given by the following expression
\begin{equation}\label{qu-Cas-12}
{C}_2={\textstyle\frac{1}{4}}\, \alpha(1{+}\alpha)\Big[(\hat{\bar z}_k \hat z^{k})^2+2\hat{\bar z}_k
\hat z^{k} +1 \Big].
\end{equation}
Thus, on the physical wave function which is subjected to the constraints~(\ref{D0-con})
\begin{equation}\label{q-con}
D^0 \Phi=\hat{\bar z}_i \hat z^i \Phi=c\,\Phi
\end{equation}
(we use the normal ordering for the even SU(2)-spinor operators, with all
operators $Z^i$ standing on the right), the Casimir~(\ref{qu-Cas-12}) takes fixed value.

The Hamiltonian (\ref{H-qu}) and the SL$(2,R)$ Casimir operator
(\ref{q-Cas-3}) can be represented in the quantum case as
\begin{equation}\label{H-qu-g}
{H} =\frac{1}{4}\,\left(\hat p^{\,2}  +\frac{\hat g}{\hat x^2} \right)\,,
\end{equation}
\begin{equation}\label{q-Cas-g}
{T}^2 = {\textstyle\frac{1}{4}}\,\hat g - {\textstyle\frac{3}{16}}\,,
\end{equation}
where
\begin{equation}\label{hat-g}
\hat g \equiv 4\alpha^2{\textstyle\frac{1}{2}}\,\hat{\bar z}_k \hat z^{k}
\left({\textstyle\frac{1}{2}}\,
\hat{\bar z}_k \hat z^{k}+1\right) - 8\alpha \hat z^{(i} \hat{\bar z}{}^{k)} \hat\psi_{(i}\hat{\bar\psi}_{k)}- 2(1{+}2\alpha)
\langle\hat\psi_{i}\hat\psi^{i}\,\hat{\bar\psi}{}^{k} \hat{\bar\psi}_{k}\rangle + {\textstyle\frac{1}{4}}\,(1{+}2\alpha)^2\,.
\end{equation}
The operators (\ref{H-qu-g}) and (\ref{q-Cas-g}) formally look as those given in
the model of ~\cite{AFF}.
However, there is an essential difference. Whereas the quantity $\hat g$ is a
constant in
the model of ~\cite{AFF}, in our case $\hat g$ is an operator which takes fixed, but different,
constant values
on different components of the full wave function.

To find the quantum spectrum of (\ref{H-qu-g}) and (\ref{q-Cas-g}), we
make use of
the realization
\begin{equation}\label{bo-re-Z}
\hat{\bar z}_i=v^+_i, \qquad \hat z^i=  \partial/\partial v^+_i
\end{equation}
for the
bosonic operators $Z^k$ and $\bar{Z}_k$, as well as
the following realization of the odd operators $\Psi^i$, $\bar\Psi_i$
\begin{equation}\label{q-re-Psi}
\hat\psi^i=\psi^i, \qquad \hat{\bar\psi}_i= {\textstyle\frac{1}{2}}\,
\partial/\partial\psi^i\,,
\end{equation}
where $\psi^i$ are complex Grassmann variables. Then, the wave function
is defined as
\begin{equation}\label{w-f}
\Phi=A_{1}+ \psi^i B_i +\psi^i\psi_i A_{2}\,.
\end{equation}

Like in the bosonic limit considered in Sect.\,2, requiring the wave function $\Phi(v^+)$ to be
single-valued results in the condition that
the constant $c$ is integer, $c\in \mathbb{Z}$. We take $c$ to be
positive in
order to ensure a correspondence with
the bosonic limit where $c$ becomes ${\rm SU}(2)$ spin. Then \p{q-con}
tells us that
the wave function $\Phi(v^+)$
is a homogeneous polynomial in $v^+_i$ of the degree $c$:
\begin{equation}\label{w-f-d}
\Phi=A^{(c)}_{1}+ \psi^i B^{(c)}_i +\psi^i\psi_i A^{(c)}_{2} \,,
\end{equation}
\begin{equation}\label{A-irred}
A^{(c)}_{i^\prime} = A_{i^\prime,}{}_{k_1\ldots k_{c}}v^{+k_1}\ldots
v^{+k_{c}} \,,
\end{equation}
\begin{equation}\label{B-irred}
B^{(c)}_i = B^{\prime(c)}_i +B^{\prime\prime(c)}_i=v^+_i B^\prime_{k_1\ldots
k_{c-1}}v^{+k_1}\ldots v^{+k_{c-1}} + B^{\prime\prime}_{(ik_1\ldots
k_{c})}v^{+k_1}\ldots
v^{+k_{c}}\,.
\end{equation}
In~(\ref{B-irred}) we singled out the ${\rm SU}(2)$ irreducible parts
$B^\prime_{(k_1\ldots
k_{c-1})}$ and $ B^{\prime\prime}_{(ik_1\ldots k_{c})}$ of the component wave
functions, with the ${\rm SU}(2)$
spins $(c-1)/2$ and $(c+1)/2$, respectively.

On the same states, the Casimir operators  (\ref{q-Cas-3})
of the bosonic subgroups ${\rm SU}(1,1)$, ${\rm SU(2)}_R$ and ${\rm SU(2)}_L$
take the values given in the Table 1.
\begin{table}[h]
\caption{The values of the Casimirs of the bosonic subgroups and $\frac{i}{4} {Q}^{ai^\prime i}{Q}_{ai^\prime i}$}
\label{tab1}
\begin{center}
\renewcommand{\arraystretch}{2}
\begin{tabular}{|c|c|c|c|c|}
\hline & ${T}^2$ & ${J}^2$ & $\!\!\!{I}^2\!\!\!$ &
$\!\!\!\frac{i}{4}  {Q}^{ai^\prime i} {Q}_{ai^\prime i}\!\!\!$ \\ \hline
$\!\!\! A^{(c)}_{k^\prime}\!\!\!$ & $\frac{\alpha^2(c+1)^2-1}{4}$ & $\frac{(c+1)^2-1}{4}$ &
$\!\!\!\frac{3}{4}\!\!\!$ & $\!\!\! 1+\alpha\!\!\!$ \\ \hline
$\!\!\! B^{\prime(c)}_{k}\!\!\!$ & $\frac{\alpha^2(c+1)^2-2\alpha(c+1)}{4} $ &
$\frac{(c+1)^2-2(c+1)}{4}$ & $\!\!\! 0\!\!\!$ & $\!\!\!\alpha(c+1)\!\!\!$ \\ \hline
$\!\!\! B^{\prime\prime(c)}_{k}\!\!\!$ & $\frac{\alpha^2(c+1)^2+2\alpha(c+1)}{4}$ &
$\frac{(c+1)^2+2(c+1)}{4}$ & $\!\!\! 0\!\!\!$ & $\!\!\! -\alpha(c+1)\!\!\!$ \\ \hline
\end{tabular}\\
\end{center}
\end{table}
For different component wave functions, the quantum numbers $r_0, j$ and
$i$, defined by
\begin{equation}
{T}^2=r_0(r_0-1)\,, \qquad {J}^2=j(j+1)\,,
\qquad {I}^2=i(i+1)\,,
\end{equation}
take the values listed in the Table 2.
\begin{table}[ht]
\caption{The ${\rm SU}(1,1)$, ${\rm SU(2)}_R$ and ${\rm SU(2)}_L$ quantum numbers}
\label{tab2}
\begin{center}
\renewcommand{\arraystretch}{2}
\begin{tabular}{|c|c|c|c|}
\hline & $r_0$ & $j$ & $i$ \\ \hline
     $A^{(c)}_{k^\prime}(x,v^+)$ & $\frac{|\alpha|(c+1)+1}{2}$ & $\frac{c}{2}$ &
$\frac{1}{2}$ \\
\hline
     $B^{\prime(c)}_{k}(x,v^+)$ & $\frac{|\alpha|(c+1)+1}{2} - \frac{1}{2}\,\rm{sign}(\alpha)$ & $\frac{c}{2}
-\frac{1}{2}$ & 0 \\ \hline
     $B^{\prime\prime(c)}_{k}(x,v^+)$ & $\frac{|\alpha|(c+1)+1}{2} + \frac{1}{2}\,\rm{sign}(\alpha)$ &
     $\frac{c}{2}+ \frac{1}{2}$& 0 \\ \hline
\end{tabular}\\
\end{center}
\end{table}
The fields $B^{\prime}_i$ and $B^{\prime\prime}_{i}$ form doublets
of SU(2)$_R$ generated by ${J}^{ik}\,$, whereas the
component fields $A_{i^\prime}=(A_{1},A_{2})$ form a doublet of ${\rm
SU(2)}_L$
generated by
${I}^{i^\prime k^\prime}$. If the super-wave function (\ref{w-f})
is bosonic
(fermionic),
the fields $A_{i^\prime}$ describe bosons (fermions), whereas the fields
$B^{\prime}_i$, $B^{\prime\prime}_{i}$
present fermions (bosons).

Each of the component wave functions  $A_{i^\prime}$, $B^{\prime}_i$,
$B^{\prime\prime}_{i}$  carries an
infinite-dimensional unitary representation of the discrete series of the
universal
covering group of the one-dimensional conformal group SU(1,1). Such representations are characterized
by positive
numbers $r_0$~\cite{Barg,Per}
(for the unitary representations of SU(1,1) the constant $r_0 >0$ must be
(half)integer).
Recall that the basis functions of these representations are eigenvectors of the compact SU(1,1) generator
\begin{equation}
T_0={\textstyle\frac{1}{2}}\,\left(m{K}+
m^{-1}{H}\right),
\end{equation}
where $m$ is a constant of the mass dimension (see the definition in \p{T-vec-def}). The corresponding eigenvalues are
$r=r_0 +n$,
$n\in \mathbb{N}$~\cite{Barg,Per,AFF}.

Let us dwell on some peculiar features of the $D(2,1;\alpha)$ quantum mechanics
constructed.
\begin{itemize}
\item As opposed to the standard ${\rm SU}(1,1|2)$ superconformal mechanics~\cite{IKL2,AIPT,W}, the
construction presented here essentially uses the variables $z_i$ (or $v^+_i$)
parametrizing the two-sphere $S^2$, in addition to the standard (dilatonic) coordinate
$x$.
\item The presence of additional ``(iso)spin'' $S^2$ variables in our construction leads to a
richer quantum spectrum. Besides, the relevant wave functions involve representations of the
two independent
SU(2) groups, in contrast to the ${\rm SU}(1,1|2)$ models of \cite{IKL2,AIPT,W,GLP2,GLP3} where only the SU(2)
realized on fermionic variables really matters.
\item In a contradistinction to the previously considered models, there naturally
emerges a quantization of the conformal coupling constant which is expressed as a SU(2)
Casimir operator,
with both integer and half-integer eigenvalues. This happens already in the bosonic
sector of the model,
and is ensured by the $S^2$ variables.
\item The variables $v^+_i$ in the expansions (\ref{A-irred}) and (\ref{B-irred})
can be identified with a half of the target space
harmonic-like variables $v^\pm_i$ (though without the standard constraint $v^{+ i}v_i^- \sim
const$).
\end{itemize}

Finally, it is worthwhile to mention that the $D(2,1;\alpha)$ invariant ${\cal N}{=}4$ superconformal mechanics
model considered in this section is the one-particle case of the $D(2,1;\alpha)$ invariant Calogero-type
multi-particle models constructed in~\cite{FIL1} (see the following subsection).
An almost identical model of $D(2,1;\alpha)$ invariant mechanics was constructed in~\cite{BK1}.
These two models differ somewhat in their treatment of the semi-dynamical isospin variables however.
In \cite{FIL1}, the {\it gauged\/} $({\bf 4, 4, 0})$ multiplet yields the additional algebraic
constraint \p{D0-con}, which is absent in~\cite{BK1}.
In the quantum theory, this constraint~\p{q-con} fixes the value of the second-order Casimir operator~$C_2$
in \p{qu-Cas-12} and so singles out only one irreducible superconformal representation in the spectrum.
Without this constraint, the space of quantum states of the model~(\ref{4N-ph}) will contain
an infinite tower of irreducible representations.

\subsubsection{Multi-particle models with $D(2,1;\alpha)$ symmetry}

Matrix extensions of the gauged model considered in the previous section lead
to multi-particle systems with ${\cal N}{=}\,4$ superconformal symmetry.
Such models are
described by the following harmonic superspace action
\begin{equation}\label{4N-gau-matrix}
S = -{\textstyle\frac{1}{4(1{+}\alpha)}}\int \mu_H  {\rm tr} \left(
\mathscr{X}^{\,-1/\alpha} \, \right) - {\textstyle\frac{1}{2}}\int
\mu^{(-2)}_A \mathcal{V}_0 \widetilde{\,\mathcal{Z}}{}^{a\,+}
\mathcal{Z}^+_a -{\textstyle\frac{i}{2}}\,c\int \mu^{(-2)}_A \,{\rm
Tr} \,V^{++} \,.
\end{equation}
The first term in~(\ref{4N-gau-matrix}) is the gauged action
of the ({\bf 1,4,3}) multiplets which are described by hermitian
$n{\times}n$-matrix superfields $\mathscr{X}=(\mathscr{X}_a^b)$,
$a,b=1,\ldots ,n$. They are in the adjoint of ${\rm U}(n)$ and are
subject to appropriate gauge-covariant constraints
\begin{eqnarray}
&&\mathscr{D}^{++} \,\mathscr{X}=0, \label{cons-X-g-V}\\
\mathscr{D}^{+}\mathscr{D}^{-} \,\mathscr{X}=0,&&
  (\mathscr{D}^{+}\bar\mathscr{D}^{-} +\bar\mathscr{D}^{+}\mathscr{D}^{-})\,
\mathscr{X}=0\,.
  \label{cons-X-g}
\end{eqnarray}

The second term in~(\ref{4N-gau-matrix}) is a
Wess-Zumino~(WZ) action describing $n$ commuting analytic superfields
$\mathcal{Z}^+_a$ which
are in the fundamental of ${\rm U}(n)$. They represent off-shell ${\cal N}{=}\,4$ multiplets ({\bf 4,4,0}) and are defined by the
constraints \begin{equation}  \label{cons-Ph-g}
\mathscr{D}^{++} \mathcal{Z}^+=0, \qquad \mathscr{D}^{+} \mathcal{Z}^+ =0\,,\qquad
\bar\mathscr{D}^{+} \mathcal{Z}^+ =0\,.
\end{equation}
The constraints~(\ref{cons-X-g-V}), (\ref{cons-Ph-g}) involves the covariant harmonic derivative
$\mathscr{D}^{++} = D^{++} + i\,V^{++}$, where the U(n) gauge matrix connection
$V^{++}(\zeta,u)$ is an analytic superfield. The gauge connections
entering the spinor covariant derivatives in~(\ref{cons-X-g}) are properly expressed
through $V^{++}(\zeta,u)$ \cite{DI1}. The parameters of the ${\rm U}(n)$ gauge group are
analytic, which implies  $\mathscr{D}^{+} = D^{+}\,,\;\bar\mathscr{D}^{+} = \bar D^{+}$. Note that
$\mathscr{X}$ is in the adjoint of ${\rm U}(n)$, so $ \mathscr{D}^{++} \mathscr{X} = D^{++}
\mathscr{X} + i\,[V^{++} ,\mathscr{X}] $, etc.

The third term in~(\ref{4N-gau-matrix}) is a
Fayet-Iliopoulos (FI) term for $V^{++}$ and the real constant $c$
is its strength. Clearly, only the trace part of $V^{++}$ (i.e. U(1) gauge connection) makes contribution to this FI term.
The superfield $\mathcal{V}_0(\zeta,u)$ is a real analytic
gauge prepotential for the ${\rm U}(n)$ singlet ({\bf 1,4,3})
superfield $\mathscr{X}_0 \equiv {\rm tr} \left( \mathscr{X}
\right)\,$. It is defined by the integral transform
\begin{equation}
\mathscr{X}_0(t,\theta_i,\bar\theta^i)=\int du \mathcal{V}_0 \left(t_A, \theta^+,
\bar\theta^+, u^\pm \right) \Big|_{\theta^\pm=\theta^i u^\pm_i,\,\,\,
\bar\theta^\pm=\bar\theta^i u^\pm_i}\,.
\end{equation}

The action~(\ref{4N-gau-matrix}) is invariant under the ${\cal N}{=}\,4$ superconformal group
$D(2,1;\alpha)$. To show this we should use the $D(2,1,\alpha)$
transformation laws given in Sect.\,4.
Once again, this invariance is ensured by the presence of the superfield multiplier $\mathcal{V}_0$ in the second term of the
action~(\ref{4N-gau-matrix}).

The local ${\rm U}(n)$ transformations leaving the action~(\ref{4N-gau-matrix}) invariant are
\begin{equation}\label{tran4}
\mathscr{X}^{\,\prime} =  e^{i\lambda} \mathscr{X} e^{-i\lambda} , \qquad
\mathcal{Z}^+{}^{\prime}
= e^{i\lambda} \mathcal{Z}^+ , \qquad
V^{++}{}^{\,\prime} =  e^{i\lambda}\, V^{++}\, e^{-i\lambda} - i\, e^{i\lambda} (D^{++}
e^{-i\lambda}),
\end{equation}
where $ \lambda_a^b(\zeta, u^\pm) \in u(n) $ is the ``hermitian'' analytic matrix
parameter, $\widetilde{\lambda} =\lambda$. Using this gauge freedom we can choose the WZ
gauge
\begin{equation}  \label{WZ-4N}
V^{++} =2i\,\theta^{+}
  \bar\theta^{+}A(t_A) .
\end{equation}

In what follows we specialize to the case $\alpha=-1/2\,$, which corresponds to the free superconformally invariant
action for $\mathscr{X}$ in (\ref{4N-gau-matrix}).

Inserting the component expressions of the superfields in the action~(\ref{4N-gau-matrix}) and
eliminating auxiliary fields by their
equations of motion we obtain, in the WZ gauge, the component action
\begin{eqnarray}\label{4N-gau-bose-a}
S_4  &=& S_b + S_f,
\\
S_b &=&  \int dt \,\Big[{\rm tr} \left( \nabla X\nabla X +c \,A \right)
+ {\textstyle\frac{n}{8}}(\bar Z^{(i} Z^{k)})(\bar Z_{i} Z_{k}) + {\textstyle\frac{i}{2}}\,X_0 \left(\bar Z_k \nabla Z^k
- \nabla \bar Z_k \, Z^k\right)
\Big],\label{4N-gau-bose-1}
\\
  S_f&=&   i\,{\rm tr} \int dt \left( \bar\Psi_k \nabla\Psi^k
-\nabla\bar\Psi_k \Psi^k
\right)  -\int dt  \,\frac{\Psi^{(i}_0\bar\Psi^{k)}_0 (\bar Z_{i}
Z_{k})}{X_0}\,,\label{4N-gau-fermi-1}
\end{eqnarray}
where
\begin{equation}
X_0:= {\rm tr} (X), \quad\Psi_0^i := {\rm tr} (\Psi^i), \quad\bar\Psi_0^i :=
{\rm tr} (\bar\Psi^i)\,.
\end{equation}

Let us consider the bosonic limit of $S_4$, i.e. the action~(\ref{4N-gau-bose-1}). We can
impose the gauge $X_a^b =0$, $a\neq b$, using the residual invariance of WZ
gauge~(\ref{WZ-4N}): $ X^{\,\prime} =  e^{i\lambda}\, X\, e^{-i\lambda} $,
$Z^{\prime}{}^{k} =  e^{i\lambda} Z^{k}$, $ A^{\,\prime} =  e^{i\lambda}\, A\,
e^{-i\lambda} - i\, e^{i\lambda} (\partial_t e^{-i\lambda})$ where $ \lambda_a^b(t) \in
u(n) $ are ordinary $d{=}1$  gauge parameters. As a result of this, and after eliminating
$A_a^b$, $a\neq b$, by the equations of motion, the action~(\ref{4N-gau-bose-1}) takes the
following form (instead of $Z^i_a$ we introduce the new fields $ Z^\prime{}^i_a =
(X_0)^{1/2}\,Z^i_a$ and omit the primes on these fields),
\begin{equation}\label{4N-bose-fix}
S_{b} = \int dt \Big\{ \sum_{a} \dot x_a \dot x_a + {\textstyle\frac{i}{2}}\sum_{a} (\bar
Z_k^a \dot Z^k_a - \dot {\bar Z}{}_k^a Z^k_a)
+  \sum_{a\neq b} \, \frac{{\rm tr}(S_a S_b)}{4(x_a - x_b)^2} - \frac{n\,{\rm
Tr}(\hat S \hat S)}{2(X_0)^2}\,\Big\}.
\end{equation}
Here, the fields $Z^k_a$ are subject to the constraints~\footnote{Here and in~(\ref{S}) we
do not sum over the repeated index $a$.}
\begin{equation}\label{4N-eq-aa}
\bar Z_i^a Z^i_a =c \qquad \forall \, a \,,
\end{equation}
and carry the residual abelian gauge $[{\rm U}(1)]^n$ symmetry, $Z_a^k \rightarrow
e^{i\varphi_a} Z_a^k\,$, with local parameters $\varphi_a(t)$. In~(\ref{4N-bose-fix}) we
use the following notation:
\begin{equation}\label{S}
(S_a)_i{}^j := \bar Z^a_i Z_a^j,\qquad
(\hat S)_i{}^j := \sum_a \left[ (S_a)_i{}^j -
{\textstyle\frac{1}{2}}\delta_i^j(S_a)_k{}^k\right].
\end{equation}
Note that at $c=0$ the constraint \p{4N-eq-aa} implies $Z^i_a =0$, i.e. a non-trivial
interaction exists only for $c\neq 0$ as in the previous cases. The new feature of the
${\cal N}{=}\,4$ case is that not all out of the bosonic variables $Z^i_a$ are eliminated by
fixing gauges and solving the constraint; there survives a non-vanishing WZ term for them
in eq. (\ref{4N-bose-fix}). After quantization these variables become purely internal
(${\rm U}(2)$-spin) degrees of freedom.

In the Hamiltonian approach, the kinetic WZ term for $Z$ in~(\ref{4N-bose-fix}) gives rise
to the following Dirac brackets:
\begin{equation}\label{DB}
[\bar Z^a_i, Z_b^j]_{{}_D}= i\delta^a_b\delta_i^j.
\end{equation}
With respect to these brackets the quantities~(\ref{S}) for each index $a$ form $u(2)$
algebras
\begin{equation}\label{su-DB}
[(S_a)_i{}^j, (S_b)_k{}^l]_{{}_D}= i\delta_{ab}\left\{\delta_i^l(S_a)_k{}^j-
\delta_k^j(S_a)_i{}^l \right\}.
\end{equation}
The quantities $(\hat S)_i{}^j$ defined in (\ref{S}) are time-independent Noether charges for the ${\rm SU}(2)$
invariance of the system~(\ref{4N-bose-fix}), so the numerator of the term $\sim
(X_0)^{-2}$ in \p{4N-bose-fix} is a constant on the equations of motion for $Z_a^i, \bar
Z^a_i\,$. So, as opposed to the ${\cal N}{=}\,1,2$ cases, the ${\cal N}{=}\,4$ gauged multiparticle action contains
a conformal potential even in the center-of-mass sector (like in \cite{GLP2,GLP3,KLP}). Modulo
this extra conformal potential (last term in~(\ref{4N-bose-fix})), the bosonic limit of the
${\cal N}{=}\,4$ system constructed is none other than the integrable U(2)-spin Calogero
model in the formulation of \cite{Poly-rev}.

It is worthwhile to note that in the considered model, as distinct from the $D(2,1;\alpha)$ invariant $n$-particle models discussed
in Sect. 4.2.2, there are $n$ independent sets of the harmonic-like target variables $Z^i_a, \bar Z^a_i$, while only one set of similar
variables is present in the models of Sect. 4.2.2. It is an open question whether it is possible, from the very beginning, to covariantly
reduce the number of the ``semi-dynamical'' superfields $\mathcal{Z}^+_a$ to one such superfield and to arrive, in this way, at the
superfield formulation of the models of Sec. 4.2.2 for which so far only the component formulation is known.

\subsection{AdS/CFT correspondence in ${\cal N}{=}\,4$ mechanics and black holes}

In the late nineties, an unexpected application of (super)conformal mechanics was discovered.

As shown in \cite{CDKKTP}, the near-horizon bosonic geometry of the extreme
Reissner--Nordstr\"{o}m black hole coincides with that of the manifold  ${\rm AdS}_2\times S^2$.
It was demonstrated in  \cite{CDKKTP}, that the motion of the relativistic
particle with mass $m$ and charge $q$
near the horizon of the extreme Reissner--Nordstr\"{o}m black hole with
large mass $M$, in the limit when the difference $(m-|q|)$ tends to zero, with $M^2(m-|q|)$ being kept
fixed, is described by AFF conformal mechanics (\ref{ac-AFF}).
The radial coordinate of ${\rm AdS}_2\times S^2$ is identified with the conformal
mechanics degree of freedom.

This result gave an explanation of some quantum properties of a particle
moving near the horizon of black hole. This concerns, in particle, the necessity to redefine the Hamiltonian $H\to
T_0$ and  the existence of an infinite number
of Hermitian quantum states \cite{Britto}. The dynamics of superparticle
moving near the extremal black holes in the same limit is in general described  by extended superconformal mechanics \cite{IKL2,AIPT}.

The fact that conformal mechanics describes the specific limit of
the ${\rm AdS}_2\times S^2$ geometry of an extremal black hole is in agreement with the general
concept of AdS/CFT correspondence. In \cite{IKNie} the meaning of this particular AdS/CFT
correspondence was further clarified for the case when the angular $S^2$ degrees of freedom are ``frozen''.
It was shown there that the standard AFF conformal mechanics and
the mechanics describing the radial motion of relativistic particle
near the horizon of extremal black hole are described by two different, but
in fact equivalent nonlinear realizations of
the $d=1$ conformal group ${\rm SO}(2,1)$. Actually, they differ only in the choice
of the parametrization of ${\rm SO}(2,1)$ group elements.
This equivalence holds for any finite value of the
black hole mass, i.e. without assuming any specific limit.

The standard conformal mechanics is based on the algebra (\ref{SL2R}), the group
parametrization (\ref{group-el-cm})
and the constraints (\ref{IKL-const}) incorporated in the action
(\ref{ac-cm-f}).
In \cite{IKNie}, for the description of the radial particle motion near the horizon
of extremal black hole,
it was suggested to use the ``AdS basis'' in the same $so(2,1)$ algebra
\begin{equation}\label{SL2R-AdS}
\hat P = H \, , \qquad \hat K=\frac{1}{\mathcal{R}}\,K-\mathcal{R}H \, ,
\qquad \hat D=\frac{1}{\mathcal{R}}\,D\,,
\end{equation}
where ${\mathcal{R}}$ is AdS radius. The commutation relations of the conformal algebra
(\ref{SL2R}) in this AdS basis  are rewritten as
\begin{equation}\label{SL2R-A}
\left[\,\hat D,\hat P\,\right] = - \frac{i}{\mathcal{R}}\,\hat P \, ,
\qquad \left[\,\hat K,\hat P\,\right]=-2i\hat D \, ,
\qquad \left[\,\hat D,\hat K\,\right]=\frac{i}{\mathcal{R}}\,\hat K
+2i\hat P\,.
\end{equation}
An element of ${\rm SO}(2,1)$ in the AdS basis is defined to be
\begin{equation}\label{group-el-cm-A}
G_0 = e^{i\tau \hat P}\, e^{i\Omega(\tau)\hat K}\, e^{i\phi(\tau)\hat D}\,.
\end{equation}
Like in the standard conformal mechanics, the field $\Omega(\tau)$ can be
eliminated by imposing the inverse Higgs constraint
\begin{equation}\label{IKL-const-A}
\hat\omega_D=0\,,
\end{equation}
and the radial motion near the horizon is described by the action
\cite{IKNie}
\begin{equation}\label{ac-cm-f-A}
S_{\rm AdS}=-m\int\,\hat \omega_P +q \int\,d\tau\, e^{-\phi/\mathcal{R}} \,.
\end{equation}
Second term in (\ref{ac-cm-f-A}) plays the role of cosmological term.
Thus, two conformal mechanics models arise as nonlinear realizations of the same spontaneously broken ${\rm SO}(2,1)$ symmetry,
with the dilaton as the only essential Goldstone field.

In \cite{IKNie}, a direct link between these two seemingly different models was established
through the following invertible change of variables:
\begin{equation}
t=\tau-\mathcal{R}e^{\phi/\mathcal{R}}\Lambda\,,\qquad
u=\frac{\phi}{\mathcal{R}} +\ln(1-\Lambda^2)\,,
\qquad z=\frac{\Lambda}{\mathcal{R}}\,,
\qquad \Lambda:=\tanh \Omega\,.
\end{equation}
Already in \cite{IKNie}, this ``radial'' ${\rm AdS}_2/{\rm CFT}_1$
correspondence was extended to the case of $\mathcal{N}=2$ ${\rm SU}(1,1|1)$ superconformal mechanics.
In \cite{BGIK}, by the explicit canonical transformations in the Hamiltonian formalism, the
equivalence of $\mathcal{N}=4$ superconformal mechanics model and
a charge massive particle propagating near the extremal black hole was shown
for any finite black hole mass and with both the radial and the angular degrees of
freedom of the particle taken into account.
A generalization of this equivalence to the case of the extremal black hole possessing  both
electric and magnetic charges was performed in \cite{Gal}.
Note that the interaction with the magnetic field of black hole is described by WZ
term, much similar to (\ref{bose0}), (\ref{bose01}), (\ref{4N-ph}).

Recently, (super)conformal mechanics was also used to describe other black hole solutions.
This is based on the fact that the isometry group of (diverse) four-dimensional extremal black hole
in the near-horizon limit contains $d{=}1$ conformal group ${\rm SO}(2,1)$ \cite{KunL}.
The extremal Reissner-Nordstr\"om was invoked in an analysis of ${\cal N}{=}\,4$ superconformal
mechanics coupled to several ${\cal N}{=}\,8$, $d{=}1$ vector multiplets, giving the component action,
supercharges and Hamiltonian with all fermionic terms included~\cite{BKSS}.
The extremal Kerr throat solution and the Kerr, Kerr-Newman and Kerr-Newman-AdS-dS black holes
were studied using superconformal mechanics in \cite{Gal10,Gal11,BK2,GalN}. In particular, in \cite{GalN}
there was proposed a general method of how to explicitly define, at the Hamiltonian level,
the canonical transformations related the conformal and AdS bases for various cases of interest.

There are other links of the black hole physics with the models of
extended superconformal mechanics. This concerns describing the set of black holes and clarifies some other
aspects  of the ${\rm AdS}_2/{\rm CFT}_1$ correspondence.
In particular, the authors of the paper \cite{GT} argued that the
large-$n$ limit of the $n$-particle
${\rm SU(1,1|2)}$ superconformal Calogero model may provide a microscopic
description of the extreme Reissner--Nordstr\"om black hole in the
near-horizon limit. This hypothesis is based on the assertion that for
a large number of particles and in the limit when all coordinates of the
Calogero model, except for one, are treated as ``small'', the Calogero model
reduces to the conformal mechanics for this ``allocated''  coordinate.
Other investigations were made in \cite{MS1,MS2,MSS,Britto,Papa}
where it was shown that the moduli spaces of $n$ black holes in four- and five-dimensional
supergravities are described by the sigma-model-type conformal quantum mechanics.
Note that the construction of a self-consistent $n$-body generalization
of black-hole quantum mechanics is a rather complicated problem
beyond the one- and two-body cases. In order to have a normalizable ground
state in the latter cases, one should apply a proper time redefinition,
just as in conformal quantum mechanics~\cite{AFF}. If the general
multi-black-hole quantum mechanics indeed amounts to supersymmetric Calogero models,
one can employ the powerful machinery developed for integrable
super-Calogero systems (see e.g.~\cite{FM,W,Vas1,Vas2,Brink,GLP1,GLP2,GLP3}).

\setcounter{equation}{0}

\section{Sketch of ${\cal N}{>}\,4$ superconformal systems }

So far, most of studies  related to the superconformal mechanics were in fact
limited to the lower ${\cal N}$ extensions, namely
to the ${\cal N}{=}\,1$, ${\cal N}{=}\,2$ and ${\cal N}{=}\,4$ ones (also,
the  ${\cal N}{=}\,3$ case was not investigated, though the corresponding off-shell multiplets for this case
likely coincide, by their component contents, with the ${\cal N}{=}\,4$ ones).
The only ${\cal N}{>}\,4$ superconformal mechanics models for which off-shell actions are known
are a few systems of ${\cal N}{=}\,8$ superconformal mechanics.
Below we give a brief outline of this class of  ${\cal N}{>}\,4$ superconformal mechanics.

Any supergroup underlying an extended superconformal mechanics should, first of all,
contain $d{=}1$ conformal group ${\rm SL}(2,\mathbb{R})$ as its subgroup.
It is the only non-compact bosonic subgroup, while other bosonic subgroups
are compact subgroups of the relevant $R$-symmetries.
The full list of simple supergroups of ${\cal N}$-extended superconformal
quantum mechanics can be found in \cite{Britto}
(more complete information can be found in \cite{Sorba}).
In contrast to the cases of ${\cal N}{\leq}\,4$ where the corresponding
superconformal groups are in fact unique,\footnote{
The degeneracy in the ${\cal N}{=}\,4$ case  is somewhat trivial, since
all possible ${\cal N}{=}\,4$ superconformal groups correspond to different choices of the parameter $\alpha$ in
the supergroup $D(2,1;\alpha)$; an exception is the supergroup ${\rm SU}(1,1|2)$, but it enters as a multiplier
in the semidirect product structure of the supergroup $D(2,1;\alpha)$ at
$\alpha=0$ and $-1$.}
there are four different ${\cal N}{=}\,8$ superconformal group: ${\rm SU}(1,1|4)$,
${\rm OSp}(8|2)$, ${\rm OSp}(4^*|4)$ and
real exceptional supergroup $F(4)$ with the $R$-symmetry subgroup ${\rm
SO}(7)$.
Until now, models of ${\cal N}{=}\,8$ superconformal mechanics have been
found for the supergroup ${\rm OSp}(4^*|4)$
in \cite{BIKL1,BIKL2} and for the supergroup $F(4)$ in \cite{DI3}.\footnote{Some further possibilities are discussed,
from the group-theoretical point of view, in a recent paper \cite{KTo}.}

Also, in ref. \cite{IKL2} on-shell component actions
for the superconformal mechanics models associated with the conformal supergroups ${\rm SU}(1,1|\,{\cal N}/2), {\cal N} \geq 4\,,$
were constructed (see also \cite{AKud,W}). Imposing the appropriate covariant constraints on MC forms which accomplish
the covariant reduction of the conformal supergroups ${\rm SU}(1,1|\,{\cal N}/2)$ to the compact supergroup
${\rm SU}(1|\,{\cal N}/2)\,$,  in \cite{IKL2} there were found the physical component contents of these models  and their equations of motion.
The on-shell action of such a $({\bf 1,{\cal N},?})$ supermultiplet is similar to the action (\ref{4N-1ph-a}),
\begin{equation}\label{N-1ph}
S_{{\cal N} \geq\, 4} = \int dt \,\Big[\dot x\dot x  +i  \left( \bar\psi_k \dot\psi^k -\dot{\bar\psi}_k \psi^k \right) -
\frac{\left(m+\bar\psi_{k} \psi^{k}\right){}^2}{x^2}\Big]  \,,
\end{equation}
where $\psi^k$ is the spinor in the fundamental representation of ${\rm SU}({\cal N}/2)\,$.
As shown in \cite{W}, the generators of  the conformal supergroups ${\rm SU}(1,1|\,{\cal N}/2)$, ${\cal N}{>}\,4$
can be obtained from the ${\cal N}{=}\,4$ generators by directly  replacing the SU(2) spinor  $\psi^k$
by the ${\rm SU}({\cal N}/2)$ spinor.
In particular, the generators of the Poincar\'{e} superalgebra are given by the expressions
\be
Q^i=\psi^i\left(p-2i\,\frac{m+\bar\psi_k\psi^k}{x} \right),\quad
\bar Q_i=\bar\psi_i\left(p+2i\,\frac{m+\bar\psi_k\psi^k}{x} \right),
\ee
\be
H=\frac{p^2}{4}+\frac{\left(m+\bar\psi_k\psi^k\right)^2}{x^2}\,.
\ee
Up to now it is unclear how to recover the system (\ref{N-1ph}) from some off-shell superfield
formalism.

Although the ${\cal N}{=}\,8$ superfield formalism was developed in
\cite{BIKL1,BIKL2}  and ${\cal N}{=}\,8$ superconformal equations of
motion (for some superfields) were obtained through nonlinear realization  in
\cite{BIKL1}, the simplest way of constructing
the ${\cal N}{=}\,8$ superconformal mechanics models is to deal with the ${\cal N}{=}\,4$ superspace formalism and
to represent irreducible ${\cal N}{=}\,8$ multiplets as direct sums of the appropriate ${\cal N}{=}\,4$
multiplets.
This expansion of the ${\cal N}{=}\,8$
superfields in terms of the ${\cal N}{=}\,4$ superfields can be schematically written as the following
splitting \cite{BIKL1,ILS}
\begin{equation}\label{8-4}
({\bf n,8,8{-}n})=({\bf n_1,4,4{-}n_1})\oplus({\bf n_2,4,4{-}n_2})\,,\qquad
{\bf n}={\bf n_1} +{\bf n_2}\,.
\end{equation}

Using the splitting
\begin{equation}\label{8-4a}
({\bf 3,8,5})=({\bf 3,4,1})\oplus({\bf 0,4,4})\,,
\end{equation}
in the paper \cite{BIKL1} there was constructed ${\cal N}{=}\,8$
superconformal Lagrangian for the supermultiplet $({\bf 3,8,5})$.
The ${\cal N}{=}\,4$ supermultiplet $({\bf 3,4,1})$ is described by the
harmonic superfield $L^{++}$ (\ref{con-lm})
whereas the supermultiplet $({\bf 0,4,4})$ is described by the analytic
superfield $\Psi^{+}(\zeta,u), \bar\Psi^+(\zeta,u)$ which
satisfies the constraint (\ref{con-hm}) but is Grassmann-odd superfield contrary
to the ({\bf 4,4,0}) superfield ${\mathcal{Z}}^+(\zeta,u)\,$.
As shown in \cite{BIKL1}, the full ${\cal N}{=}\,8$  superconformal
invariance requires the values $\alpha=1$ in
the ${\cal N}{=}\,4$  superconformal symmetry $D(2,1;\alpha)$ and so it is ${\rm OSp}(4^*|2)$.
It is easy to see that only ${\rm OSp}(4^*|4)$ supergroup can contain this ${\cal N}{=}\,4$
superconformal group ${\rm OSp}(4^*|2)$ as a subgroup.

The bosonic part of the component action,
after  making some field redefinitions and elimination of the auxiliary fields, can be written in the following concise form
\be
S_B^{(3,8,5)} = \int dt \left[e^q (\dot{q})^2 - m^2 e^{-q} + 4 e^q \frac{\dot\lambda\dot{\bar\lambda}}{(1 + \lambda\bar\lambda)^2}
- 2im \frac{\lambda\dot{\bar\lambda} - \dot\lambda\bar\lambda}{1 + \lambda\bar\lambda}\right], \lb{3851}
\ee
where the parameter $m$ firstly appears in the ${\cal N}{=}\,4$ superfield action of the model as the coefficient before the WZ term
of the multiplet $({\bf 3,4,1})$. This term should be necessarily added for ensuring ${\cal N}{=}\,8$ supersymmetry of the total action.
The action \p{3851} is the sum of two conformally invariant actions, that of conformal mechanics (the first two terms), and that of a charged
particle moving in the Dirac monopole background (the last two terms, the first one being a conformally-covariantized action
of the $d{=}1$ SU(2)$/$U(1) sigma model and the second one the relevant $d{=}1$ WZ term). Thus the bosonic sector of this ${\cal N}{=}\,8$
superconformal mechanics coincides with the bosonic sector of the ${\cal N}{=}\,4, D(2,1;\alpha)$ superconformal mechanics of refs.
\cite{IKLech,IL} for the choice $\alpha = 1\,$.

Another ${\cal N}{=}\,4$ superfield splitting which yields the same ${\cal N}{=}\,8$ superconformal model is
\begin{equation}\label{8-4c}
({\bf 3,8,5})=({\bf 1,4,3})\oplus({\bf 2,4,2})\,,
\end{equation}
where the first supermultiplet is described by the ${\cal N}{=}\,4$ superfield $v$ subjected to the $\textrm{SU}(1,1|2)$
covariant constraints \p{cons-X-rel1} (with $m \rightarrow -2m$)
and the second one is the ${\cal N}{=}\,4$ chiral multiplet described by a complex superfield $\varphi\,,$ $D_i\varphi
= \bar D^i \bar\varphi = 0\,$. The ${\cal N}{=}\,4$ superfield action in this case has the form \cite{BIKL1}
\be
S^{(3,8,5)} = -\frac{1}{4}\int dt d^4\theta \left[v \log \left(v +\sqrt{v^2 + 4\varphi\bar\varphi}\right) - \sqrt{v^2 +
4\varphi\bar\varphi} \right].
\ee
It has the manifest ${\rm SU}(1,1|2)$ superconformal symmetry and hidden ${\cal N}{=}\,8$ Poincar\'e supersymmetry which close on the same ${\cal N}{=}\,8$
superconformal group ${\rm OSp}(4^*|4)$ which includes ${\rm SU}(1,1|2)$ as one of its subgroups, along with ${\rm OSp}(4^*|2)\,$.
The bosonic component action of course coincides with \p{3851} (after field redefinitions and elimination of auxiliary fields). The constant
$m$ which specifies the strength of both the conformal potential and WZ term comes now from the constraint on $v\,$, while in the
previous splitting option it explicitly enters he superfield action.

In the same paper \cite{BIKL1}, there also was considered a model of ${\cal N}{=}\,8$ superconformal mechanics based on the off-shell
supermultiplet ({\bf 5,8,3}). This model respects the same superconformal group ${\rm OSp}(4^*|4)$ and have two equivalent
${\cal N}{=}\,4$ superfield descriptions based upon the splittings
\be
(\mbox{a}) \;\;({\bf 5,8,3})=({\bf 1,4,3})\oplus({\bf 4,4,0})\,, \qquad (\mbox{b}) \;\;({\bf 5,8,3})=({\bf 3,4,1})
\oplus({\bf 2,4,2})\,.
\ee
The action is most simple in the second case, with the constituent ${\cal N}{=}\,4$ multiplets being described, respectively, by the
real SU(2) vector superfield $W^{(ik)}\,,$ $D^{(i} W^{kl)} = \bar D^{(i} W^{kl)} = 0\,,$ and by chiral superfields $\Phi\,, \bar\Phi\,,$
$D^i\Phi = \bar D^i \bar\Phi = 0\,$. The superconformal action reads
\be
S^{(5,8,3)} = 2 \int dt d^4\theta \frac{\log \left(\sqrt{W^2} + \sqrt{W^2 + \frac{1}{2}\Phi\bar\Phi} \right)}{\sqrt{W^2}}\,.
\ee
In the bosonic sector it yields a particular (conformally invraiant) type of the SO(5) invariant $d{=}1$ sigma models of ref. \cite{diac}:
\be
S^{(5,8,3)}_B = \int dt \frac{\dot{W}^{ik}\dot{W}_{ik} + \frac{1}{2}\dot\Phi\dot{\bar\Phi}}{\left(W^2 + \frac{1}{2}\Phi\bar\Phi \right)}\,,
\ee
where the involved quantities are first components of the original superfields.\footnote{After field redefinitions, this action
can be represented as a sum of the standard conformal mechanics action for the dilaton (the radial part of the 5-vector
$(W^{ik}, \Phi, \bar\Phi)$) without potential term and some conformally invariant sigma model action for the remaining four angular variables.}
The ${\cal N}{=}\,4$ superconformally invariant WZ action for the multiplet ({\bf 3,4,1}) can be extended to ${\cal N}{=}\,8$
superconformally invariant action, thus producing conformal potentials for the bosonic fields.

An example of ${\cal N}{=}\,8$ superconformal mechanics based on a different ${\cal N}{=}\,8$ superconformal group was constructed in \cite{DI3}.
It is based on the ${\cal N}{=}\,8$ off-shell supermultiplet ({\bf 1,8,7})~\footnote{
The general ${\cal N}{=}8$ supersymmetric
component action of this multiplet was constructed in \cite{KRT}.}
with the following ${\cal N}{=}\,4$ superfield splitting
\begin{equation}\label{8-4b}
({\bf 1,8,7})=({\bf 1,4,3})\oplus({\bf 0,4,4})\,.
\end{equation}
The supermultiplet $({\bf 1,4,3})$ is described by the real harmonic
superfield ${\mathscr{X}}(t,\theta^\pm, \bar\theta^\pm, u)$
(\ref{cons-X-g-V-1}), (\ref{cons-X-g-1}) or (\ref{cons-X-g1}).
The supermultiplet $({\bf 0,4,4})$ is described by the odd analytic
superfield $\Psi^{+}(\zeta,u)$
as above. It was shown that the $D(2,1;\alpha)$-invariant action is
\begin{equation}\label{8N}
S^{N=8} =-\textstyle{\frac{1}{4(1{+}\alpha)}}\,\displaystyle{\int} \mu_H \,
\mathscr{X}^{\,-1/\alpha} + {\textstyle\frac{1}{2}}\,\int \mu^{(-2)}_A \,
\mathcal{V}\,
\bar{\Psi}{}^+\, {\Psi}^+\, ,
\end{equation}
and it consists of the superconformal action (\ref{4N-X1}) of the ({\bf 1,4,3}) multiplet
and Wess--Zumino-type action of the form (\ref{4N-WZ-1}). This action is
invariant under full
${\cal N}{=}\,8$  superconformal transformations only for $\alpha=-1/3$.
The only ${\cal N}{=}\,8$  superconformal group into which one can embed
$D(2,1;-1/3)$ is the exceptional ${\cal N}{=}\,8$  superconformal group $F(4)$.
It is still unclear whether it is possible to generate a conformal potential for the physical
scalar field without breaking this underlying $F(4)$ superconformal symmetry.

\setcounter{equation}{0}

\section{Conclusion}

In this paper we have presented a particular view
of the current state of superconformal mechanics.
We acknowledge that our view is biased and incomplete
due to the vast amount of literature on modern research in this area.
We have tried to combine two approaches employed in the study of these models.
The first one starts with a {\it classical\/} (super)field description,
based on the Lagrangian formulation of the system.
In this approach, the superconformal symmetry is transparent but the quantization
often faces significant challenges due to the generic nonlinear character of the systems.
The second approach starts from a {\it quantum\/} description,
studying superconformal quantum systems algebraically.
In this case, the geometrical interpretation becomes obscure,
even after a final reconstruction of the relevant quantum superconformal algebras.
It must be admitted that most articles on superconformal mechanics settle only on one
of the two approaches.
However, to overcome the problems manifest in either treatment, it is desirable
to analyze systems with superconformal symmetry by combining both approaches,
i.e.~starting from the analysis of the symmetries of the classical action
and ending with the quantum picture, with a construction of the quantum superconformal
algebra and uncovering the physical spectrum.
Such a strategy will allow for a better understanding of conformal and superconformal mechanics
and perhaps facilitate a more widespread application of these nice theories.

Let us try to predict the research development on superconformal mechanics in the near future.
In our opinion, the following (approximate) problems will be addressed predominantly:
\begin{itemize}
\item
Construction of extended superconformal (quantum) mechanics with isospin degrees of freedom
and with diverse dynamical supermultiplets
\item
Study of various many-particle systems with superconformal symmetry,
including those with isospin degrees of freedom;
obtaining versions of spin-Calogero systems and entirely new ones; analysis of the integrability properties
of these systems
\item
Deeper understanding of the role of superconformal systems, beyond the simplest cases,
in the AdS/CFT correspondence and in the physics of black holes
\item
Understanding of the significance of the (iso)spin variables in superconformal mechanics
(are they just angular degrees of freedom in physical problems, or truly spin ones?)
\item
Constructing new superconformal systems with ${\cal N}\,{>}\,4$ supersymmetry and seeking
out their applications.
\end{itemize}

\section*{Acknowledgements}

\noindent
The authors would like to thank Sergey Krivonos for valuable remarks.
E.I. {\&} O.L. thank the co-authors of their papers on (super)conformal mechanics.
We acknowledge support from a grant of the Heisenberg-Landau Programme and RFBR
grants 09-01-93107, 09-02-01209, 11-02-90445 (S.F. {\&} E.I.), as well as
from the DFG-RFBR collaboration grant 436 RUS 113/669 (new: LE 838/12).
S.F. {\&} E.I. thank the Institute of Theoretical Physics at Leibniz Universit\"at Hannover
for warm hospitality. E.I. thanks SUBATECH, University of Nantes, for kind hospitality
at the final stages of this work.

\end{document}